\documentclass[12pt]{article}

\usepackage{scicite}
\usepackage{graphicx}
\usepackage{enumitem}
\usepackage[labelsep=quad]{caption}

\usepackage{times}

\newcommand{\Cal}[1]{\mathcal{#1}}
\makeatletter
\renewcommand{\fnum@figure}{{\bf Fig. \thefigure}}
\renewcommand{\fnum@table}{{\bf Table \thetable}}
\makeatother

\newenvironment{sciabstract}{%
\begin{quote} \bf}
{\end{quote}}

\title{Tunable chiral anomalies and coherent transport on a honeycomb lattice}

\author
{Vasil~A.~Saroka,$^{1,2,3\ast}\dagger$ Fanmiao~Kong,$^{4}$ Charles~A.~Downing,$^{5}$\\ Renebeth~B.~Payod,$^{6}$ Felix~R.~Fischer,$^{7,8,9,10}$ Xiankai~Sun,$^{11}$\\ Lapo~Bogani,$^{4}$\\
\\
\normalsize{$^{1}$Center for Quantum Spintronics, Department of Physics,}\\ 
\normalsize{Norwegian University of Science and Technology}\\
\normalsize{NO-7491, Trondheim, Norway}\\
\normalsize{$^2$  Institute for Nuclear Problems, Belarusian State University,}\\
\normalsize{Bobruiskaya 11, 220030 Minsk, Belarus}\\
\normalsize{$^3$ TBpack Ltd., 27 Old Gloucester Street,}\\
\normalsize{London, WC1N 3AX, United Kingdom}\\
\normalsize{$^{4}$ Department of Materials, University of Oxford,}\\
\normalsize{16 Parks Road, OX1 3PH Oxford, United Kingdom}\\
\normalsize{$^{5}$  Department of Physics and Astronomy, University of Exeter,}\\
\normalsize{Exeter EX4 4QL, United Kingdom}\\
\normalsize{$^{6}$  Institute of Mathematical Sciences and Physics, University of the Philippines,}\\
\normalsize{Los Ba\~{n}os, Laguna 4031, Philippines}\\
\normalsize{$^7$ Department of Chemistry, University of California,}\\
\normalsize{Berkeley, CA 94720, USA}\\
\normalsize{$^8$ Materials Sciences Division, Lawrence Berkeley National Laboratory,}\\
\normalsize{Berkeley, CA 94720, USA}\\
\normalsize{$^9$ Kavli Energy NanoSciences Institute at the University of California Berkeley}\\
\normalsize{and the Lawrence Berkeley National Laboratory,}\\
\normalsize{Berkeley, California 94720, USA}\\
\normalsize{$^{10}$ Bakar Institute of Digital Materials for the Planet,}\\
\normalsize{Division of Computing, Data Science, and Society, University of California,}\\
\normalsize{Berkeley, CA 94720, USA.}\\
\normalsize{$^{11}$ Department of Electronic Engineering, The Chinese University of Hong Kong,}\\
\normalsize{Shatin, New Territories, Hong Kong SAR, China}\\
\\
\normalsize{$^\ast$To whom correspondence should be addressed; E-mail:  40.ovasil@gmail.com.}\\
\normalsize{$\dagger$Present address:  Department of Physics, University of Rome Tor Vergata and INFN,}\\
\normalsize{Via della Ricerca Scientifica 1, 00133 Roma, Italy}
}

\date{}

\begin{document} 


\baselineskip24pt


\maketitle


\begin{sciabstract}
 The search for energy efficient materials is urged not only by the needs of modern electronics but also by emerging applications in neuromorphic computing and artificial intelligence. Currently, there exist two mechanisms for achieving dissipationless transport: superconductivity and the quantum Hall effect. Here we reveal that dissipationless transport is theoretically achievable on a honeycomb lattice by rational design of chiral anomalies tunable without any magnetic fields. Breaking the usual assumption of commensurability and applying an external electric field lead to electronic modes exhibiting chiral anomalies capable of dissipationless transport in the material bulk, rather than on the edge. As the electric field increases, the system reaches a cubic-like dispersion material phase. While providing performance comparable to other known honeycomb lattice-based ballistic conductors such as an armchair nanotube, zigzag nanoribbon and hypothetical cumulenic carbyne, this scheme provides routes to a strongly correlated localization due to flat band dispersion and to exotic cubic dispersion material featuring a pitchfork bifurcation and a critical slowing down phenomena. These results open a new research avenue for the design of energy efficient information processing and higher-order dispersion materials.
\end{sciabstract}


\section*{Introduction}

Dissipationless charge transport is possible through superconductivity~\cite{Onnes1911} and the quantum Hall effect~\cite{Klitzing1980}. Both phenomena have been reported for honeycomb lattice materials. Mott-insulating and superconducting states have recently been found in a moir{\'e} superlattices of graphene bilayers at the magic angle of $1.1^{\circ}~$ \cite{Cao2018a,Cao2018b}. The half-integer quantum Hall effect has been reported for graphene~\cite{Novoselov2005}. The former boils down to flat band and it carries a lot of similarities with strongly correlated high temperature superconductors, while the latter relies on gapless edge mode generation via time-reversal symmetry breaking with magnetic field. These phenomena experimentally work at low temperatures. Superconducting Cooper pairs are fragile not only with respect to temperature ($>30$~K), but also magnetic field ($> 0.2-20$~T), magnetic impurities~\cite{Davydov1990} and non-magnetic disorder~\cite{Gantmakher2010}. The Hall conductance quantization is impressively stable against disorder, but its observation still requires low temperatures ($\sim 4$~K) or very high magnetic fields ($\sim 30$~T)~\cite{Novoselov2007}.

One alternative way to achieve dissipationless charge transport is suggested by topological quantum matter. It has been realized that Haldane’s model of the anomalous quantum Hall effect, when combined with spin-orbit interactions, allows ballistic transport via helical states, that makes the quantum spin Hall (QSH) effect possible even at zero magnetic field~\cite{Qi2011a}, see Fig.1A. This problem has been also approached from a different route by making use of the internal valley degree of freedom produced by the honeycomb lattice~\cite{Ren2016}.
The reciprocal space of the honeycomb lattice features two non-equivalent Dirac points conventionally denoted as {\bf K} and {\bf K}$^{\prime}$. These regions of the reciprocal space with linear energy bands centered at  {\bf K} and {\bf K}$^{\prime}$ points are called valleys. 
The zigzag edges of the honeycomb lattice host the edge states that are well-separated in the reciprocal space. However, these edge states do not transverse the bulk energy gap. This can be seen by applying a staggered potential that opens the band gap between the edge states. Nevertheless, one can make the edge states to cross the bulk energy gap by creating a kink in the staggered sublattice potential~\cite{Wang2021}. By creating two kinks via back gate potential transversing the ribbon, a valley Hall regime can be achieved~\cite{Liu2021a}. In this regime, for any given edge state and valley, a counter propagating state is found only at the opposite edge of the ribbon. Hence, in the absence of the short range scatters causing intervalley mixing, the backscattering is suppressed. The longitudinal electrical current along the ribbon causes a transverse valley current -- the quantum valley Hall (QVH) effect, see Fig.~1A. The property of the edge modes which locks their direction of propagation to the edge, spin or valley is referred to as chirality or helicity. Engineering gapless chiral edge modes without external magnetic field is an important milestone for the adaptation of quantum Hall dissipationless transport to low-powered electronic circuits.

With respect to the transport properties and energy bands projected on the transport direction, solids can be described as either gapped or gapless, including the recent topological materials~\cite{BookAshcroft1976}. Gapped materials like the semiconducting Si and GaAs and the insulating diamond and SiO$_2$, have a common trait where their projected conduction and valence bands exhibit extrema at the band edges. Around the band edge, an adequate expansion of the energy band function requires at least second order terms, therefore the energy band dispersion is quadratic for gapped structures. On the other hand, gapless materials such as the typical metals and the interfaces of topological insulators have projected bands that cross  the Fermi level so that they possess zero-energy states (ZES) pinned at the Fermi energy. In the ZES vicinity, the energy band function has at least a first order expansion resulting to a linear energy band dispersion. Thus, we can classify solids into two major classes having either quadratic or linear dispersions of the projected bands. Non-conventional cases aside from the ones described above include the strongly correlated Mott insulators like the NiO and MnO which exhibit dispersionless flat bands~\cite{Rodl2009,Gebhardt2023}, the Group~V semimetals such as bismuth with quadratic dispersion~\cite{Issi1979}, and the Dirac~\cite{Novoselov2004} and Weyl~\cite{Lv2015} semimetals that exhibit a linear dispersion relation in the vicinity of the Fermi level. Materials with well-defined cubic or higher order projected dispersions have not yet been reported.

In this paper, we search for metallic nanostructures favorable for the dissipaitonless quantum transport and find an example of a completely flat band that is an eigenmode of a chiral symmetry operator and that can be transformed into a valley chiral charge conserving gapless mode exhibiting cubic dispersion. This transformation to a gapless chiral mode is driven by an external electric field allowing for continuous switching between flat band, i.e. strongly correlated insulator-like, and gapless, i.e. topological insulator-like dissipationless transport, regimes. In strike difference to topological insulators, however, the reported dissipationless transport is realized by bulk rather than edge states, see Fig.~1A.
The onset of a new material phase with cubic dispersion relation is manifested by a transitory state of the valley chiral mode that demonstrates distinctive signatures of pitchfork bifurcation.
These phenomena offer unique possibility to investigate a broad variety of quantum transport regimes in unconventional topological materials, such as a half-bearded graphene nanoribbon (hbGNR).
We unveil that within the graphene nanoribbons (GNRs) incommensurate with the bulk graphene the said ribbon is an iconic representative of a unique class exhibiting dissipationless transport and cubic dispersion phase. Next, we introduce the notion of incommensurability from the point of graph theory and then we design the features outlined above.

\section*{Main results}
\paragraph*{Incommensurability and strong pairing.} The hbGNR is incommensurate with the bulk graphene. This means the hbGNR unit cell cannot be “tiled” with an integer number of graphene unit cells. This statement can be formulated in terms of graph theory starting from a celebrated Su-Schrieffer-Heeger (SSH) model. A periodic SSH chain presented in Fig.~1B consists of a bipartite unit cell containing two atoms that are denoted as A (white) and B (gray). The chain is characterized by two interatomic distances $d_1$ and $d_2$ and the corresponding hopping integrals that are functions of $d$'s. The ratio between $d$'s determines the topological phase of this system. The trivial and non-trivial topological phases are characterized by 1D ${\bf Z}_2$ invariant that takes two distinct values $0$ (trivial) and $1$ (non-trivial)~\cite{BookAsboth2015,Fuchs2021}. When $t_1(d_1) > t_2(d_2)$, the infinite SSH chain is a trivial insulator with $\nu = 0$, but for $t_1(d_1) < t_2(d_2)$ it is a topological insulator with $\nu = 1$. The bulk-boundary correspondence implies that a boundary ZES forms at the interface between regions with different $\nu$'s. It is easy to verify that, there are no ZESs, when a strong paring occurs between intracell atoms ($t_1 > t_2$) and the SSH chain is truncated between the bulk unit cells. However, as shown in Fig.~1C, a pair of ZESs localized at the edges of the chain arises when $\nu = 1$ and $t_1 < t_2$. However, this consideration, that is valid for an {\it even} number of atoms in the SSH chain, fails for an {\it odd} number of atoms. As seen from Fig.~1D, independently of the $\nu$, the odd SSH chain features a sole {\it ubiquitous} ZES that localizes either on one edge or another. Here the system with two edges hosts a single ZES. Both {\it even} and {\it odd} SSH chains featuring ZESs can be interpreted as incommensurable structures since a pair of strongly bonded atoms chosen as a SSH model unit cell cannot fully tile these structures. At the same time, the even SSH chain, that lacks ZESs, can be grouped into an integer number of strongly bonded pairs representing a structure that is commensurable with the periodic SSH unit cell.

\paragraph*{Perfect matchings and topological phases.} An intuitive notion of commensurable and incommensurable SSH chains allow for a precise description in terms of graph theory \textit{matchings}~\cite{BookLovasz1986}. Consider a finite SSH chain as an undirected regular multigraph $G(V,D)$, with $V$ and $D$ being vertixes and edges, where parallel edges between the vertexes are used to represent strong pairing between the atoms, while single edges are used to represent inter pair connections (see Fig.~1E). The graph is regular if each vertex has the same the number of connected edges -- the vertex degree. Hence, we assign an alternating pattern of single and parallel edges so that to preserve the fixed vertex degree for all vertexes. This is, however, impossible at the end vertixes of the graph. Therefore, we introduce {\it ghost edges} that are the triangular features in Fig.~1E. A ghost edge attaches to a sole vertex. It contributes into the vertex degree, but does not contribute into the {\it adjacency matrix} of the graph. Thus, an even SSH chain corresponds to a graph with each vertex end being supplemented with a ghost edge as presented in Fig.~1E so that the regular vertex degree is $3$, i.e. the vertex degree is $n+1$, where $n$ is the number of nearest neighbors in the interior of the system. If both end vertexes of the even SSH chain graph are supplemented with $2$ ghost edges, as shown in Fig.~1E for the incommensurate case, the regular vertex degree $=3$ can be preserved only if the non-adjacent parallel edges change their pattern. The two patterns of the parallel edges are, in fact, {\it matchings}~\cite{BookLovasz1986} of the simple underline graph shown in Fig.~1E. The commensurate even SSH chain has a parallel edge distribution following a {\it perfect matching} of the simple underline graph and it corresponds to a trivial insulator of periodic SSH chain. In contrast, in the incommensurate even SSH chain, the parallel edges are distributed according to a {\it maximal matching} of the underline simple graph (but neither to a {\it maximum} nor to {\it perfect matching}) and this corresponds to a 1D ${\bf Z}_2$ topological insulator of periodic SSH chain.
In other words, a {\it perfect matching} in the graph is an indicator of the \textit{trivial topological phase}. 
For any finite simple graph with even number of vertixes the sufficient and necessary condition for the existence of a perfect matching is given by Tutte’s theorem~\cite{Tutte1947}. This can be applied to identify if a topological phase transition is possible in any complex graphs, which includes those beyond condensed matter systems such as neural networks in deep learning and neuromorphic computing~\cite{Schuman2022}. The given graph-theoretic scheme naturally covers the case of the {\it odd} SSH chain. Since graphs with odd number of vertexes lack {\it perfect matchings}, the odd SSH chains are intrinsically incommensurate or, for the brevity of notation, simply {\it incommensurate}. The given theoretical formulation bridges 1D ${\bf Z}_2$ topological insulators with topologically frustrated polyaromatic hydrocarbons~\cite{Mishra2020a}. It sheds light on a recent tailoring of topological order to obtain metallic structures~\cite{Cirera2020,Rizzo2020a,McCurdy2023}.

\paragraph*{Nullity theorem.} Based on the above consideration we can suggest an analog of the bulk-boundary correspondence in counting the number of in-gap states in non-trivial topological phases. For simple bipartite hexagonal graphs, the number of zero-energy eigenvalues of the graph adjacency matrix is given by the nullity theorem~\cite{Fajtlowicz2005}: $\eta = \alpha - \beta$, where $\alpha$ and $\beta$ are the maximum numbers of pairwise nonadjacent vertices and edges of $G$, respectively. While the nullity theorem is valid for even SSH chain in the commensurate case, it fails to count the number of zero-energy eigenvalues for the even SSH chain in the incommesurate case presented in Fig.~1E. On the other hand, the number of unsaturated vertices for any maximal matching (i.e. it covers maximum and perfect matchings too) is observed to be a better criterion: 
{\it the eigenspectrum of the adjacency matrix of a multigraph contains in-gap zero or quasi-zero-energy eigenvalues if the pairwise non-adjacent parallel edges follow a non-perfect matching pattern. The number of such in-gap eigenvalues is equal to the number of unsaturated vertexes in the multigraph.} 
The generality of this principle is demonstrated for two multigraphs in Fig.~1F hosting the two different matching patterns -- perfect and maximal ones -- which are found in graphene and phosphorene quantum dots (QDs). 
This result is in agreement with sophisticated multi-parameter calculations for monolayer phosphorene QDs~\cite{Saroka2017f}.

\paragraph*{Incommensurate graphene nanoribbons.} Now we apply the above principles for analysis of one-dimensional honeycomb lattice structures -- GNRs. An infinite regular multigraphs with the vertex degree $= 4$ can be constructed for zigzag and bearded GNRs. The only way to get a multigraph with a perfect matching pattern for zigzag GNR (ZGNR) is to introduce a parallel double edge within the unit cell by attaching a single ghost edge to the outmost vertex of ZGNR, as shown in Fig.~1G, and then to continue the pattern of alternating single and parallel double edges until the opposite boundary vertex is reached for which a single ghost edge is introduced. A similar pattern of perfect matching can be obtained for a bearded GNR. However, one needs to attach two ghost edges to the outmost vertexes on both sides of the graph for a bearded GNR as depicted in Fig.~1G. Since perfect matchings are constructed for both graphs, the in-gap states are not guaranteed in both zigzag and bearded GNRs so that both structures can be classified as commensurate ones. To ensure stable in-gap states, a non-trivial topology of the system is needed. As shown in Fig.~1G, the combination of zigzag and bearded edges breaking the {\it inversion symmetry} in a hbGNR makes perfect matching impossible. Hence, the hbGNR is intrinsically incommensurate. In other words, a topological in-gap mode is expected by considering only the graph topology.

\paragraph*{Flat zero-energy modes.} As shown in Fig.~2A, for the hbGNR with $N_r = 51$ atoms in the unit cell, a perfectly flat zero-energy mode (ZEM) is observed at the Fermi level (see also supplementary text~S1 and~S2 and fig.~S1). The flat character is favorable for strong many-body effects. It shall drive this system into a strongly correlated regime since any weak Coulomb interaction is large compared to the width of the perfectly flat band. This ZEM also possesses topological stability such that it does not depend on the details of the physical model and it cannot be removed from the gap unless the ribbon  structure is modified. In a graph theoretical sense presented above, the ZEM is topological. However, being obtained from pure graphene, it originates neither from the Chern nor 2D ${\bf Z}_2$ topological invariant. The valley Chern number is not its reason, too, since the staggered sublattice potential is absent. In fact, incommensurate hbGNRs exhibit distinct behaviour even when driven into the said topological phases (see supplementary text~S3). For all the three flavors of topological insulators, the bulk boundary correspondence should be interpreted with care when dealing with incommensurate systems. In particular, the standard phase diagrams need accounting for the chemical potential (see figs.~S2-S4 and their discussion in supplementary text~S3). Nevertheless, the wave function of hbGNR ZEM does exhibit several peculiar similarities to the states that are characterized by the topological invariants mentioned above.

The first peculiarity of ZEM is its chirality. By chirality we mean the asymmetric distribution of the electron density for the band state wave functions. In Fig.~2A, both the inverse participation ratio (IPR) and the electron density show the shift of localization of ZEM from the bearded to zigzag edge as $k$ changes from $0$ to $\pi$  (see methods for the IPR definition). Thus, ZEM possesses a well-defined chirality everywhere except at the Dirac {\bf K} point, $k = 2\pi/3$, where the chirality is ill defined since the wave function extends over the whole width of the ribbon. As can be seen from the IPR color scheme of Fig.~2A, the localization on a zigzag edge is higher than that on the bearded. The hbGNR ZEM density behaviour differs from that of a ZGNR having a unit cell with $N=52$ atoms, where the wave function is highly localized on both edges, as presented in Fig.~1B. The hbGNR ZEM localization is similar to the quantum valley Hall regime of the ZGNR presented for comparison in Fig.~1C. Thus, the ZEM of the hbGNR is chiral. This result is further supported by the continuum ${\bf k}\cdot {\bf p}$-model. The ZEM is
the Dirac equation solution that is an eigenstate of the chiral symmetry operator given by the Pauli matrix $\sigma_z$: either $\Psi_{+}(x) = (1, 0)^{\mathrm{T}} \sqrt{\kappa_y/\sinh (\kappa_y L)}\, \exp\left(-\kappa_y x\right)$ or $\Psi_{-}(x) = (0, 1)^{\mathrm{T}} \sqrt{\kappa_y/\sinh (\kappa_y L)}\, \exp\left(\kappa_y x\right)$, where $\kappa_y$ is the electron wave vector relative to the Dirac point and $L$ is the ribbon width (see supplementary text~S4 and fig.~S5). In contrast to a celebrated fermion-$1/2$ soliton~\cite{Jackiw1976}, the ZEM solution is unbounded to its topological defect which can be seen as a 1D analog of a 2D vortex of the Kekul{\'e} pattern~\cite{Hou2007}. This peculiarity is acquired because ZEM is normalized in a finite space determined by the ribbon width rather than in the whole space, as required for the soliton. We stress that such solutions are perfectly fine in any bounded systems, therefore, under certain conditions they might even result in some skin-effects for astrophysical superconducting strings~\cite{Witten1985}.

The second peculiarity of hbGNR ZEM is the sublattice (pseudo-spin) polarization throughout the whole Brillouin zone as clearly seen in Fig.~2A. These numerical results are perfectly described by the given continuum model solutions and are further supported by the analytical tight-binding model (see supplementary text~S2). The continuum model considered from the non-linear dynamics and index theory viewpoint also reveals an unconventional topological nature of this flat sublattice polarized mode. From this point of view, however, the mode is characterized by a zero winding number but unlike the bulk modes it resides in the vicinity of the unstable saddle fixed point of the Dirac equation vector field  (see supplementary text~S4.3 and figs.~S6 and~S7).

\paragraph*{From flat modes to chiral anomalies.} Although the flatness of the mode is favorable for strong correlations and for macroscopic coherent effects, such as superconductivity or ferromagnetism, it is detrimental for the conventional electron transport. Hence, we speculate the possibility of a dispersive in-gap mode arising form the sublattice mixing introduced via an external electric field~\cite{Saroka2015a}. Such a system can be considered as the long-sought after 1D metal stable against the Peierls and Jahn–Teller distortions. The external in-plane field causes a dispersion of the ZEM around {\bf K} and {\bf K}$^{\prime}$ points, as shown in the electronic band structure of hbGNR in Fig.~2D (see also supplementary text~S4.4 and fig.~S8). The Fermi level can always be adjusted to these points by the electrostatic doping through the back gate voltage. As seen in Fig.~2E, the opposite direction of the electric field flips the signs of the in-gap mode slopes in both valleys. The slopes are also observed to be equal in magnitude but opposite in signs in the {\bf K} and {\bf K}$^{\prime}$ valleys. The reciprocal space configuration of the ZEM is favorable for a single channel dissipationless ballistic transport. The ZESs with the opposite group velocities are far enough from each other and from the time reversal (TR) invariant points in the $k$-space. 
Hence, the elastic backscattering is suppressed, and this suppression is protected by the TR symmetry in a way similar to the quantum valley Hall regime and topological kink states~\cite{Wang2021} (see further continuum model topological accounts and how the chirality and TR symmetry coexist here in supplementary text~S4.5 and figs.~S9-S10). Also, we encounter here a momentum-valley locking: the charge current is valley polarized so that this mode works as a perfect valley filter~\cite{Rycerz2007}. Contrary to the quantum valley Hall regime, the charge transport here does not necessarily stick to the edges. Likewise, this system has no electrostatic domain walls required for the topological kink states~\cite{Wang2021}.

Applying a sufficiently large in-plane electric field, the dispersion of the in-gap mode fills the bulk energy gap, thereby connecting the valence and conduction bands. This regime is reminiscent of the edge modes in the topological insulators as well as of the chiral anomaly of the Weyl semimetals (WS)~\cite{Yan2017} or chiral anomaly bulk states~\cite{Wang2022}. However, in contrast to the topological insulators edge states, the conductive mode here can be bulk which can be advantageous for managing heat gradients in electronic devices or more efficient use of the gain medium in photonic laser devices. In contrast to chiral anomaly of WS, in the proposed scheme magnetic field is not needed, see Fig.~1A, which opens the way to compact devices. The chiral anomaly bulk states~\cite{Wang2022} are engineered solely by the edge geometry, while in our case the similar structure arises from a flat band due to an external in-plane electric field thereby offering fine tunability. The electric field required for filling the hbGNR bulk energy gap can be reduced by increasing the hbGNR width. Figure~2F shows that the ZEM band width is a linear function of the applied in-plane field in a wide energy range. This linear dependence is also analytically justified within the continuum model (see supplementary text~S4.4 and fig.~S8).

\paragraph*{Cubic dispersion and pitchfork bifurcation.} A non-trivial behaviour is observed for wide ribbons and high external fields, such as the $N_r = 51$ with an in-plane electric field  $e\varepsilon/t_1 > 0.005$~\AA$^{-1}$. Once dispersive in-gap mode hybridizes with the bulk subbands, its dispersion at the Dirac points changes from a linear to cubic-like. Upon this hybridization, the dispersive in-gap mode breaks down into three parts: two partly flat remnant modes and a new chiral gapless mode. The gapless mode has a cubic dispersion without extrema and it features an inflection point centered at the Dirac {\bf K} point as shown in Fig.~3A. Upon further increase of the external field and admixture of additional bulk bands, the gapless mode transforms from having no extrema to having two local extrema: one maximum (to the left/right) and one minimum (to the right/left) in {\bf K}/{\bf K}$^{\prime}$ valley, as shown in Fig.~3B. 
A detailed ZEM evolution towards a gapless mode with cubic-type dispersion is presented in Fig.~3C. Thus, around the {\bf K} point, a smooth transformation from a completely flat to a linear and finally a cubic character is observed. Here, the external field works as control parameter which causes a pitchfork bifurcation. The dispersion curve behaves as $E(k) = a k^3 + c k$, which has two extrema for $c<0$ ($a > 0$) and none of them for $c>0$ ($a > 0$); in general, only odd powers of momentum contribute into the dispersion relation around the Dirac points as follows from the analytical solution and dispersion equation in the supplementary text~4.6 and fig.~S11. Not only does this breeding of the ZES with opposite group velocities take similarity after the tripling of the fixed points in the non-linear dynamical systems but it also replicates such effects as critical slowing down that is characteristic for merging of the vector field fixed points characterized by the opposite values of the topological index (see supplementary texts~4.7 and fig.~S12). This analogy with bifurcation, however, is approximate since, in effect, the hbGNR in the external field maps to a non-stationary dynamical system featuring turbulent vector field (see supplementary text~S4.5 and Movie~S1).
The cubic nature of the gapless mode dispersion is robust and supported by fitting the dispersion curve with a cubic functional (see supplementary text~S5 and fig.~S13). Such a dispersion is quite uncommon even for plasma waves even though it can be easily obtained from the Korteweg–De Vries equation by neglecting its non-linearity (see supplementary text~S6). The only similar example is the dispersion of fast-magnetosonic whistlers in the Earth's bow shock frame~\cite{Krauss-Varban1991}. Similarly, tripling of the Fermi levels has been only hypothesized for the edge states in a peculiar context of the fractional quantum Hall effect~\cite{Chamon1994}. Recently, the analogous reversing of the group velocity within the cubic-like dispersion has been proposed for an effectively unrolled armchair nanotube subjected to a step electrostatic potential~\cite{Xia2023a}.

The chiral character of the gapless mode persists despite all the drastic transformations of the mode. In particular, the chiral charge defined as the difference between the left and right movers at the Fermi level in each valley is well-defined and preserved before and after the bifurcation (see supplementary text~4.5). The chiral character of the gapless mode is also seen from the electron density distributions before and after the bifurcation presented in Figs.~3A and~3B. The characteristic feature of the bifurcation is the change of the wave function localization from the zigzag to bearded edge for $0 < k < 2
\pi/3$ and from the bearded to zigzag edge for $2 \pi/3 < k < \pi$ as compared to Fig.~2A,D, and E. The reported above rich properties of hbGNR ZEM cannot be reproduced on a square lattice (see supplementary text~S7 and fig.~S14).

\paragraph*{Resilience to static charged impurity disorder.} The honeycomb lattice is reach for dissipationless ballistic transport schemes. The most notable examples of ballistic conductors are carbon nanotubes~\cite{McEuen1999}, electrostatically doped zigzag graphene nanoribbons~\cite{Wakabayashi2007} and graphene itself~\cite{Ferreira2015}. We could also add to this list cummulenic carbyne though disorder effects seems have been systematically omitted for it~\cite{Chen2009,Garner2018}.
All of these systems exhibit resilience to a static impurity elastic backscattering. It can be shown that the reported scheme is of comparable efficiency to these well-known ballistic conductors (see supplementary text~S8 and table~S1). The fundamental difference of the reported scheme is its unique tunability together with combination of immediately accessible adjacent phenomena. 

\paragraph*{Experimental detection.} The non-trivial behaviour exhibiting pitchfork bifurcation have consequences in the density of states and quantum transport measurements. Figures~3D and~3E present evolution the density of states (DOS) and transmission coefficients of a hbGNR in the increasing in-plane electrostatic field. 

The DOS tracks the flat bands and the pitchfork bifurcation. As seen from Fig.~3D, the initial DOS maximum from the fully flat ZEM splits into two lower amplitude peaks. These peaks move towards the bulk bands and they eventually mix with these bulk bands but do not lose their magnitude which indicates a maintenance of the flatness feature in the band structure. The DOS between the two peaks forms a nearly constant background. Upon reaching a critical in-plane electric field a peak in the DOS arises at $E/t_1=0$ atop of the uniform in-gap background. This peak corresponds to a cubic dispersion, $E(k) \sim k^3$, presented in Fig.~3A. Further increase of the external field splits the bifurcation DOS peak into two peaks with an inward looking asymmetry, thereby marking the onset of the pitchfork bifurcation in Fig.~3B. The two new peaks move into opposite directions, leaving behind an increased DOS background between them.

The single channel electron transmission through the hbGNR changes from zero to unity and finally reaches a multi-channel non-coherent transport regime. At a zero external field, the transmission coefficient is zero throughout the bulk gap of the hbGNR as shown in Fig.~3E. The absence of the transmission is an indication of a Mott insulator-like regime. As the external field gradually increases, the system develops a widening region of the perfect single channel transmission originating from the dispersive in-gap mode discussed above. Once the external field increases to the critical value corresponding to the pitchfork bifurcation, a new transmission plateau appears on top of the in-gap transmission pedestal which is proceeded by a short period when $\partial_k E(k)_{k=\mathbf{K}} = 0$ corresponding to a critical slowing down in the quantum transport. The plateau is twice higher than the pedestal signifying the onset of the multi-channel regime. As schematically presented in Fig.~3E, the width of the plateau is bounded from left and right by the two bifurcation peaks in DOS (dashed vertical lines in both Figs.~3D and~3E). The width of the plateau determines the region of the cubic dispersion where the energy within the valley is non-single valued, as schematically presented by the dashed folded curve in Figs.~3D and~3E.

\paragraph*{Classes of candidate GNRs.} The synthesis of one-dimensional organic metals remains challenging due to the common characteristic of a polyradical ground state and consequently high chemical reactivity~\cite{Rizzo2020a,McCurdy2023}. In effect, a free-standing hbGNR which supports a gapless chiral mode may be diffcult to experimentally prepare due to the intrinsic high reactivity of localized radical states lining the edges. Thus, we apply our prediction to a synthetically more realistic system that recreates the hbGNR electronic structure -- an asymmetrically hydrogenated ZGNR~\cite{Kusakabe2003}. In this case, one zigzag edge of the ZGNRs is lined by trigonal planar C–H groups while the opposite edge features tetrahedral methylene (CH$_2$) groups. This asymmetric substitution excludes one carbon atom per unit cell from the $\pi$-network giving rise to the characteristic $\pi$-system of hbGNRs. A ZGNR with $N = 52$ featuring the asymmetric hydrogenation along the zigzag edges is shown in Fig.~3F. Within this system, all the fundamental results including the pitchfork bifurcation and electron density edge-to-edge transfer for ZEM are perfectly replicated using density functional theory (DFT) (see methods). 
The only difference between the DFT and tight-binding results is an increase of the critical electric field by one order of magnitude in the DFT calculations, which can be attributed to the screening caused by the electric polarization of the metallic structure~\cite{gava2009ab}. Despite this slight discrepancy, it is evident that the hbGNR $\pi$-orbital network can in principle be modelled within the structure of asymmetrically hydrogenated ZGNR.

Applying our basic concepts other families of incommensurate GNRs capable of dissipationless coherent transport can be revealed. Namely, combining zigzag and cove edges or zigzag and gulf edges in a single ribbon results in the incommensurate zigzag-cove-edged GNRs (zcGNRs) and zigzag-gulf-edged GNRs (zgGNRs), respectively. Even though the development of a practical synthesis is still under investigation for zcGNRs and zgGNRs, recent advances in the bottom-up synthesis of GNRs have shown that both classes of incommensurate GNRs are synthetically accessible. One crucial difference between zcGNRs and zgGNRs is in their translation period $T$. While zcGNRs possess a period $T = 2a$, the unit cell in zgGNRs is characterized by $T = 3a$, where $a$ is the graphene lattice constant. By using the cutting line method on the graphene ﬁrst Brillouin zone (BZ) together with the zone folding, it easy to show that the difference in the size of the unit cell between the ribbons will give rise to two valleys centered around projected {\bf K} and {\bf K}$^{\prime}$ points. In the case of zcGNRs these valleys are shifted away from the TR invariant points $\Gamma$ and $X$ as demonstrated in Fig.~4A. In contrast, in the case of zgGNRs both {\bf K} and {\bf K}$^{\prime}$ points project to the TR invariant point $\Gamma$ as depicted in Fig.~4A so that these ribbons must exhibit only a single valley in their band structures. These distinct features are evident in Fig.~4B, where DFT band structures are presented for wide $N>50$ zcGNRs and zgGNRs. Moreover, both zcGNRs and zgGNRs demonstrate flat ZEMs along with the above described valley structures. In the presence of external in-plane electric field, both zcGNRs and zgGNRs become metallic as seen in Fig.~4B. It is however only the zcGNRs that exhibit the chiral anomaly structure similar to hbGNRs and are capable of the dissipationless coherent transport. In general, the electronic band structure of hbGNR-based incommensurate ribbon exhibits only a single valley at $\Gamma$ if the translation period of an incommensurate ribbon is $T = 3a n$, where $n$ is a natural number. Hence, such ribbons are ill-suited candidates for the realization of a dissipationless conductor. Alternate families of incommensurate GNRs with larger translation periods  $T \neq 3a n$ are schematically shown in Fig.~4C. Wide ribbons of these GNR families featuring the depicted edge geometries have three common properties: (i) two valley band structures, (ii) a sufficiently large bulk energy gap, and (iii) a flat ZEM. When subjected to an external in-plane electric ﬁeld, wide ribbons exhibit chiral anomalies at low values of the ﬁeld while a pitchfork bifurcation can be seen at higher values of the electric field. We have to note, however, that further increase of the superlattice period of such incommensurate GNRs results in a reduction of the two valley separation which eventually makes inter-valley scattering more probable and is not favorable for dissipationless transport.

\paragraph*{Metamaterials perspective.} An alternative experimental verification can be obtained on analogous ultracold atom lattices~\cite{Jotzu2014}, photonic ~\cite{Plotnik2014,Ozawa2019,Xia2023} and polaritonic~\cite{Milicevic2015,Whittaker2019,Whittaker2021} crystals, and microwave metamaterials~\cite{Bellec2014,Dautova2017,Dautova2018}. Such artificially designed systems do not have limitations of the synthetic chemistry, but the main obstacle may be the imitation of the external electric field effect.
The proposed phenomena, however, can be readily verified in photonic and nanomechanical metamaterials. In photonic crystals the pseudo-electric fields can be created by the side-to-side laser writing~\cite{Liu2021}. The second extremely promising experimental playground is the platform of “nanomechanical graphene”—a honeycomb lattice of free-standing Si$_3$N$_4$ nanomechanical membranes with a  modified lattice structure~\cite{Xi2021}. Such nanomechanical graphene can be fabricated with high precision by using electron-beam lithography and subsequent dry and wet etching processes. The geometries of the nanomechanical membranes are uniquely determined by the relative positions of the etched holes $(r_1, r_2) = (r_0 - \delta_{b}, r_0 + \delta_{b})$, as shown in Fig.~1C of Ref.~\cite{Xi2021}. The influence of an external in-plane electrostatic field to the graphene honeycomb lattice can be imitated by tuning the on-site potential wihtin the unit cell of a nanomechanical graphene ribbon. This can be realized during the lithographic patterning by setting $\delta_{b} = 0$ and linearly varying $r_0$ across the width of the nanomechanical ribbon. As for experimental characterization, one can electrically actuate the elastic waves of the nanomechanical membranes and measure their propagating behaviors by using optical interferometry. The energy band diagrams can be obtained by first recording the real-space distribution of elastic waves along the desired direction and then performing Fourier transform to project the signal to the momentum space.

\paragraph{Conclusions.}In summary, a general graph-theoretic approach predicting in-gap zero-energy modes is proposed to engineer dissipationless tranport regimes on the honeycomb lattice. The bulk dissipationless transport is achievable in such structures merely with an external electrostatic field, which demonstrates the possibility to realize transport features of topological insulators and Weyl semimetals under much more facile conditions. 
The identified dispersion transformation at around the intrinsic Fermi level, ranging from fully flat to cubic-like, paves the way to access a broad variety of notable quantum transport regimes in a single experimental device.
These results also show the possibility of designing electric field tunable cubic materials suitable for non-linear electronics and photonics. The cubic materials with the inflection point in their dispersion available at the Fermi level may be promising candidates for replacing complex semiconductor superlattices in observation of Bloch oscillators and for implementation of submillimeter wavelength emitters~\cite{Esaki1970}. Further studies may include other types of crystallographic system such as kagome and Lieb lattices~\cite{Yan2019} or graphyne-based structures~\cite{Kang2019}.
Since  the reported flat band is intimately related to a fermion-$1/2$ soliton, it may exhibit a fractional fermion number, which also should be a subject of future research.

\section*{Methods}
\subsection*{Tight-binding calculations}
In the tight-binding calculations, we follow a single-parameter tight-binding (TB) model. All numerical calculations are performed in a $p_z$-orbital approximation with TBpack {\it Mathematica} package~\cite{TBpackSaroka}. Since even for the nearest neighbour hopping $t_1$ a wide range of phenomenological values is available~\cite{Payod2020}, we make results independent of this varying quantity by defining all energies in terms of $t_1$.

A homogeneous in-plane electric field applied across the ribbon width is modeled as on-site potentials~\cite{Chang2006}: $U = -e {\bf \varepsilon}\cdot {\bf r}/t_1$, where $e$ is the elementary charge, ${\bf \varepsilon}$ is the strength of the field and ${\bf r}$ is the position of the lattice site. The hopping integral, $t_1$, is set to be independent of the external field, since the largest field applied to the ribbon ($e \varepsilon/t_1 = 0.1$~\AA$^{-1}$) is much less than the atomic field $1/a_0 \approx 0.7$~\AA$^{-1}$, where $a_0=1.42$~\AA~is the nearest neighbor distance in the graphene lattice. Here we omit the electrostrictive deformation and screening effects. While the electrostrictive deformation in graphene nanoribbons is small ($2.6\%$) and nanoribbon structures can be considered as rigid ones~\cite{Chang2006}, the screening for gapless metalic ribbons is significant. Hence, the external field shall be interpreted as an effective field experience by the on-site electron after screening. While solving the eigenproblem for the matrix Hamiltonian, we keep tracking the eigenenergy branches by sorting the eigenvalues with respect to maximum overlap of the eigenvectors at the neighboring $k$-points. The sorting starts from the center of the Brillouin zone, $k=0$, where ZEM is flat and easily identifiable.

To visualize the electron wave function localization in the energy band diagram we use a color scheme based on the inverse participation ratio (IPR)~\cite{MartinezAlvarez2018,Wegner1980,Thouless1974,Bell1970}: $\mathrm{IPR} = \sum_{{\bf r}}\left|\Psi({\bf r})\right|^4$, where $\int_V \left|\Psi({\bf r})\right|^2\,d{\bf r}=1$. The IPR characterizes the volume of space, where the absolute value of the wave function is essentially non-zero and as such can be imagined as the size of the eigenstate. For a normalized  eigenvector $C_{j}(k) = \left(C_{j1}(k), C_{j2}(k), \ldots, C_{jN}(k)\right)^{\mathrm{T}}$, which corresponds to the $j$-th eigenstate $E_j(k)$ of an $N \times N$ TB Hamiltonian, the IPR is given by
\begin{equation}
    \mathrm{IPR}_j(k) = \sum_{i=1}^{N} \left|C_{ji}(k)\right|^4 \, .
\end{equation}
This quantity is equal to one for a perfectly localized state in which an electron density occupies only one atomic site. However, the IPR it is much less than unity for a fully delocalized state, and it further approaches zero in the thermodynamic limit ($N \to \infty$) as the size of the system increases.

\subsection*{Quantum transport}
The coherent quantum transport through a hbGNR is modeled within the non-equilibrium Green’s function formalism. We partition the full system Hamiltonian as in Ref.~\cite{Munoz-Rojas2006}:
\begin{equation}
H = \left(
\matrix{
H_L & H_{LC} & 0 \cr
H_{CL} & H_C & H_{CR} \cr
0 & H_{RC} & H_{R}
}
\right),
\end{equation}
where $H_C$ is a square matrix with size equal to the number of atoms in the central scattering region of the device, while $H_{L,R}$ and $H_{LC},H_{CL}, H_{CR}, H_{RC}$ are semi-infinite square and rectangular matrices describing the semi-infinite electrodes and their coupling to the central scattering region, respectively. Here, the scattering region is a single hbGNR unit cell, while the leads are represented by semi-infinite hbGNRs [cf. with Ref.~\cite{Wakabayashi2001}].
The Green's function defined as 
\begin{equation}
    (H-IE)\Cal{G}(E) = 1\,,
\end{equation}
where $I$ is the identity matrix, can be partitioned as
\begin{equation}
    \Cal{G} = \left(\matrix{\Cal{G}_L & \Cal{G}_{LC} & \Cal{G}_{LR} \cr
\Cal{G}_{CL} & \Cal{G}_C & \Cal{G}_{CR} \cr
\Cal{G}_{RL} & \Cal{G}_{RC} & \Cal{G}_{R}}\right)\, .
\end{equation}
In terms of partition blocks, the central region Green's function is
\begin{equation}
    \Cal{G}_C = \left[EI - H_C - \Sigma_L - \Sigma_R\right]^{-1}\, ,
\end{equation}
where the self-energies $\Sigma_{X}$ ($X=L,R$) are given by
\begin{equation}
    \Sigma_X(E) = H_{CX}g_X(E)H_{XC}
\end{equation}
with $g_X(E) = \left(EI - H_X\right)^{-1}$ being Green's functions of the semi-infinite leads. We calculate the semi-infinite lead Green’s functions via the surface Green’s functions by defining them in self-consistent equations:
\begin{equation}
\label{eq:surfaceGF}
    g_{\mathrm{surf}}(E) = \left((E + i \eta) I - H_0 - V^{\dagger}g_{\mathrm{surf}}(E)V\right)^{-1}\, ,
\end{equation}
where $H_0$ and $V$ are diagonal and first off-diagonal blocks of the lead region, respectively, and $\eta$ is a small real and positive parameter that sets $g_{\mathrm{surf}}(E)$ as a retarded Green's function.

Finally, we define the coupling matrices
\begin{equation}
    \Gamma_{X} = i \left(\Sigma_X - \Sigma^{\dagger}_X\right) = - 2 \mathrm{Im}(\Sigma_X)\, ,
\end{equation}
so that the transmission coefficient can be obtained as
\begin{equation}
    T(E) = \mathrm{Tr}\left[\Gamma_L(E)\Cal{G}_C(E)\Gamma_R(E)\Cal{G}_C^{\dagger}(E)\right]\, .
    \label{method:Transmission}
\end{equation}
Equation~(\ref{method:Transmission}) can be reduced to
\begin{equation}
    T(E) = \mathrm{Tr}\left[\Gamma_{L,11}(E)\Cal{G}_C(E)\Gamma_{R,NN}(E)\Cal{G}_C^{\dagger}(E)\right]\, ,
    \label{method:TransmissionReduced}
\end{equation}
if the full Hamiltonian tri-block diagonal structure is taken into account~\cite{Compernolle2003}. This approach speeds up the calculations of the transmission coefficient. In Eq.~(\ref{method:TransmissionReduced}), $\Cal{G}_C$ must be replaced with a block $\Cal{G}_{C,1N}$ when the scattering region consists of $N$ unit cells. In practice, the Hamiltonian containing two unit cells for each lead and one unit cell as a scattering region was constructed by TBpack~\cite{TBpackSaroka} to extract $H_0$, $V$ and $H_C$ as initial parameters needed in the calculations.

We noticed that the convergence of the self-consistent procedure is guaranteed only if a small real $\eta > 0$ is added not only to the initial approximation $g_{\mathrm{surf},0}(E) = \left((E + i \eta) I - H_0\right)^{-1}$ but also to the repeatedly applied function~(\ref{eq:surfaceGF}). Also, a narrow peak can be obtained at $E/t_1=0$ in the calculations if large values of $\eta = 10^{-3}$ are used. However, this peak disappears for small $\eta = 10^{-5}$, and as such it is deemed to be a numerical artefact rather than a physical feature. We have verified that the given algorithm reproduces the results reported in Ref.~\cite{Wakabayashi2001}.

The static impurity potentials are modeled by Gaussian bell-shaped function following Ref.~\cite{Wakabayashi2007}. The resulting disordered potential is
\begin{equation}
    V(x, y, z) = \sum_{i=1}^{N} u \exp \left[- \frac{\left(x-x_i\right)^2 + \left(y-y_i\right)^2 + z^2}{d^2} \right] \, ,
     \label{method:DisorderedPotential}
\end{equation}
where $u$ is the impurity strength, $d$ is the impurity potential scattering range in \AA, $x_i$ and $y_i$ are positions of $N$ impurities distributed randomly with $xOy$-plane so that their density is $n_{\mathrm{im}} = N/S$, where $S$ is the area in~\AA$^2$ of the central scattering region. The impurity strength is uniformly distributed within the range $|u| < u_{\mathrm{M}}$, where $u_{\mathrm{M}}$ is the normalized value given by $u_{\mathrm{M}} = u_0 t_1/(\pi d^2)$. The disordered potential~(\ref{method:DisorderedPotential}) is used to update the on-energies of the Hamiltonian.

\subsection*{Density functional theory calculations}
The spin-restricted DFT calculation was carried out in Siesta~\cite{soler2002siesta} with SISL~\cite{NickPapior2021} as postprocessing tool. We employed Perdew–Burke–Ernzerhof (PBE) generalized gradient approximation (GGA) exchange and correlation functional~\cite{PhysRevLett.77.3865}, and Monkhorst-Pack k-mesh of $21\times1\times1$, and energy cut-off of 400~Ry. The structure was relaxed until the force was below $0.01$~eV/\AA. Electron charge density was visualized in XCrySDen~\cite{kokalj2003computer}.



\bibliography{scibib}

\bibliographystyle{Science}

\section*{Acknowledgments}
The authors thank K.~Batrakov, E.~Thingstad, J.~Zheng, V.~Demin, A.~Qaiumzadeh, A.~Ferreira, K.~Yonekura, C.~Shang, Z.~Gao, K.~Wakabayashi and C.~Beenakker for useful stimulating discussions and J.~Danon for providing computational facilities at NTNU. 
\paragraph*{Funding:}
X.S. acknowledges funding support by the Research Grants Council of Hong Kong (project no. 14209519). C.A.D. is supported by the Royal Society University Research Fellowship (URF/ R1/ 201158). F.R.F. was partly supported by the Heising-Simons Faculty Fellows Program at the UC Berkeley. V.A.S. was partly supported by the Research Council of Norway Center of Excellence funding scheme (project no. 262633, “QuSpin”) and EU HORIZON-MSCA-2021-PF-01 (project no. 101065500, TeraExc).

\paragraph*{Authors contributions:}
V.A.S. and C.A.D. conceived the initial idea. V.A.S. developed graph theoretical description and performed TB calculations. R.B.P. verified the results. V.A.S and C.A.D. interpreted TB results within the continuum effective mass model. X.S. proposed metamaterial platform and experimental arrangement for testing predictions. F.R.F. proposed a zcGNR family for investigation and realization by synthetic chemistry methods. F.K., V.A.S. and L.B. performed DFT calculations and interpreted the results. V.A.S., R.B.P., F.K, X.S. and L.B. wrote the first draft of the manuscript. All authors contributed to the further editing of the manuscript.

\paragraph*{Competing interests:}
The authors declare no competing interests.

\paragraph*{Data and materials availability:}
All data are available in the manuscript or the supplementary materials. Additional data supporting the findings of this study are available from the corresponding authors upon reasonable request. The code developed for this study is available under MIT License from GitHub~\cite{TBpackSaroka}.

\section*{Supplementary materials}
Supplementary Text\\
Figs. S1 to S14\\
Table S1\\
References (\textit{75-110})\\
Movie S1


\clearpage

\begin{figure}
    \includegraphics[width=0.99\textwidth]{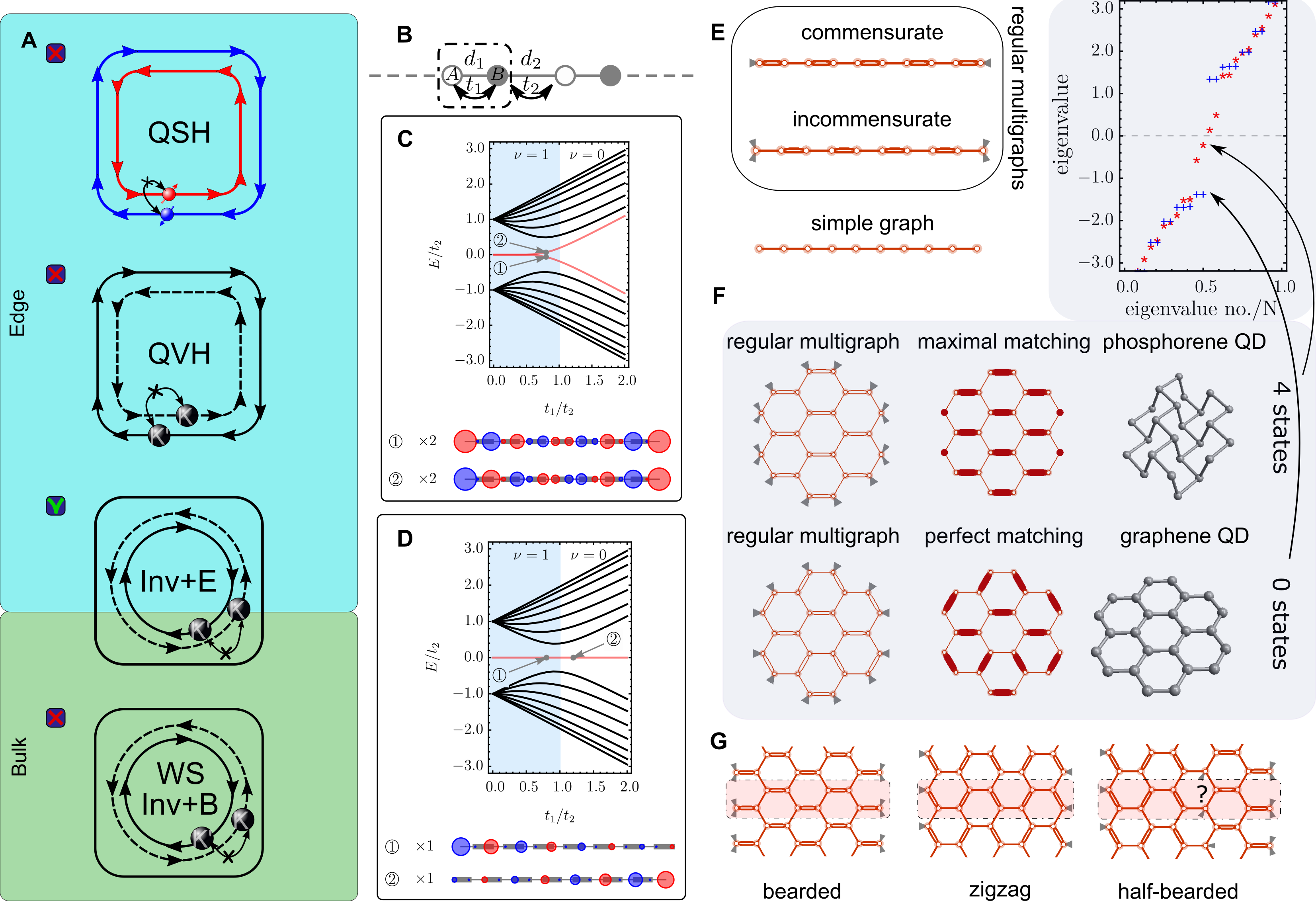}
    \caption{{\bf Zero-energy states and graph topology}. ({\bf A}) Dissipationless transport schemes. Ivn: breaking of the inversion symmetry; E and B: electric and magnetic fields. ({\bf B}) Periodic SSH model. $t_{1,2}$ are hopping integrals, $d_{1,2}$ are inter site distances. ({\bf C}) Energy levels of the finite even SSH chain as function of $t_1/t_2$  and the wave functions of $E=0$ states (red curves). The light blue and white backgrounds mark regions of topological and trivial phases, respectively. The negative and positive on-site weights of the wave function are denoted as blue and red. For visibility the weights are scaled by the factor shown to the left. The thin and thick gray lines connecting the SSH chain sites visualize strengths of $t_{1}$ and $t_2$. ({\bf D}) Same as (C) but for the finite odd SSH chain. ({\bf E}), The regular multigraphs for a commensurate and incommensurate finite even SSH chain and a corresponding simple graph. ({\bf F}) Counting the in-gap states for two regular multigraphs and matching their patterns corresponding to two chemical systems. Plot: eigenvalues of adjacency matrices versus the normalized eigenvalue counting number. ({\bf G}) Regular multigraphs for various GNRs with translation invariance along to the zigzag crystallographic direction. The dot dashed rectangle in (B) and (G) marks the unit cells of periodic structures.
    }
\end{figure}

\begin{figure}
     \includegraphics[width=\textwidth]{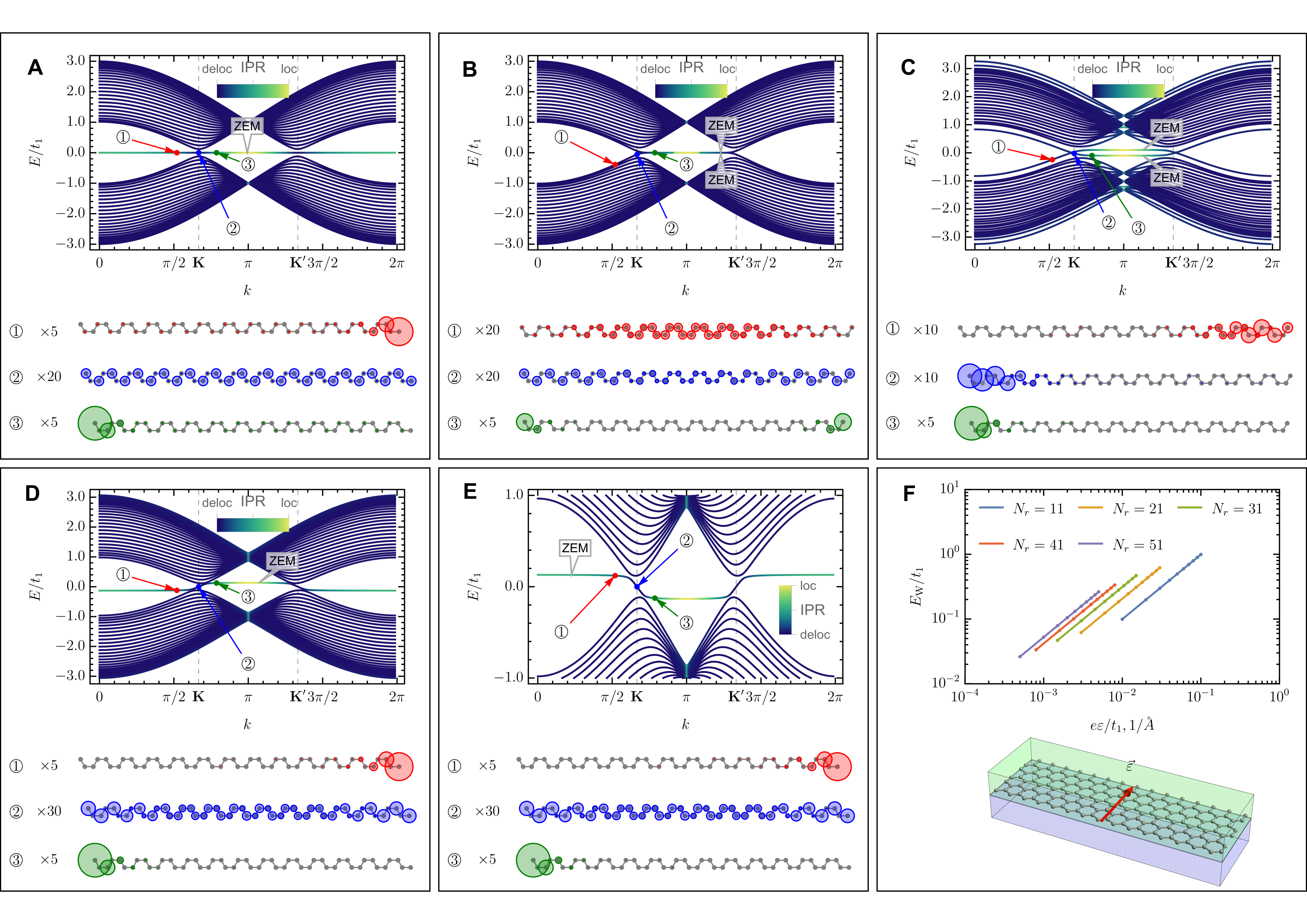}
        \caption{{\bf Converting a flat ZEM into a dispersive in-gap mode.} ({\bf A}) Energy bands and electron density distribution over the unit cell of a ribbon at $3$ depicted $k$-points in the {\bf K} valley of the hbGNR $N_r = 51$. Colour bars: the inverse participation ratio (IPR) of each eigenstate. Circles: on-site electron densities scaled by a factor shown to the left. ({\bf B}) Same as (A), but for ZGNR $N = 52$. ({\bf C}) Same as (B), but for the quantum valley Hall regime: $t_1 = 1$; $\Delta = 0.2 t_1$; $U = 0.3 t_1$; $U_l = - U$; $U_r = U$, where $\Delta$ is the amplitude of the on-site staggered potential, $U$ is the side back gate potential amplitude, $U_{l,r}$ are the left and right side back gate potential that is applied to the $12$ atoms on each side of the ribbon. ({\bf D}) Same as (A), but for hbGNR $N_r = 51$ in the in-plane electric field $e\varepsilon/t_1 = 0.005 $~\AA$^{-1}$. ({\bf E}) Same as (D), but for the opposite direction of the applied field and focused at the zero energy of the band structure. ({\bf F}) A zero-energy band width as a function of the in-plane electrostatic field for hbGNR $N_r = 11$, $21$, $31$, $41$ and $51$. Scheme: the in-plane electric field applied across the ribbon.
        }
\end{figure}

\begin{figure}
     \includegraphics[width=\textwidth]{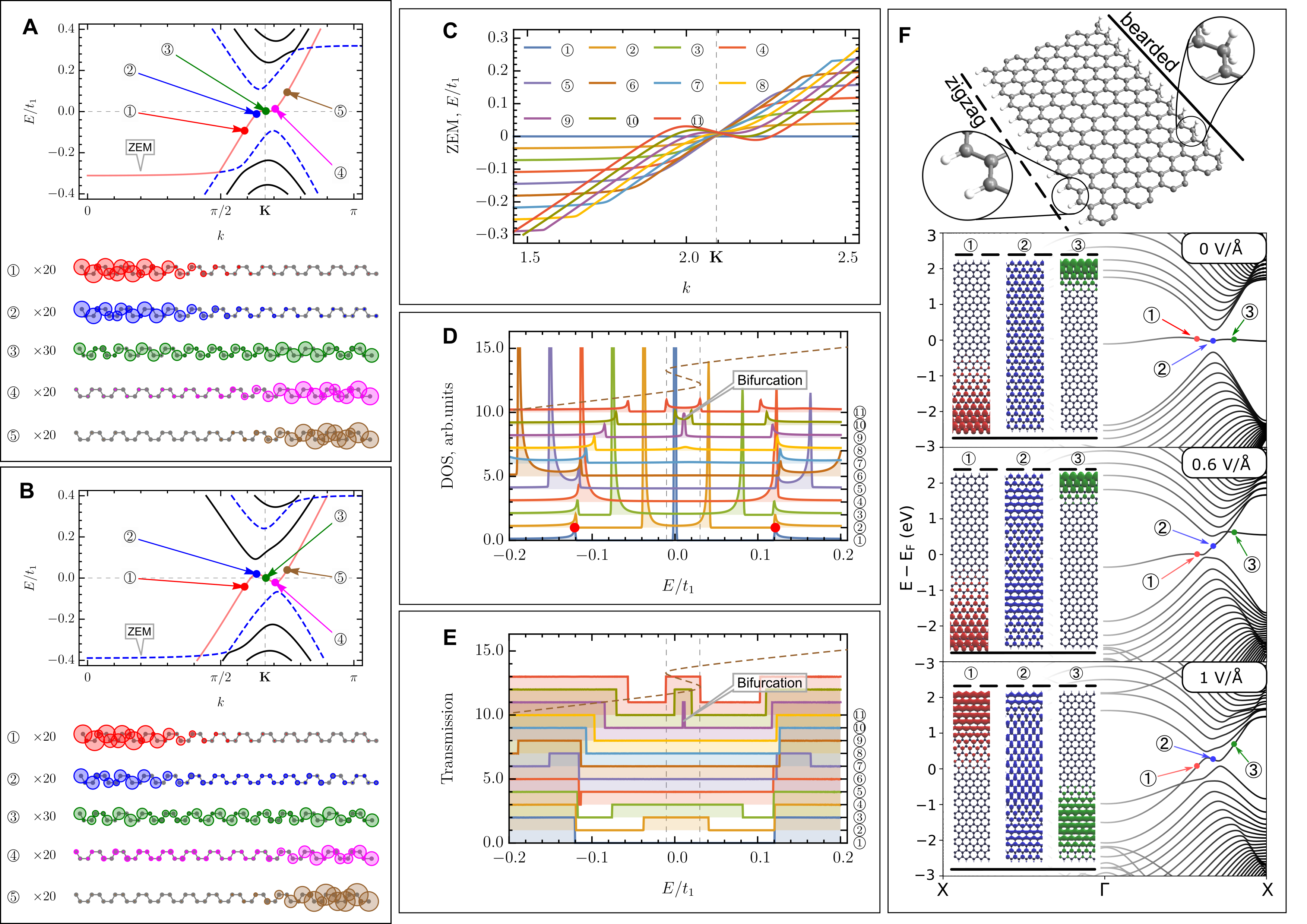}
    \caption{{\bf Chiral gapless mode pitchfork bifurcation}. ({\bf A}) The energy bands of hbGNR $N_r = 51$ in the electric field $e\varepsilon/t_1 = 0.012$~\AA$^{-1}$ together with the electron density distribution over the ribbon unit cell for the selected zero-energy mode energy levels at several $k$-points. ({\bf B}) Same as (A) but for $e\varepsilon/t_1 = 0.015$~\AA$^{-1}$. The circles representing on-site electron densities are scaled by factors shown to the left. In (A) and (B) dashed blue dispersion curves are the remnant partially flat bands and light red curves are the chiral gapless modes. ({\bf C}) hbGNR $N_r = 51$ ZEM evolution in the electric fields from $e\varepsilon/t_1 = 0.0$ to $0.015$~\AA$^{-1}$ with the step $0.0015$~\AA$^{-1}$ marked with circled numbers $1$-$11$, respectively. ({\bf D}), ({\bf E}) The density of states and the transmission coefficient in the electric fields in (C). Red dots in (D) mark the peaks used for normalization of all the curves. The spectra are shifted vertically for clarity. The natural horizontal drift to the left is also retained for clarity. The wiggly dashed brown curve represents the dispersion of the chiral gapless mode for $e\varepsilon/t_1 = 0.015$~\AA$^{-1}$. The vertical dashed gray lines mark the energies of the peaks in the density of states arising after the bifurcation onset. ({\bf F}) DFT energy bands and electron densities in a few $k$-points for ZGNR $N=52$ asymmetrically hydrogenated as shown in the scheme above the plots and subjected to $0$, $0.6$ and $1$~V/\AA~in-plane electric fields. The same isovalue is used for imaging the charge densities.}
\end{figure}

\begin{figure}
     \includegraphics[width=\textwidth]{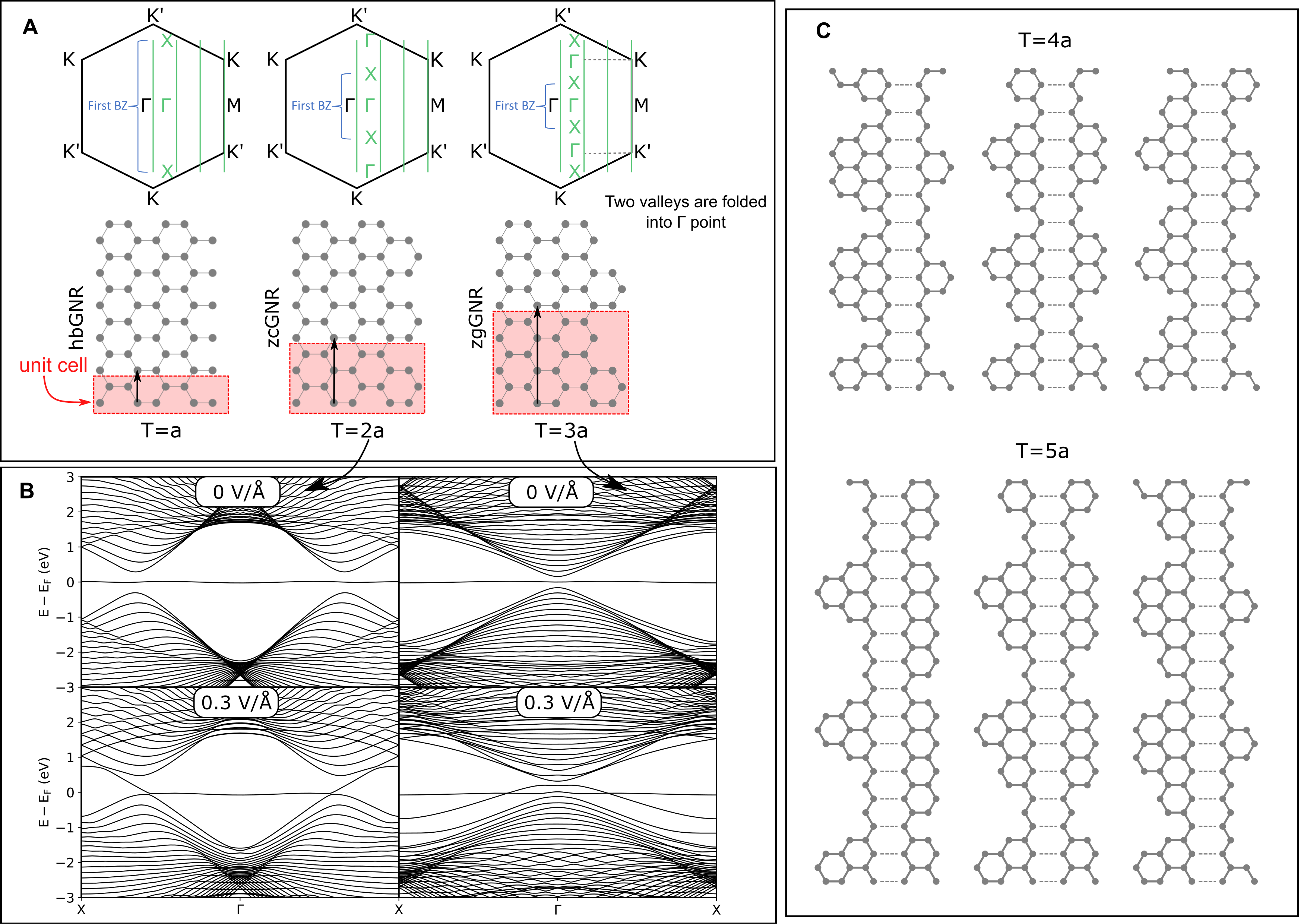}
    \caption{{\bf Dissipationless conductors}. ({\bf A}) The folding scheme explaining $3a$-periodicity of the single valley appearance in intrinsically incommensurate GNRs. $a=2.46$~\AA~is the graphene lattice constant. Black hexagons represent the first Brillouin zone (BZ) of graphene. Light green lines are the cutting lines resulting from the transverse momentum quantization in a GNR. BZs of GNRs with different translation period $T$ are denoted by light blue curly brakets. ({\bf B}) DFT energy bands for zcGNR and zgGNR  species with $25$ zigzag chains in the backbone ZGNR and subject to the in-plane electric fields. ({\bf C}) Schematic illustrations of intrinsically incommensurate GNRs with the dissipationless transport properties in the presence of an external in-plane electric field.}
\end{figure}

\clearpage

\renewcommand\thesection{{S\arabic{section}}}
\renewcommand\thesubsection{\thesection.\arabic{subsection}}
\renewcommand\thesubsubsection{\thesubsection.\arabic{subsubsection}}
\renewcommand{\theequation}{S\arabic{equation}}
\renewcommand{\thefigure}{S\arabic{figure}}
\renewcommand{\thetable}{S\arabic{table}}
\setcounter{equation}{0}
\setcounter{figure}{0}
\setcounter{table}{0}

{\center
{\huge Supplementary Materials for}\\

\vspace{12pt}

{\Large \bf Tunable chiral anomalies and coherent transport on a honeycomb lattice}
\vspace{3pt}

Vasil~A.~Saroka, Fanmiao~Kong, Charles~A.~Downing, Renebeth~B.~Payod, Felix~R.~Fischer, Xiankai~Sun, Lapo~Bogani

}
{\center Corresponding author: Vasil Saroka, 40.ovasil@gmail.com

}

\paragraph*{The PDF file includes:}
\begin{itemize}[topsep=8pt,itemsep=0pt,partopsep=0pt, parsep=0pt]
\item[] Supplementary Text
\item[] Figs. S1 to S14
\item[] Table S1
\item[] References
\end{itemize}

\paragraph{Other Supplementary Materials for this manuscript include the following: }
\begin{itemize}[topsep=8pt,itemsep=0pt,partopsep=0pt, parsep=0pt]
\item[] Movie S1
\end{itemize}

\clearpage
\section*{Supplementary Text}

\section{\label{subsec:ZEM}Zero-energy mode and iso-spectral properties of hbGNRs and carbon nanotubes}
Let us take a closer look at the electronic properties of the half-bearded graphene nanoribbons (hbGNRs). Figure~S1A shows the energy band structure of hbGNRs compared with that of an armchair single-wall carbon nanotube (aSWCNT). An aSWCNT is a good reference since its energy band curves contour the full spectrum of a bulk graphene. This feature is conveniently used to track peculiar edge states that evade the bulk graphene spectrum~\cite{Saroka2017}. Indeed, for the hbGNR a perfectly flat zero-energy band is seen in fig.~S1A. This flat band is fully outside the bulk graphene spectrum, which is a deep contrast to the picture when zigzag and bearded graphene nanoribbons (GNRs) are compared to aSWCNTs and partly flat bands only partially escape outside the bulk graphene spectrum~\cite{Klein1994,Nakada1996,Saroka2018}. 
A couple of other observations shall be mentioned with respect to the energy bands of hbGNRs and aSWCNTs. The bulk energy bands of the hbGNR are isospectral with the double degenerate energy bands of the chosen single-wall carbon nanotube. This effect is of the same type as that for $(1,4)$-divinyl benzene and $2$-phenyl butadiene and is rooted in the spectral graph theory~\cite{Cvetkovic1974}. The observed effect for hbGNRs differs from that of both zigzag and bearded GNRs, thus making hbGNRs more similar to armchair GNRs. (cf. Fig.~3 in Ref.~\cite{White2007a}, Fig.~2 in Ref.~\cite{Saroka2018}, and Fig.~6 in Ref.~\cite{Hartmann2019}). It follows from fig.~S1A that for an aSWCNT and a pair of hbGNRs there is a rule of decomposition presented in fig.~S1B. Namely, $N_t = 2N_r +2$, where $N_t (N_r)$ is the number of atoms in the aSWCNT (hbGNR) unit cell. This rule is distinct from the known rules for zigzag and armchair GNRs/carbon nanotubes, $N_t = 2N_r + 4$, and bearded GNRs/aSWCNTs, $N_t = 2 N_r$~\cite{Saroka2018}, thereby making hbGNRs a special class of material as is expected from their unique structural properties. Later we shall use this iso-spectral property to find a correct description of the hbGNR zero-energy mode within the continuum model.

\section{Analytical theory of hbGNRs\label{app:TBtheoryofhbGNRZEM}}
This section provides a detailed analytical description of the zero-energy mode (ZEM) of hbGNR when an external electric field is absent. On the other hand, it demonstrates the persistent character of the zero-energy solution in the hbGNR when an external electrostatic field is applied in-plane across the ribbon width.

\subsection{Pristine hbGNR\label{app:TBtheoryofhbGNRZEMwithoutEfield}}
In this subsection, we shall treat the problem of hbGNR energy bands and wave functions with the transfer matrix method from first principles without making assumptions about the zero-energy states. This treatment is based on that for zigzag GNR in Ref.~\cite{Saroka2017}. We start from the tight-binding Hamiltonian that can be obtained from the Hamiltonian of a zigzag (Z) or bearded (B) GNR by its extension to the odd size:
\begin{eqnarray}
    H_{\mathrm{ZGNR}} &=& \left(\matrix{
    0 & t_1 q & 0 & 0 \cr
    t_1 q & 0 & t_1 & 0 \cr
    0 &  t_1 & 0 & t_1 q\cr
    0 & 0 & t_1 q & 0 
    }\right) \, , \\
    H_{\mathrm{BGNR}} &=& \left(\matrix{
    0 & t_1 & 0 & 0 \cr
    t_1 & 0 & t_1 q & 0 \cr
    0 &  t_1 q & 0 & t_1\cr
    0 & 0 & t_1 & 0 
    }\right) \, , \\ 
    H_{\mathrm{hbGNR}} &=& \left(\matrix{
    0 & t_1 q & 0 & 0 & 0 \cr
    t_1 q & 0 & t_1 & 0 & 0 \cr
    0 &  t_1 & 0 & t_1 q & 0\cr
    0 & 0 & t_1 q & 0 & t_1 \cr
    0 & 0 & 0 & t_1 & 0
    }\right) \, , 
    \label{eq:hbGNRTBHamiltonian}
\end{eqnarray}
where $q = 2 \cos \left(k/2\right)$, $k\equiv k a$ with $a = 2.46$~\AA~being graphene lattice constant, and $t_1$ is the nearest-neighbor hopping integral. The eigenproblem for hbGNR Hamiltonian~(\ref{eq:hbGNRTBHamiltonian}) can be presented in the transfer matix form with the two transfer matrices $T_1$ and $T_2$:
\begin{equation}
    \left(\matrix{
    c_N \cr
    c_{N+1}
    }\right) = T_1 T^{(N_r-1)/2} \left(\matrix{
    c_0 \cr
    c_1
    }\right) \, ,
    \label{eq:TransferMatrixEquation}
\end{equation}
where $N_r$ is the number of atoms in the hbGNR unit cell, $T = T_1 T_2$, and
\begin{eqnarray}
    T_1 &=& \left(\matrix{
    0 & 1 \cr
    -1/q & \alpha/q 
    }\right) \, , \\ T_2 &=& \left(\matrix{
    0 & 1 \cr
    -q & \alpha
    }\right)\, ,
\end{eqnarray}
with $\alpha = E/t_1$ being the dimensionless energy. Since $N_r$ is odd for hbGNR, the power $\left(N_r-1\right)/2$ is actually the parameter $w = N/2$ for ZGNR in Ref.~\cite{Saroka2017}. Hence, the power of the transfer matrix $T$ can be evaluated by using Eq.~18 in Ref.~\cite{Saroka2017} and later substituting $w = \left(N_r-1\right)/2$. For $T^w$, we explicitly have
\begin{equation}
    T^{w} = \frac{(-1)^w}{q \sin \theta} \left(\matrix{
    - \left\{\sin\left(w \theta\right) + q \sin\left[\left(w+1\right) \theta \right]\right\} & \alpha \sin\left(w \theta\right) \cr
    -\frac{1+q^2-2q\cos \theta}{\alpha} \sin\left(w\theta\right) & \sin\left(w \theta \right) - q \sin \left[\left(w+1\right)\theta\right]
    }\right)\, ,
    \label{eq:TmatrixPower}
\end{equation}
where the new parameter $\theta$ reduces eigenvalues of $T$ to the exponential form $\lambda_{1,2} = - e^{\mp i \theta}$, i.e. it results from $(\alpha^2 - q^2 - 1)/(2 q) = - \cos \theta$ that corresponds to a direct band numbering in terminology of Ref.~\cite{Saroka2017}. Symbolically, the full transfer matrix $\mathbf{T} = T_1 T^w$ is
\begin{eqnarray}
    \mathbf{T} &=& \left(\matrix{
    0 & 1 \cr
    -\frac{1}{q} & \frac{\alpha}{q}
    }\right) \left(\matrix{
    \left(T^w\right)_{11} & \left(T^w\right)_{12} \cr
    \left(T^w\right)_{21} & \left(T^w\right)_{22}
    } \right)\, ,\nonumber \\
    &=&\left(\matrix{
    \left(T^w\right)_{21} & \left(T^w\right)_{22} \cr
   \left(-\left(T^w\right)_{11} + \alpha \left(T^w\right)_{21}\right)/q & \left[-\left(T^w\right)_{21} + \alpha \left(T^w\right)_{22}\right]/q
    }\right)\, ,
\end{eqnarray}
whence the secular equation for the hard-wall boundary condition $\mathbf{T}_{22} = 0$ is
\begin{eqnarray}
    \mathbf{T}_{22} &=& \frac{1}{q} \left( - \frac{\alpha (-1)^w}{q \sin \theta} \sin\left(w\theta\right) + \frac{\alpha (-1)^w}{q \sin \theta} \left\{\sin\left(w \theta \right) - q \sin \left[\left(w+1\right)\theta\right] \right\}\right) \, , \nonumber \\
    &=& \frac{(-1)^{(w+1)}\alpha}{q \sin \theta} \sin \left[\left(w+1\right)\theta\right] = 0\, ,
    \label{eq:SecularEquation}
\end{eqnarray}
where we have used Eq.~(\ref{eq:TmatrixPower}). As seen from Eq.~(\ref{eq:SecularEquation}), the secular equation is satisfied if either (i) $\alpha = 0$ or (ii) $\theta = \pi j/(w+1)$. The (i) option yields the ZEM that is independent of $k$ and fully flat. The (ii) option describes the bulk bands that are
\begin{equation}
    E = \pm t_1 \sqrt{4 \cos^2\left(\frac{k}{2}\right) - 4 \cos\left(\frac{k}{2}\right) \cos \theta + 1}\, .
    \label{eq:hbGNRBulkBands}
\end{equation}
One can notice that Eq.~(\ref{eq:hbGNRBulkBands}) is identical to that for aSWCNTs (see Appendix C of Ref.~\cite{Saroka2017}) if 
\begin{eqnarray}
\theta_{t,j} &=& \theta_{r,j} \, , \nonumber \\
\frac{2 \pi j}{N_t/2}  &=& \frac{\pi j}{(N_r -1)/2 + 1} \, , \nonumber \\  
N_t &=& 2 N_r + 2 \, ,
\label{eq:MatchingTransverseMomenta}
\end{eqnarray}
where we have used $w = (N_r - 1)/2$. Equations~(\ref{eq:hbGNRBulkBands}) and~(\ref{eq:MatchingTransverseMomenta}) explain the isospectral properties of hbGNRs and aSWCNTs presented in fig.~S1A. The found $\theta$'s allow us to obtain the eigenvectors of the hbGNR bulk bands. The components of the eigenvectors can be found from the following recursive relations [cf. with Eq.~(\ref{eq:TransferMatrixEquation})]:
\begin{eqnarray}
    \left(\matrix{
    c_j \cr
    c_{j+1}
    }\right) &=& T_1 T^{(j-1)/2} \left(\matrix{
    c_0 \cr
    c_1
    }\right) \, , \qquad \mathrm{odd}\, j \nonumber \\
    \left(\matrix{
    c_j \cr
    c_{j+1}
    }\right) &=& T^{j/2} \left(\matrix{
    c_0 \cr
    c_1
    }\right) \, , \qquad \mathrm{even}\, j
    \label{eq:RecursiveRelations4evecscomp}
\end{eqnarray}
where $c_0 = 0$ is required by the hard-wall boundary condition and $c_1$ can be found from the eigenvector normalization. The latter means that $c_j = \left(T_1 T^{(j-1)/2}\right)_{12} c_1$, when $j = 2p + 1$, and $c_j = \left(T^{j/2}\right)_{12} c_1$, when $j = 2 p$, while $p=0, \ldots, w$ in both cases. After reordering $c_{2p+1} \rightarrow c_{2p-1}$, the components can be explicitly written as
\begin{eqnarray}
    c^{(j)}_{2 p -1} &=& N_{j} \left\{\sin\left[\left(p-1\right)\theta_j\right] - q \sin\left(p\theta_j\right)\right\} \,, \nonumber \\
    c^{(j)}_{2p}&=& N_{j} \alpha_{j,\pm} \sin\left(p\theta_j\right), \qquad p = 1, \ldots, w+1 \, ,
    \label{eq:hbGNRBulkEigenVectors}
\end{eqnarray}
where $\theta_j = \pi j / (w+1)$, $w = (N_r - 1)/2$, $\alpha_{j,\pm} = \pm \sqrt{q^2 - 2 q \cos \theta_j + 1}$ and 
\begin{eqnarray}
    N_j &=& \left(\sqrt{S^{(1)}_{\theta_j,w} - 2 q S^{(2)}_{\theta_j,w} + \left(2 q^2 - 2 q \cos \theta_j + 1\right) S^{(3)}_{\theta_j,w}}\right)^{-1} \, , 
    \label{eq:hbGNRevecsNormConstbulk}
\end{eqnarray}
with $S^{(1,2,3)}_{\theta_j,w}$ given by
\begin{eqnarray}
    S^{(1)}_{\theta_j,w} &=& \frac{1}{4} \left(1 + 2 w - \frac{\sin\left[\left(2 w + 1\right)\theta_j\right]}{\sin \theta_j}\right) \, , \\
    S^{(2)}_{\theta_j,w} &=& \frac{1}{4} \left(2 w \cos \theta_j + \frac{\sin 2\theta_j}{\sin \theta_j} - \frac{\sin\left[2\left(w + 1\right)\theta_j\right]}{\sin \theta_j}\right)\, , \\
    S^{(3)}_{\theta_j,w} &=& \frac{1}{4} \left(3 + 2 w - \frac{\sin\left[\left(2 w + 3\right)\theta_j\right]}{\sin \theta_j}\right) \, .
\end{eqnarray}
It should be noted that $\theta_j$ does not depend on $k$.

The ZEM case is more tricky, since we do not have $\theta_j$ to substitute into the recursive relations given by Eq.~(\ref{eq:RecursiveRelations4evecscomp}). In principle, one can notice that the secular equation solution $\alpha = 0$ formally defines $\theta$ from $\pm \sqrt{q^2 - 2 q \cos \theta + 1} = 0$ as $\theta = \arccos \left((q^2+1)/2q\right)$. However, $\left|(q^2+1)/2q\right|\geq 1$ for any $q \in \left(-\infty,+\infty\right)$, making the obtained formal result meaningless. Similar to the treatment of ZGNR edge states in Ref.~\cite{Saroka2017}, the found pathology can be cured by the analytic continuation: $\theta \rightarrow i \beta$. This substitution necessarily changes the trigonometric functions in Eqs.~(\ref{eq:hbGNRBulkEigenVectors}) to the hyperbolic ones, thereby enforcing the localization of eigenvector on a few or even one component. Noticing also that $\alpha = 0$ sets all even components of the eigenvector to zero, we finally have
\begin{eqnarray}
    c^{(j)}_{2 p -1} &=& N_{0} \left\{\sinh\left[\left(p-1\right)\beta_j\right] - q \sin\left(p\beta_j\right)\right\} \, \nonumber \\
    c^{(j)}_{2p}&=& 0, \qquad p = 1, \ldots, w+1 \, ,
    \label{eq:hbGNRZEMEigenVectors}
\end{eqnarray}
where $\beta =  \mathrm{arcosh} \left( (q^2+1)/2q\right)$, and the normalization constant is 
\begin{equation}
    N_0 = \left(\sqrt{\sum^{w+1}_{p=1}\left\{\frac{\sinh\left[(p-1)\beta\right]}{\sinh\left(p\beta\right)} - q\right\}^2 \sinh^2 \left(p \beta\right)}\right)^{-1}\, .
    \label{eq:hbGNRevecsNormConstedge}
\end{equation}
A simplified expression of Eq.~(\ref{eq:hbGNRevecsNormConstedge}) with summations similar to the Eq.~(\ref{eq:hbGNRevecsNormConstbulk}) can be obtained. Those expressions, however, are not practical from a numerical point of view because of the need to operate with large numbers resulting from the hyperbolic functions when $q \rightarrow 0$ and $\beta \rightarrow \pm \infty$. In this case, the lost of precision leads to a non-regular oscillatory behavior in the region close to $k = \pi$. In addition, for an increasing width of the ribbon this numerically non-stable region promptly expands from $k=\pi$ towards the Dirac point $k = 2 \pi /3$, thereby making the primary region of interest inaccessible. The procedure bypassing this obstacle is to keep the summation in Eq.~(\ref{eq:hbGNRevecsNormConstedge}) and to set its terms to zeros whenever the expression in $\left\{\ldots\right\}$ is small. Equations~(\ref{eq:hbGNRZEMEigenVectors}) explain the sublattice polarization of the flat ZEM wave function seen in Fig.~2A of the main text.

\subsection{hbGNR with an in-plane external field\label{app:TBtheoryofhbGNRZEMwithEfield}}
The purposes of this subsection are twofold. Firstly, to provide an insight into the topological stability of the hbGNR ZEM. Here by topological stability we mean an impossibility to remove this mode from the gap by application of an external electrostatic potential applied in-plane of the GNR across its width. Secondly, to give an approximate analytical description of the hbGNR ZEM in the external potential. This analytical description implies obtaining expressions for the ZEM electron wave function as function of the external potential at the Dirac point as the main point of interest.

In order to achieve these goals, we slightly change the applied external potential in such a way that it ensures the same increment, $\Delta$, of the potential between the atomic sites. This approach is an accurate model of the finite non-dimerized  Su-Schrieffer-Heeger chain subjected to the external electrostatic field. For GNRs, however, such a potential configuration may seem more challenging since it requires a precise engineering of the potential applied to the ribbon within its unit cell length scale. Nevertheless, this configuration is not a hypothetical model, since the precise potential engineering has been demonstrated on nanomechanical systems~\cite{Xi2021}. It can be easily checked numerically that the modified potential leads to the same results as those presented in the main text of the manuscript, where the potential is generated by a homogeneous electrostatic field. This insensitivity to the specific details of the applied external potential already indicates the topological stability of the hbGNR ZEM.

The Hamiltonian of a hbGNR in the model field is
\begin{eqnarray}
    H_{\mathrm{hbGNR}} &=& \left(\matrix{
    \Delta & t_1 q & 0 & 0 & \ldots & 0 \cr
    t_1 q & 2 \Delta & t_1& 0 & \ldots & 0 \cr
    0 &  t_1 & 3 \Delta & \ddots & \ddots & \vdots\cr
    \vdots &  \ddots & \ddots & \ddots & t_1 q & 0\cr
    0 & \ldots & 0 & t_1 q & (N_r-1) \Delta & t_1 \cr
    0 & \ldots & 0 & 0 & t_1 & N_r \Delta
    }\right) \, , \label{eq:hbGNRTBHamiltonianInEfield}
\end{eqnarray}
where $N_r$ is the number of atoms in the ribbon unit cell that is an odd integer and $\Delta$ is the difference of the electrostatic potentials induced on the neighboring sites by an external field. By specialising to the case of the Dirac point, $k=2 \pi /3$, we get $q = 1$, so that the Hamiltonian
\begin{eqnarray}
    H_{\mathrm{hbGNR}} &=& \left(\matrix{
    \Delta & t_1 & 0 & 0 & \ldots & 0 \cr
    t_1 & 2 \Delta & t_1& 0 & \ldots & 0 \cr
    0 &  t_1 & 3 \Delta & \ddots & \ddots & \vdots\cr
    \vdots &  \ddots & \ddots & \ddots & t_1  & 0\cr
    0 & \ldots & 0 & t_1 & (N_r-1) \Delta & t_1 \cr
    0 & \ldots & 0 & 0 & t_1 & N_r \Delta
    }\right) \, \label{eq:hbGNRTBHamiltonianInEfieldAtDiracPoint}
\end{eqnarray}
has a simpler diagonal band structure. By placing the coordinate system origin in the center of the ribbon unit cell, the main diagonal of Eq.~(\ref{eq:hbGNRTBHamiltonianInEfieldAtDiracPoint}) can be made symmetric:
\begin{eqnarray}
    H_{\mathrm{hbGNR}} &=& \left(\matrix{
    - w\Delta & t_1 & 0 & 0 & \ldots & 0 \cr
    t_1 & -(w-1) \Delta & t_1& 0 & \ldots & 0 \cr
    0 &  t_1 & -(w-2) \Delta & \ddots & \ddots & \vdots\cr
    \vdots &  \ddots & \ddots & \ddots & t_1  & 0\cr
    0 & \ldots & 0 & t_1 & (w-1) \Delta & t_1 \cr
    0 & \ldots & 0 & 0 & t_1 & w \Delta
    }\right) \, , \label{eq:hbGNRTBHamiltonianInEfieldAtDiracPointSym}
\end{eqnarray}
where $w = \left(N_r -1\right)/2$. The eigenspace problem for the Hamiltonian given by the Eq.~(\ref{eq:hbGNRTBHamiltonianInEfieldAtDiracPointSym}) can be re-scaled. Since any diagonalizable matrices $M$ and $\widetilde{M}$, such that $\widetilde{M} = M/s$, where $s \neq 0$, have the same set of eigenvectors and are characterized by eigenvalues that are related as $\widetilde{\lambda} = \lambda / s$, the said eigenspace problem can be substituted by the eigenspace problem for the Hamiltonian
\begin{eqnarray}
    \widetilde{H}_{\mathrm{hbGNR}} &= \left(\matrix{
    - 2 w & \tau & 0 & 0 & \ldots & 0 \cr
    \tau & -2 (w-1) & \tau& 0 & \ldots & 0 \cr
    0 &  \tau & -2 (w-2) & \ddots & \ddots & \vdots\cr
    \vdots &  \ddots & \ddots & \ddots & \tau  & 0\cr
    0 & \ldots & 0 & \tau & 2 (w-1) & \tau \cr
    0 & \ldots & 0 & 0 & \tau & 2 w
    }\right) \, , \label{eq:hbGNRTBHamiltonianInEfieldAtDiracPointSymRes}
\end{eqnarray}
where $\tau = t_1/(\Delta/2)$ is a single variable of this problem. The factor of $2$ in the $\tau$ definition and on the main diagonal of Eq.~(\ref{eq:hbGNRTBHamiltonianInEfieldAtDiracPointSymRes}) is introduced for the purpose of further analytical treatment. We are going to show that $\widetilde{H}_{\mathrm{hbGNR}}$ always admits a zero eigenvalue as valid solution with a well-defined eigenvector. 

The set of simultaneous equations for the eigenspace problem in question is
\begin{eqnarray}
    \left(\matrix{
    - 2 w & \tau & 0 & 0 & \ldots & 0 \cr
    \tau & -2 (w-1) & \tau& 0 & \ldots & 0 \cr
    0 &  \tau & -2 (w-2) & \ddots & \ddots & \vdots\cr
    \vdots &  \ddots & \ddots & \ddots & \tau  & 0\cr
    0 & \ldots & 0 & \tau & 2 (w-1) & \tau \cr
    0 & \ldots & 0 & 0 & \tau & 2 w
    }\right) \left(\matrix{
    c_1 \cr
    c_2 \cr
    c_3\cr
    \vdots\cr
    c_{N_r -1} \cr
    c_{N_r} 
    }\right) &=& E \left(\matrix{
    c_1 \cr
    c_2 \cr
    c_3\cr
    \vdots\cr
    c_{N_r -1} \cr
    c_{N_r} 
    }\right)\, , \label{eq:hbGNRTBHamiltonianInEfieldAtDiracPointSymResEP}
\end{eqnarray}
or equivalently
\begin{eqnarray}
    \left(-2 w - E\right) c_1 + \tau c_2 &=&0\, , \label{eq:hbGNRTBHamiltonianInEfieldAtDiracPointSymResEP1} \\ 
    \tau c_{j-1} + \left[-2 \left(w - j + 1\right) - E\right] c_j + \tau c_{j+1} &=&0\, , \qquad j = 2 \ldots N_r - 1 \, ,\label{eq:hbGNRTBHamiltonianInEfieldAtDiracPointSymResEP2} \\
    \tau c_{N_r-1} + \left(2 w - E\right) c_{N_r} &=& 0 \, . \label{eq:hbGNRTBHamiltonianInEfieldAtDiracPointSymResEP3}
\end{eqnarray}
If $E=0$, then the set of equation for $j=2, \ldots, N_r - 1$ in Eqs.~(\ref{eq:hbGNRTBHamiltonianInEfieldAtDiracPointSymResEP2}) is automatically satisfied using a well-known Bessel function tri-term recurrence relation~\cite{BookGradsteyn2000}: $z J_{n-1}(z) + z J_{n+1}(z) = 2 n J_n (z)$. Namely, we shall choose $c_j = J_{w - j + 1}(\tau)$, where $j =  1, 2, \ldots, N_r$. Note that as $c_j$'s index $j$ goes from $1$ to $N_r$, the diagonal index on the main diagonal of Eq.~(\ref{eq:hbGNRTBHamiltonianInEfieldAtDiracPointSymResEP}) changes from $-w$ to $w$, while the order of the Bessel function solution of $c_j$'s changes from $w$ to $-w$. This choice of $c_j$ transforms each of the two remaining Eqs.~(\ref{eq:hbGNRTBHamiltonianInEfieldAtDiracPointSymResEP1}) and~(\ref{eq:hbGNRTBHamiltonianInEfieldAtDiracPointSymResEP3}) to the following equation: $-(t/s)\, J_{(N_r+1)/2}(\tau) = 0$. The found equation imposes restriction on the validity of the suggested eigenvector. Ideally, one would prefer an eigenvector that converts both Eqs.~(\ref{eq:hbGNRTBHamiltonianInEfieldAtDiracPointSymResEP1}) and~(\ref{eq:hbGNRTBHamiltonianInEfieldAtDiracPointSymResEP3}) into identities for any value of $\tau$. The suggested eigenvector, however, is an accurate solution only for such $\tau_i$'s that are zeros of $J_{(N_r+1)/2}(z)$: $J_{(N_r+1)/2}(\tau_i)=0$. Physically, this means that the proposed eigenvector is valid for only a discrete set of values of the external field. Owing to this limitation, the solution shall be referred to as {\it quasi-exact}. Since the above mentioned re-scaling of the eigenproblem by $\Delta/2$ does not affect the zero-energy, the eigenvalue is insensitive to the  presence of an external field. However, the corresponding eigenvector shows evidences that exposing hbGNR to the external field eliminates the sublattice polarization of the pristine ZEM. These observations are in agreement with the results in Fig.~2D of the main text and with considerations of the continuum ${\bf k} \cdot {\bf p}$-model in the supplementary text~\ref{app:LowEnergyTheoryDiracEquation}.

\subsection{Summary}
In this section, we have shown that hbGNR ZEM always contains a state that is pinned to the zero-energy at the Dirac point, which is unaffected by the presence of an external field. This insensitivity to the external field originates from the ZEM topological stability. The sublattice polarization of the zero-energy state is sensitive to the external field and it can be tuned by changing the applied field.

\section{Haldane and Kane-Mele models in commensurate and incommensurate ribbons\label{app:BulkBoundaryCorrespondenceCommVsIncomm}}
In this section, we analyse the classical topological edge states and their transport properties in commensurate and incommensurate systems. We consider two standard models: Haldane model of a Chern insulator~\cite{Haldane1988} and Kane-Mele model of a $\mathbf{Z}_2$ topological insulator~\cite{Kane2005a}. Also, our discussion is supplemented with the case of topological kink states that arise from the quantum valley Hall effect and are attributed to valley Chern insulators~\cite{Liu2021a}. In each case, we also pay particular attention to the bulk-boundary correspondence principle, the phase diagrams and an edge state dispersion transformation as a function of the topological transition driving parameter.

Before moving forward, let us first recollect the standard definition of a topological insulator and define three main types of the topological insulators in terms of their corresponding topological invariants. An integer quantum Hall effect is a regime, where a material becomes a topological insulator, that is, it is insulating in the bulk but conducting at the surface due to the presence of the edge states. These states are explained by the topological band theory and the bulk-boundary correspondence principle~\cite{Hasan2010}. The bulk-boundary correspondence says that the interface between a topological insulator and normal insulator produces a chiral gapless mode. The difference between left and right moving gapless modes, is given by the difference between the values of a topological invariant calculated for the bulk systems to the left and right from the interface. The proper topological invariant for the quantum Hall effect is the Chern number~\cite{Hasan2010}: $\Cal{C}_n = (2\pi)^{-1} \int_{\Omega} \nabla_{{\bf k}}  \times A_n({\bf k})\, d{\bf k}$, where $A_n({\bf k}) = i \langle u_n({\bf k}) \left|\nabla_{{\bf k}} \right| u_n({\bf k})\rangle$ is the Berry connection defined on the periodic parts of Bloch functions $\left| u_n({\bf k})\right.\rangle$ and the integration is carried out over the whole Brillouin zone $\Omega$. For the quantum spin Hall effect, the Chern invariant is zero. The proper topological invariant is the 2D $\mathbf{Z}_2$ invariant~\cite{Kane2005,Fu2006}: $\nu = (2\pi)^{-1} \left( \oint_{\partial \Omega_{1/2}} A({\bf k})\, d{\bf k} - \int_{\Omega_{1/2}} \nabla_{{\bf k}}  \times A({\bf k})\, d{\bf k}\right) \mathrm{mod}\; 2$, where $\Omega_{1/2}$ is the half of the Brillouin zone connected to another half by the time reversal symmetry, $\partial \Omega_{1/2}$ is the edge of $\Omega_{1/2}$ and $A({\bf k}) = \sum_n A_n({\bf k})$, where the sum runs over all occupied states. 
For the quantum valley Hall effect the proper topological invariant is a topological charge~\cite{Yao2009} or the valley Chern number~\cite{Wang2021}: $\Cal{C}_{n,{\bf K}({\bf K}^{\prime})} = (2\pi)^{-1} \int_{\Omega_{{\bf K}({\bf K}^{\prime})}} \nabla_{{\bf k}}  \times A_n({\bf k})\, d{\bf k}$, where $\Omega_{{\bf K}({\bf K}^{\prime})}$ stand for the {\bf K} and {\bf K}$^{\prime}$ valley regions. 
Thus, all gapless edge modes used as channels for dissipationless transport can be associated with a distinct topological invariant. For the clarity of the following discussion, we remind that term mode is used here as a synonym of a band rather than that of a state. The mode has a dispersion. We refer to the mode as the zero-energy one if it contains a zero-energy state which is a state at the Fermi level. The flavor of the topological insulator is determined by its topological invariant.

\subsection{Chern insulator}
Let us now switch on Haldane's $t_2$ term adiabatically and observe the changes in zigzag (commensurate) and half-bearded (incommensurate) ribbons. From the phase diagram of Haldane's model presented in Fig.~2 of Ref.~\cite{Haldane1988}, we expect the system to maintain two topological edge modes, since we use zero staggered potential. The bulk system is topologically non-trivial and characterized by Chern number $\mathcal{C}=1$. Therefore, when two interfaces with vacuum are formed, each of two edges must carry a ``one way" moving edge-localized mode. Both modes cross the bulk gap and pass through the zero energy. In agreement with the bulk-boundary correspondence, the difference between the number of left and right moving modes for each edge is equal to the difference of the Chern numbers across the interface: $N_L - N_R = \Delta \mathcal{C}  = \mathcal{C} - \mathcal{C}_{\mathrm{vac}} = 1$, where $\mathcal{C}_{\mathrm{vac}} = 0$ is the Chern number of the vacuum. 

\subsubsection{Commensurate ribbon}
Figure~S2 presents the energy bands and electron densities for the chosen $k$-points for small, $t_2 = 0.01 t_1$, and large, $t_2 = 0.1 t_1$, values of Haldane's parameter for zigzag and half-bearded GNRs. In a zigzag GNR, introducing a small next-to-nearest neighbor hopping parameter, $t_2$, results in a splitting of the partially flat energy bands at the Dirac points {\bf K}  and {\bf K}$^{\prime}$, while band degeneracy is preserved in the Kramer's time reversal (TR) invariant point $k = \pi$. In this latter point, the conduction band from {\bf K} ({\bf K}$^{\prime}$) valley connects to the valence band from {\bf K}$^{\prime}$({\bf K}) valley, forming a pair of ZEMs filling the bulk energy gap. As seen from the distribution of the electron densities in fig.~S2A, the wave functions of these two ZEMs are sharply localized at the opposite edges of the ribbon. This is exactly what is expected from the bulk-boundary correspondence principle. We should also notice that in $k$-space both ZEMs extend throughout the whole Brilloiun zone. Moreover, the TR symmetry links $k = \pi - q$ state in each mode to $k = \pi + q$ states in another mode, where $q$ is a relative momentum measured form $k = \pi$. In the vicinity of the TR invariant point $k = \pi$, the two modes have linear dispersion with opposite group velocities. Since the Haldane's term breaks the TR symmetry of the Hamiltonian, the scattering amplitude for any pair of TR symmetry related states is not vanishing;  see Chapter~4 of Ref.~\cite{BookBernevig2013} and our further discussion of $\mathbf{Z}_2$ topological insulators. Thus, an elastic backscattering is theoretically possible. However, the wave functions of the TR symmetry related states, circled 2-5 and circled 3-4, are spatially distant. This means that wider ribbons have wave functions of pair states, which do not overlap. Hence, the elastic backscattering is suppressed in such systems and dissipationless ballistic transport is realized. However, if the ribbon is narrow enough, then there is no physical reason forbidding the backscattering. Hence, dissipative transport shall be observed for narrow ribbons. For the commensurate ribbons, the wave functions spatial separation is a crucial factor that enables dissipantionless transport. This picture does not change for larger values of $t_2$ as shown in  fig.~S2C.

\subsubsection{Incommensurate ribbon}
In a half-bearded GNR, the same Haldane's term makes the fully flat ZEM dispersive, as shown in fig.~S2B. In this case, the initial ZEM represented by a straight horizontal line turns into a broken line due to two factors: (i) the shifting of energy states at {\bf K} and {\bf K}$^{\prime}$ point towards conduction and valence bands, respectively, and (ii) the pinning of the energy states at the TR invariant points $k=0$ and $\pi$. In contrast to the commensurate ZGNR, the broken line in hbGNR does not link valence and conduction bands for small values of $t_2$.
We arrive at similar conclusions when using larger values of $t_2$, as shown in fig.~S2D. The inset of fig.~S2D shows that for large $t_2$ such as $t_2 = 0.1 t_1$, the ZEM is still detached from the conduction and valence bands by a small avoided crossing. This means that this system can be turned into a true insulator as a result of doping by positioning the Fermi level within this avoided crossing. We have checked that the avoided crossing persists even for $t_2$ as large as $t_1$, though for such extra large values the bands overlap at different $k$'s so that the positioning of the Fermi level within avoided crossing gap is not possible. In all given cases, the ZEM is subdivided by {\bf K} and {\bf K}$^{\prime}$ points into two submodes, where wave functions are localized at the opposite edges of the ribbon. The bulk-boundary correspondence predicts two edge modes in a system with two edges, therefore we have to associate it with the two edge localized submodes. Another important observation is that there is a threshold value for $t_2$ to achieve an overlap between the ZEM and bulk bands. This implies that below the threshold $t_2$ the standard phase diagram of Haldane's model must have an electrostatic doping level as one of its dimensions, when dealing with incommensurate systems. Although such doping cannot change the Chern number and, therefore, the topological class of the material, it significantly affects the way a finite piece of topological material manifests itself in the transport. For instance, when Haldane's model hexagonal lattice is doped by applying an out-of-plane electric field, then trimming this lattice into an hbGNR can result in a system that is not conductive neither in the bulk nor at the edges. In this case, despite being originating from a topological insulator the system is physically equivalent to a trivial insulator. Concurrently, choosing a different level of doping involves the edges of the system into electronic transport again, which from the physical point of view corresponds to a standard well-known presentation of a topological insulator.

In the vicinity of the two TR invariant points, $k = 0$ and~$\pi$, the dispersion of the ZEM is linear. Contrary to the commensurate ZGNR, the states of the ZEM in hbGNR do not have TR partners. The ZEM states at $k = \pi-q$ lack the same energy partners located at $\pi + q$, as seen from the points circled 2-5 and circled 3-4 in Figs.~S2B and~S2D. Hence, the elastic backscattering is by default forbidden. In this case, the edge localization of the ZEM states wave functions and the width of the ribbon do not play any role in achieving the ballistic transport. Therefore, we infer that dissipationless transport in incommensurate ribbons can be realized via bulk modes too.

\subsection{{\bf Z}$_2$ topological insulator}
Now that we have revealed the main features of Chern insulators with respect to commensurate and incommensurate systems, we proceed to $\mathbf{Z}_2$ topological insulators. We adiabatically switch on the Kane-Mele spin-orbit term via the $t_2$ parameter. To highlight the difference between spin up and spin down states, we introduce a finite staggered on-site potential $\Delta = 2 t_2$ while we set the Rashba spin-orbit term to zero. Figure~S3 presents the energy bands and electron spin polarization densities for the chosen $k$-points in zigzag and half-bearded GNRs. According to the phase diagram of the Kane-Mele model presented in Fig.~1 of Ref.~\cite{Kane2005}, the bulk system is a 2D $\mathbf{Z}_2$ topological insulator characterized by $\nu = 1$ as the spin-orbit term is switched on. For the $\mathbf{Z}_2$ topological insulator, the bulk-boundary correspondence says that $\Delta \nu = (N_K\mathrm{mod}\; 2)$, where $\Delta \nu$ is the change of $\mathbf{Z}_2$ index across an interface and $N_K$ is the number of the Kramers pairs at the Fermi level for the edge states.

Since by default $\mathbf{Z}_2$ topological insulators deal with spinful particles, we briefly recall what Kramers pairs are and how they participate in elastic scattering. For half-integer spin particles, the TR symmetry operator $\mathcal{T}$ acting on a Bloch state $\psi$ flips not only the particle momentum but also its spin. Hence, Kramers pairs are formed by $\psi_{k,\uparrow} \leftrightarrow \psi_{-k,\downarrow}$ and $\psi_{k,\downarrow} \leftrightarrow \psi_{-k,\uparrow}$. Note that for one-dimensional periodic structures the momentum flip around $k=0$ is the same as the momentum flip around $k = \pi$; for the simplicity of the following discussion we will always refer to $k\leftrightarrow-k$ flip even though fig.~S3 shows Brillouin zones that are more adapted for $\pi+q \leftrightarrow \pi-q$ notation, where $q$ is the momentum shift from the $k=\pi$ point. For $\mathcal{T}$-symmetric Hamiltonians, $[\mathcal{T},H(k)] = 0 \Rightarrow \mathcal{T}H(k)\mathcal{T}^{-1} = H(-k)$. Thus, the scattering matrix element between the TR invariant Kramers pairs, i.e. when $k = -k$, is vanishing:
\begin{eqnarray}
    \langle \psi \left| H(k) \mathcal{T} \right| \psi \rangle &=& \langle \psi \left| \mathcal{T}H(k)\mathcal{T}^{-1} \mathcal{T} \right| \psi \rangle = \langle \psi \left| \mathcal{T} H(k) \right| \psi \rangle = \nonumber\\
    &=& \langle \psi \left| H^{\dagger}(k) \mathcal{T}^{\mathrm{T}} \right| \psi \rangle \stackrel{\left\{T^{\mathrm{T}} = - T\right\}}{=} - \langle \psi \left| H^{\dagger}(k) \mathcal{T} \right| \psi \rangle \stackrel{\left\{H^{\dagger} = H\right\}}{=} \nonumber \\
    & \stackrel{\left\{H^{\dagger} = H\right\}}{=}& - \langle \psi \left|H(k)  \mathcal{T} \right| \psi \rangle \, ,
    \label{eq:ScatteringMatrixElementBetweenKramersPairStates}
\end{eqnarray}
where we have used the fact that $H$ is self-adjoint (Hermitian) operator and $\mathcal{T}^{\mathrm{T}} = - \mathcal{T}$. The latter relation follows from the definition of TR operator 
\begin{equation}
    \mathcal{T} = U K \, ,
\end{equation}
where $U$ is a unitary matrix, i.e. $UU^{\dagger} = U^{\dagger}U = I$, and $K$ is the complex conjugate. Since $U$ is antisymmetric, $-U = U^{\mathrm{T}}$, the transposition of the TR operator yields 
\begin{equation}
    \mathcal{T}^{\mathrm{T}} = (UK)^{\mathrm{T}} = U^{\mathrm{T}}K = -UK = - \mathcal{T} \, .
    \label{eq:TransposeOfTimeReversalOperator}
\end{equation}
The antisymmetric character of $U$ comes from the condition imposed on $\mathcal{T}$ by the half-integer spin: $\mathcal{T}^{2} = -1 = (UK)(UK) = UU^{\ast}KK = UU^{\ast}$. Accounting for the unitarity of $U$, we then arrive at $-U^{\ast} = U^{-1} = U^{\dagger}$. Taking the complex conjugate from the left- and right-hand sides of the latter equality leads to the antisymmetric condition $-U = U^{\mathrm{T}}$ used in Eq.~(\ref{eq:TransposeOfTimeReversalOperator}). 

The fundamental property of a particle with a half-integer spin is that $\mathcal{T}^{2} = -1$. It originates from the fact that $\mathcal{T}$ flips the spin of the particle thereby rotating it around some axis by angle $\pi$. This rotation for a particle with the spin equal to $1/2$ can be realized by choosing $U = i \sigma_y$, where 
$$\sigma_y = \left(\matrix{
  0 & -i\cr
  i & 0
}\right)$$ is Pauli matrix. This yields $\mathcal{T}^2 = (i \sigma_y K) (i \sigma_y K) = i \sigma_y (-i) \sigma^{\ast}_y KK = \sigma_y \sigma^{\ast}_y = - \sigma_y^2 = -1$.
The proven relation in Eq.~(\ref{eq:ScatteringMatrixElementBetweenKramersPairStates}) shows that the elastic backscattring is forbidden between all Kramers pairs of a given pair of bands so that only a spin-conserving scattering is allowed: $\psi_{k,\uparrow} \leftrightarrow \psi_{-k,\uparrow}$ or $\psi_{k,\downarrow} \leftrightarrow \psi_{-k,\downarrow}$. This is an important corollary that we will use for analysis of the transport properties of $\mathbf{Z}_2$ topological insulators with respect to the commensurate and incommensurate GNRs.

\subsubsection{Commensurate ribbon}
For a ZGNR, the effect of the spin-orbit term is similar to that in Haldane's model. For both spin polarized pairs of partially flat bands, we observe splittings at the Dirac points {\bf K} and {\bf K}$^{\prime}$, while their degeneracy is preserved at the TR invariant point $k=\pi$. When the staggered potential is zero ($\Delta = 0$), a fourfold degeneracy is found at $k = \pi$ point, since ZEM~$1$, ZEM~$2$, ZEM~$3$, and ZEM~$4$ pass through zero energy in this point. As seen from fig.~S3A, a finite $\Delta = 2 t_2$ shifts the crossing between ZEM~$1$ and ZEM~$4$ from $k = \pi$ towards the {\bf K} valley. The crossing between ZEM~$2$ and ZEM~$3$ is shifted towards {\bf K}$^{\prime}$ valley. 
As expected from the bulk boundary correspondence, each edge hosts two modes with opposite spin polarizations. Having opposite spin polarizations, ZEM~$1$ and ZEM~$3$ form a Kramers pair.
As seen from spin polarization densities in points circled~3 and circled~4 in fig.~S3A, the ZEM~$1$ and ZEM~$3$ are both localized at the same left edge. They cross the Fermi level $E_f = 0$ at the $k$ points related by the TR symmetry. Thus, the number of Kramers pairs for the left edge localized ZEMs is $N_K = 1$. Then $(N_K \mathrm{mod}\; 2) = 1$, which is exactly equal to the change of the $\mathbf{Z}_2$ index across the left ribbon edge: $\Delta \nu = \nu - \nu_{\mathrm{vac}} = 1 - 0 = 1$, where $\nu$ and $\nu_{\mathrm{vac}}$ are the $\mathbf{Z}_2$ index of the bulk Kane-Mele system and the vacuum, respectively. Similar reasoning applies to the right edge of the ribbon, where ZEM~$2$ and ZEM~$4$ are localized as seen from spin polarization densities in points circled~2 and circled~5 in fig.~S3A. Thus, four ZEMs are expected for the system with two edges, which is indeed observed. 

The four topological ZEMs have peculiar transport properties due to the preserved TR symmetry. Equation~(\ref{eq:ScatteringMatrixElementBetweenKramersPairStates}) states that the backscattering between Kramers pairs is strictly forbidden as long as TR symmetry is respected by the Hamiltonian. Thus, no scattering must be present between ZEM~$1$ and ZEM~$3$ as well as between ZEM~$2$ and ZEM~$4$. Equation~(\ref{eq:ScatteringMatrixElementBetweenKramersPairStates}), however, does not forbid spin-conserving elastic scattering between ZEM~$1$ and ZEM~$4$ as well as ZEM~$2$ and ZEM~$3$. For $\Delta = 0$, the same energy states in these pairs of modes are symmetrically positioned at $k$ and $-k$. Thus, the momentum and energy conservation laws in the elastic backscattering between these pairs of modes is respected. The probability of this process, however, is given by the overlap of the wave functions of the corresponding states. It is seen from points circled~4 and circled~5 in fig.~S3A that the wave functions in these spin-conserving pairs are localized at the opposite edges of the ribbon, so that the overlap is negligible. Hence, the elastic backscattering does not occur. In summary, for commensurate ZGNR, the TR symmetry and wave function spatial distancing are important in achieving a reliable dissipationless ballistic transport.

\subsubsection{Incommensurate ribbon}
For the case of a hbGNR, the Kane-Mele spin-orbit term shifts the spin up and down states at {\bf K} ({\bf K}$^{\prime}$) points in the opposite directions while preserving their degeneracy at the TR invariant Kramers points $k = 0$ and $\pi$. This spin state shifting breaks both spin down ZEM~$1$ and spin up ZEM~$2$ into two regions: region~I located between {\bf K} and {\bf K}$^{\prime}$ points, and region~II that is a complement of the region~I to the full Brillouin zone. In each region, the wave functions of the spin up and down ZEM states are localized at the same ribbon edge as shown by the spin polarization density in points circled~1, circled~3 and circled~4 of fig.~S3B. At the boundary of regions~I and~II the wave functions are extended throughout the whole width of the ribbon. This picture differs from the hbGNR case in the Haldane's model, where the wave function preserves the localized character while being distributed between the opposite ribbon edges in equal fractions. In the given hbGNR case, the ZEM localized at the edges of the ribbon do not connect the valence and conduction bands. As can be seen in the inset of fig.~S3B, there is an avoided crossing between ZEM~$1$ and the valence bands. Despite considerable spin-orbit interaction that drives the bulk system to the non-trivial $\mathbf{Z}_2$ topological regime with $\nu = 1$, the hbGNR being cut from the bulk system can be converted to a true insulator by positioning the Fermi level within the avoided crossing. 

Similar to the considered case in Haldane's model, the bulk-boundary correspondence can be formally satisfied if we subdivide both ZEM $1$ and ZEM $2$ into submodes residing within the above mentioned regions~I and~II. Within the region~I, we have a Kramers pair formed by points circled~3 and circled~4 in fig.~S3B. These points correspond to states with different spin polarization, and they are positioned at TR symmetry related momenta $k$ and $-k$. The wave functions of the corresponding states are localized at the zigzag (left) edge of the ribbon. Hence, $N_K = 1$, therefore $(N_K \mathrm{mod}\; 2) = 1 = \Delta \nu$. In the same way, we could consider a pair of states in region~II, thereby obtaining two additional submodes and a total of four $\mathbf{Z}_2$ topological edge submodes for the system with two edges. Also, we have checked that ZEMs do not touch any of the valence and conduction bands even for as large spin-orbit coupling as $t_2 = t_1$. The dispersion of the ZEMs at the TR invariant points $k = 0$ and~$\pi$ is almost linear. In contrast to the Haldane's model, hbGNR ZEMs in Kane-Mele model have TR partners thereby forming Kramers pairs. However, as we have shown in Eq.~(\ref{eq:ScatteringMatrixElementBetweenKramersPairStates}), the elastic backscattering is forbidden between the Kramers pairs as long as the TR symmetry is respected. On the other hand, the considered ZEMs lack spin-conserving partners for which the backscattering would be allowed. Again, we come to the conclusion that the edge localization is not important for achieving ballistic transport in this incommensurate GNR. Thus, the dissipationless transport could be realized by bulk modes.

\subsection{Quantum valley Hall effect}
Similar considerations can be given to the case of the quantum valley Hall (QVH) regime and topological kink states in the commensurate zigzag and incommesurate half-bearded GNRs. In analysing these systems, one should take into account that TR symmetry operator switches valleys, and the suppression of elastic backscattering comes from valley distancing in $k$-space rather than from the vanishing of the scattering matrix elements between the Kramers pairs. The case of commensurate GNR is presented in main text Fig.~2C, while the case of incommensurate ribbon is shown in fig.~S4. Two ingredients are essential in achieving QVH regime in monolayer ribbons: (1) the breaking of inversion symmetry by introducing a staggered sublattice potential leading to the energy gap opening, and (2) creating two kinks in the staggered potential by the side gate voltages. The former induces a non-zero Berry curvature within each of the two valleys of graphene lattice, making the valley Chern number a well-defined topological invariant. The side gate voltage pushes the partially flat bands within the bulk gap, thereby creating channels for ballistic transport. Notice that for the hbGNR there is only one band that is localized on a single edge. Thus, we again arrive at the same conclusion above that dissipationless transport in incommensurate systems can be realized via bulk bands. The in-plane electric field applied to the hbGNR in the main text is a single ingredient that achieves the same ballistic regime. Moreover, when combined with an electrostatic doping, the in-plane electric field allows tuning between the bulk and edge mode character of the dissipationless transport.

\subsection{Summary}
To conclude, we have seen that incommesurate systems possess peculiarities in both electronic and transport properties. The behaviour of incommensurate systems considerably deviates from that of commensurate systems for all standard flavours of the topological insulators: Haldane's, Kane-Mele's and Liu's models representing standard Chern, $\mathbf{Z}_2$ and valley Chern topological insulators. Not only can incommensurate systems provide dissipationless tranport via edge mode, but also dissipationless transport via bulk modes. In all the considered above cases, incommensurate systems can be driven into true insulators by electrostatic doping, making the standard phase diagrams less practical without accounting for doping dimension. For all above-considered systems, the dispersion of ZEMs at the intrinsic Fermi level, $E_{\mathrm{F}} = 0$, is quasi-linear for all values of the primary model parameter $t_2$ [Haldane's and Kane-Mele's models] or $U$ [Liu's model of the QVH effect in monolayer ribbons~\cite{Liu2021a}]. Moreover, no cubic-type dispersion is observed.

\section{Low-energy theory of hbGNR in continuum model\label{app:LowEnergyTheoryDiracEquation}}

In condensed matter systems, the Dirac equation arises from the low-energy limit of the Schr{\"o}dinger equation around special degeneracy points of the energy bands. These special points represent topological objects acting as the Berry curvature sources in the $k$-space and resembling Dirac monopoles. The graphene {\bf K} and {\bf K}$^{\prime}$ points are classical examples of Berry curvature monopoles referred to as Dirac points. The low-energy description of graphene has been known for the long time~\cite{DiVincenzo1984a}. When supplemented by hard-wall boundary conditions, it has also been used to describe the electronic properties of graphene nanoribbons~\cite{Brey2006,Akhmerov2008} and finite length metallic single-wall carbon nanotubes~\cite{McCann2004}. In particular, the two best known GNR geometries with zigzag and armchair edges have been considered and the derivation of partially flat zero-energy bands featuring edge localized states has been reported~\cite{Brey2006}. However, a similar systematic description is currently missing for the hbGNRs fully flat zero-energy band. Therefore, here we develop a relevant analytic treatment of hbGNRs zero-energy bands as well as deepen understanding of their fundamental behavior in an external in-plane electric field and supporting claims of the main text with respect to the intimate relation of ZEM to the quantum field theory, topology and chaos and non-linear dynamics.

\subsection{Bulk energy bands\label{app:LowEnergyTheoryDEBulkBands}}
Following the case of zigzag GNR in Ref.~\cite{Brey2006}, we start from the Dirac equation and amend the boundary condition in order to get a secular equation for the hbGNR. For a GNR oriented along $y$-axis, the electron momentum along this axis is a good quantum number. Hence, in {\bf K} valley the Dirac equation for a two-component spinor reads
\begin{eqnarray}
    - i \partial_x \phi_B + i \kappa_y \phi_B &=& \varepsilon \phi_A \, , \label{eq:DiracEquation4GNR1} \\
    - i\partial_x \phi_A - i \kappa_y \phi_A &=& \varepsilon \phi_B \, , \label{eq:DiracEquation4GNR2} 
\end{eqnarray}
where $\kappa_y$ is the electron momentum along the ribbon measured with respect to the Dirac point. Expressing $\phi_A$ from Eq.~(\ref{eq:DiracEquation4GNR1}) as $\phi_A = - \left(i/\varepsilon\right)\partial_x \phi_B + \left(i \kappa_y / \varepsilon \right) \phi_B$ and putting it into Eq.~(\ref{eq:DiracEquation4GNR2}), we arrive at a single second order equation:
\begin{equation}
    - \partial_{xx} \phi_B + \kappa_y^2 \phi_B = \varepsilon^2 \phi_B \, .
    \label{eq:DiracEquation4GNR2ndOrder} 
\end{equation}
Equation~(\ref{eq:DiracEquation4GNR2ndOrder}) has a general solution $\phi_B = C_1 e^{zx} + C_2 e^{-zx}$, with $z = \sqrt{\kappa_y^2 - \varepsilon^2}$ and $C_{1,2}$ as constants determined from the boundary and normalization conditions. For a zigzag GNR, the outermost atomic sites at the opposite edges belong to different sublattices. The wave function must vanish on the next auxiliary atomic sites~\cite{Akhmerov2008}, therefore the boundary condition for the ribbon of width $L$ is $\phi_B(0) = \phi_A(L) = 0$. In contrast, the outermost edge atoms of a hbGNR belong to the same sublattice, say sublattice $A$. Hence, the wave function must vanish on the next auxiliary atomic sites belonging to sublattice $B$. Therefore, we set for the hbGNR
\begin{equation}\phi_B(0) = \phi_B(L) = 0\, . \label{eq:DiracEquationhbGNRbc}\end{equation}
This condition yields the following set of equations:
\begin{eqnarray}
    C_1 e^{z \cdot 0} + C_2 e^{-z \cdot 0} &=& 0\, , \nonumber \\ C_1 &=& - C_2 \,, \label{eq:DiracEquationBC1} \\
    C_1 e^{z L} + C_2 e^{-z L} &=& 0 \, , \nonumber \\ 
    2 C_1 \sinh \left(z L\right)&=& 0 \, , \label{eq:DiracEquationBC2}
\end{eqnarray}
Equation~(\ref{eq:DiracEquationBC1}) links the two initially independent constants $C_1$ and $C_2$, leaving only one degree of freedom represented by either constant $C_1$ or $C_2$. This remaining degree of freedom is fixed by normalization condition. Equation~(\ref{eq:DiracEquationBC2}) leads to the secular equation $\sinh \left(zL\right) = 0$. This secular equation has only one solution $z=0$, whence it follows that $\varepsilon = \pm \left|\kappa_y\right|$. It is obvious that the chosen method of solving equation does not provide the correct description of the ZEM. The method leads to two principle problems: (i) there are two bands marked by `$+$' and `$-$' that normally attributed to the conduction and valence bands, (ii) the obtained dispersion is linear. In contrast to the derived result, the ZEM in question is a sole mode that pertains to neither conduction nor valence band. Since the external electric field is absent, the ZEM must also be independent of $\kappa_y$ rather than a mode which exhibits linear dispersion. Thus, a different approach is needed even for the most basic description of the mode not mentioning its spectacular transformations in the external field from fully flat to the cubic dispersion. The failure in finding a correct description of the ZEM, however, does not mean that the initial method is not valid for the bulk bands of hbGNRs. By substituting $z = i \kappa_x$, one can transform the above-mentioned secular equation to $\sin \left(\kappa_x L\right) = 0$ [cf. with Eq.~(\ref{eq:SecularEquation})], which leads to quantization of the transverse electron momentum $\kappa_x$: $\kappa_x = \pi n / L$, with $n$ being integer. Using this momentum quantization, the valence and conduction energy bands are described by $\varepsilon = \pm \sqrt{\kappa_y^2 + \left(\pi n / L\right)^2}$, where $n =  \pm 1, \pm 2, \ldots$. The option $n = 0$ shall be excluded from the energy band consideration for it leads to the already discussed $\varepsilon = \pm \left|\kappa_y\right|$. 

While the failure to achieve a valid description of the ZEM within the given formalism can make us skeptical on the presented bulk states, we shall notice the following. One more argument supporting the validity of the bulk bands description of hbGNRs can be given based on their isospectral feature with respect to the energy bands of the tubes. Indeed, imposing the periodic boundary conditions on Eqs.~(\ref{eq:DiracEquation4GNR1}) and~(\ref{eq:DiracEquation4GNR2}), $\phi_B(0)=\phi_B(L)$ and $\partial_x \phi_B(0) = \partial_x \phi_B(L)$, one can derive the following secular equation for the circumferencial momentum of electrons, $\kappa_x$: $e^{i \kappa_x C} = 1$, where $C$ is the tube circumference. The secular equation leads to $\kappa_x = 2 \pi n / C$, with $n$ being integer, which results in the tube energy bands that are identical to hbGNR bulk energy bands once $L = C / 2$. The bulk bands derived from the low-energy description of hbGNR are, indeed, isospectral with those of the corresponding armchair tube. Also, the same structural relation takes place as in the tight-binding model between the width and circumference of the isospectral ribbon and tube (see fig.~S1B).

To find normalized spinor wave functions $\Psi = \left(\phi_A, \phi_B\right)^{\mathrm{T}}$ for the bulk bands of the hbGNR, it is convenient to consider the ribbon to be positioned symmetrically with respect to $x=0$. For this case, the boundary conditions are $\phi_B(-L/2) = \phi_B (L/2) = 0$ [cf. with Eqs.~(\ref{eq:DiracEquationhbGNRbc})]:
\begin{eqnarray}
    C_1 e^{- z L/2} + C_2 e^{z L/2} &=& 0 \,, \nonumber \\
    C_1 e^{z L/2} + C_2 e^{- z L/2} &=& 0 \, . \label{eq:DiracEquationbc2}
\end{eqnarray}
The set of Eqs.~(\ref{eq:DiracEquationbc2}) is a homogeneous system of linear equations with respect to $C_1$ and $C_2$. Non-trivial solutions of such system can be found only if the determinant of its matrix is zero. This condition defines the quantization of $\kappa_x$, which holds the same form as before: $\kappa_x = \pi n / L$, with $n$ being positive integer.
From the second equation in Eqs.~(\ref{eq:DiracEquationbc2}), we get $C_2 = - C_1 e^{i \kappa_x L}$ so that $\phi_B = - 2 i C_1 e^{i \kappa_x L/2} \sin \left[\kappa_x \left(L/2 - x\right)\right]$ and $ \phi_A = \pm 2 C_1 e^{i \kappa_x L/2} \cos\left[\kappa_x \left(L/2 - x\right) + \varphi \right]$, where `$+$' and `$-$' in $\phi_A$ are used for the conduction and valence band, respectively, and $\varphi = \arcsin \left( \kappa_y/\sqrt{\kappa_y^2+\kappa_x^2}\right)$. Hence,
the normalized spinor is
\begin{equation}
    \Psi = \left(\matrix{
    \phi_A \cr
    \phi_B
    }\right) =  \frac{1}{\sqrt{L}} \left(\matrix{
    \pm \cos \left[\kappa_x \left(L/2 - x \right) + \varphi \right]    \cr
    -i \sin \left[\kappa_x \left(L/2 - x \right) \right]
    }\right)  \, ,
    \label{eq:DiracEquation4GNRSpinorPsi}
\end{equation}
where we have disregarded the global phase $e^{i \kappa_x L/2}$ obtained in the derivation.

\subsection{Zero-energy mode\label{app:LowEnergyTheoryDEZEM}}
Since the seminal paper by Jackiw and Rebbi~\cite{Jackiw1976}, the zero-energy modes (ZEMs) of the Dirac fermions have been a subject of the intensive research in the quantum field theory due to their relation to purely mathematical index theorem~\cite{Witten2020} and exotic concepts such as charge fractionalization~\cite{Niemi1986,Jackiw2012}, Majorana fermions~\cite{Jackiw2012,Yanagisawa2020} and superconducting strings~\cite{Witten1985}. Taking this into account, it is quite natural to make the conjecture that the ZEM of hbGNR can be described in a similar approach (see also Ref.~\cite{BookShen2012}).

The Dirac equation itself is invariant with respect to the TR operation. The hbGNR ZEM in the tight-binding model is also TR symmetric. Namely, the two valleys around {\bf K} and {\bf K}$^{\prime}$ points are related by the $\mathcal{T}$-operator. In a single valley description of ZEM, the TR symmetry is locally broken while globally it is preserved. Breaking of the TR symmetry implies the mode chirality. The chiral mode can be described within the Dirac equation formalism 
when the ZEM is an eigenspace mode of the particle conjugation or chiral symmetry operator represented by the Pauli matrix $\sigma_z$ (see Refs.~\cite{Jackiw1981,Ryu2002}). 
In this case, the two Eqs.~(\ref{eq:DiracEquation4GNR1}) and~(\ref{eq:DiracEquation4GNR2}) collapse into one. Following this prescription, we can rewrite Eqs.~(\ref{eq:DiracEquation4GNR1}) and~(\ref{eq:DiracEquation4GNR2}) in a matrix form:
\begin{equation}
    \left( - i \sigma_x \partial_x - \sigma_y \kappa_y \right) \Psi = \varepsilon \Psi \, ,
    \label{eq:DiracEquation4GNRPauliMatrices}
\end{equation}
where $\Psi = \left(\phi_A, \phi_B \right)^\mathrm{T}$. By setting $\varepsilon = 0$ and multiplying the left and right hand sides by $\sigma_x$, we arrive at
\begin{equation}
    \partial_x \Psi = - \sigma_z \kappa_y \Psi \, .
    \label{eq:DiracEquation4GNRPauliMatricesZEM}
\end{equation}
Equation~(\ref{eq:DiracEquation4GNRPauliMatricesZEM}) has a standard form of the Jackiw-Rebbi ZEM equation: $\partial_x \Psi = m (x) \sigma_z \Psi$, where $m(x)$ is the spatial mass-kink profile~\cite{Jackiw2012,BookShen2012,Cayssol2021}. However, in contrast to the standard Jackiw-Rebbi equation, see Refs.~\cite{Jackiw2012,BookShen2012,Cayssol2021} and references therein, in Eq.~(\ref{eq:DiracEquation4GNRPauliMatricesZEM}) the gap generating mass does not depend on the spatial coordinates but rather on some external parameter. In other words, we deal with the case $\partial_x \Psi = m (\eta) \sigma_z \Psi$, where $m(\eta)$ is the mass term profile in a parameter space. For solids, the reciprocal space is a natural parameter space.  When $m \equiv k $ in a 1D system, the reciprocal space origin provides a sign change in the mass profile .
Thus, in Eq.~(\ref{eq:DiracEquation4GNRPauliMatricesZEM}) we deal with a $k$-dependent mass. This mass is somewhat similar to the Haldane mass~\cite{Hasan2010}, but it does not need inversion symmetry to have a sign change.

Equation~(\ref{eq:DiracEquation4GNRPauliMatricesZEM}) reduces to a scalar equation if spinor $\Psi$ is an eigenvector of $\sigma_z$: $\sigma_z V_{\lambda} = \lambda V_{\lambda}$, with $\lambda_{+}=1, V_{+} = \left(1, 0\right)^{\mathrm{T}}$ and $\lambda_{-}=-1, V_{-} = \left(0, 1\right)^{\mathrm{T}}$. Hence, $\Psi_{\lambda} = V_{\lambda} \, \chi (x)$, where $\chi(x)$ is the scalar function to be found from 
\begin{equation}
    \partial_x \chi (x) = - \lambda \kappa_y \chi(x) \, .
    \label{eq:DiracEquation4GNRZEMscalar}
\end{equation}
The solution of this first order equation is $\chi_{\pm}(x) = A \exp\left(-\lambda_{\pm} \kappa_y x\right)$, carrying only one constant that can be fixed by the normalization condition. Thus, the spinor is \begin{equation}
    \Psi_{\pm}(x) = V_{\pm} A \exp\left(- \lambda_{\pm} \kappa_y x\right)\, .
    \label{eq:DiracEquation4GNRPauliMatricesZEMSolution}
\end{equation}
The spinor is exponentially decaying or growing for $\kappa_y \neq 0$ and it is constant for $\kappa_y = 0$, which aligns well with the behaviour of ZEM known from the tight-binding model. Moreover, eigenvectors of $\sigma_z$ exhibit the two-component polarization which can be attributed to the sublattice polarization of the ZEM. There are, however, substantial difficulties in interpretation if we construct the ZEM as the original Jackiw-Rebbi solution in Ref.~\cite{Jackiw1976}. As follows from the tight-binding results, the sublattice polarization of ZEM is preserved even when the localization of the wave function changes form zigzag to bearded edge of the ribbon, i.e. when $\kappa_y > 0$ changes to $\kappa_y<0$ in {\bf K} valley. However, to produce a normalizible Jackiw-Rebbi solution, $\lambda$ should be chosen for $\kappa_y > 0$ and $\kappa_y<0$ so that to ensure that spinors $\Psi_{\pm}(x\rightarrow \pm \infty) \rightarrow \mathbf{0}$:
\begin{eqnarray}
    \kappa_y > 0: \; \Psi_{+} &=& A V_{+} \exp\left(-\kappa_y x\right)\,, \nonumber \\ 
    \kappa_y < 0: \; \Psi_{-} &=& A V_{-} \exp\left(\kappa_y x\right) \, .
    \label{eq:DiracEquation4GNRPauliMatricesZEMsol}
\end{eqnarray}
As one can see from Eqs.~(\ref{eq:DiracEquation4GNRPauliMatricesZEMsol}), neither the sublattice polarization nor the wave function localization can be properly described by thus constructed solution. According to Eq.~(\ref{eq:DiracEquation4GNRPauliMatricesZEMsol}) the sublattice polarization changes as $\kappa_y$ changes sign. Also, the wave function is always localized at the same interface $x=0$ and does not shift upon $\kappa_y$ variation. Furthermore, the boundary condition  $\Psi_{\pm}(x \rightarrow \pm \infty) \rightarrow \mathbf{0}$ is incompatible with that given by Eq.~(\ref{eq:DiracEquationhbGNRbc}).

The above reasoning is true only within the original Jackiw-Rebbi framework from Ref.~\cite{Jackiw1976} that disregards the exponential solutions on the basis of their non-normalizability. But this is true only for an infinite system. The exponential solutions are normalizable on a finite region of space. Therefore, if one accounts for the finite transverse size of the hbGNR, then the counterpart of the Jackiw-Rebbi solution is a valid physical entity found in real-world systems. For fully polarized modes provided by  Eq.~(\ref{eq:DiracEquation4GNRPauliMatricesZEMSolution}) the boundary conditions in Eq.~(\ref{eq:DiracEquationhbGNRbc}) are automatically satisfied. As shown in fig.~S5A, all outer atoms of hbGNR (blue) belong to the same sublattice $A$, while the next lattice nodes (red) all belong to another sublattice $B$. The hardwall boundary condition demands the wave function being zero on all red sites that belong to $B$ sublattice in  fig.~S5A. This requirement is automatically fulfilled once the spinor $\Psi$ is proportional to the eigenvector of $\sigma_z$ with zero $B$ component: $V_{+} = \left(1, 0\right)^{\mathrm{T}}$. In this way, we correctly choose the solution between the two spinors provided by Eq.~(\ref{eq:DiracEquation4GNRPauliMatricesZEMSolution}). It becomes obvious now that the arbitrary constant from the general solution of the first order Eq.~(\ref{eq:DiracEquation4GNRZEMscalar}) corresponds only to the normalization degree of freedom. 
The proper solution is 
\begin{equation}
    \Psi_{+}(x) = \left(\matrix{1 \cr 0}\right) C_1 \exp\left(-\kappa_y x\right) \, .
    \label{eq:DiracEuationhbGNRJackiwRebbiExponentialGrowth}
\end{equation}
This solution looks similar to the localized boundary mode in Ref.~\cite{Witten2020} (cf. with Eq.~(2.5) in there), but is is not propagating along the boundary. It is convenient to normalize the solution given by Eq.~(\ref{eq:DiracEuationhbGNRJackiwRebbiExponentialGrowth}) assuming that the ribbon is placed symmetrically with respect to $x=0$. Owing to the limited width of the nanoribbon, the integration has a finite result. Hence, the normalization constant is readily found:
\begin{eqnarray}
    \int_{-L/2}^{L/2} \Psi_{+}^{\dagger}(x) \Psi_{+}(x) d\, x &=& 1 \, , \nonumber \\
    \left|C_1\right|^2 \int_{-L/2}^{L/2} \exp\left(-2 \kappa_y x\right) d\, x &=& 1\, , \nonumber \\
    \left|C_1\right| &=& \sqrt{\frac{\kappa_y}{\sinh (\kappa_y L)}} \, .
\end{eqnarray}
Thus, the wave function is
\begin{equation}
    \Psi_{+}(x) = \left(\matrix{1 \cr 0}\right) \sqrt{\frac{\kappa_y}{\sinh (\kappa_y L)}}\, \exp\left(-\kappa_y x\right) \, .
    \label{eq:DiracEuationhbGNRJackiwRebbiExponentialGrowthNorm}
\end{equation}
Equation~(\ref{eq:DiracEuationhbGNRJackiwRebbiExponentialGrowthNorm}) describes the behaviour of the ZEM wave function familiar from the tight-binding model in the main text Fig.~2A. As seen in fig.~S5B, the wave function for $\kappa_y > 0$ is exponentially growing towards the left edge which shows the high localization on the zigzag edge. A similar localization is seen for the bearded edge as shown in  fig.~S5C, when $\kappa_y < 0$. Finally, the wave function is constant throughout the ribbon interior when $\kappa_y = 0$ as shown in fig.~S5D. To summarize, the variation of $\kappa_y$ around the Dirac point ($\kappa_y = 0$) shifts the ZEM wave function from one edge to another while preserving the sublattice polarization and its chiral character. Only in the vicinity of the Dirac point the wave function loses localization. In all of these cases, the wave function is still the eigenfunction of the chiral symmetry operator $\Cal{S} = \sigma_z$. Thus, the asymmetric electron density distribution attributed to the ZEM chirality in Fig.~2 is well-described in the continuum model and further supported by a deeper argument based on the $\Cal{S}$ operator eigenfunctions. Note that once $B$ sublattice provides outer atoms of the hbGNR, the bearded and zigzag edges swap their positions as seen in fig.~S5A. This is also accounted for by Eq.~(\ref{eq:DiracEquation4GNRPauliMatricesZEMSolution}). The $B$ sublattice analog of Eq.~(\ref{eq:DiracEuationhbGNRJackiwRebbiExponentialGrowthNorm}) carries not only a different eigenvector but also an additional minus sign in the exponent:
\begin{equation}
    \Psi_{-}(x) = \left(\matrix{0 \cr 1}\right) \sqrt{\frac{\kappa_y}{\sinh (\kappa_y L)}}\, \exp\left(\kappa_y x\right) \, ,
    \label{eq:DiracEuationhbGNRJackiwRebbiExponentialGrowthNormA}
\end{equation}
that swaps the wave function localization from the left (right) to right (left) side for $\kappa_y>0$ ($\kappa_y < 0$). The swapping of the wave function localization corresponds to the switching of zigzag and bearded edges.

The solutions given by Eqs.~(\ref{eq:DiracEuationhbGNRJackiwRebbiExponentialGrowthNorm}) and~(\ref{eq:DiracEuationhbGNRJackiwRebbiExponentialGrowthNormA}) are equivalent to Eq.~9 describing isolated Majorana bound states in Ref.~\cite{Fu2008} but they do not describe Majorana fermions. For the $2 \times 2$ Dirac equation $i\partial_t\Psi = H \Psi$, with $H$ given by Eq.~(\ref{eq:DiracEquation4GNRPauliMatrices}), the particle-hole (charge conjugation) symmetry operator is $\Xi = \sigma_x K$, where $K$ denotes complex conjugation. Acting with this $\Xi$ operator on $\Psi_{\pm}$ in  Eqs.~(\ref{eq:DiracEuationhbGNRJackiwRebbiExponentialGrowthNorm}) and~(\ref{eq:DiracEuationhbGNRJackiwRebbiExponentialGrowthNormA}), we see that $\Xi \Psi_{\pm} \neq \Psi_{\pm}$, which means the found ZEM is not a Majorana fermion. 

It is known that chiral states appear whenever TR symmetry $\Cal{T}$ is broken. Usually, these chiral states require presence of an external magnetic field or an intrinsic net zero magnetic flux like in the Haldane model. However, the breaking of $\Cal{T}$ occurs for the ZEM of a hbGNR due to the peculiar hbGNR edges and boundary conditions. For the above $2 \times 2$ Dirac equation  $\Cal{T} = i \sigma_y K$. Hence, $\Cal{T} \Psi_{+}(\kappa_y) = - \Psi_{-}(-\kappa_y)$ which means that $\Psi_{+}$ mode does not have a TR partner. Although $\Cal{T} \Psi_{+}(\kappa_y)$ is an eigenstate of $H(-\kappa_y) \equiv H_{\mathbf{K}^{\prime}}(\kappa_y)$, with $H_{\mathbf{K}^{\prime}} = - i \sigma_x \partial_x + \sigma_y \kappa_y$ being Hamiltonian in {\bf K}$^{\prime}$ valley, this $\Cal{T} \Psi_{+}(\kappa_y)$ eigenstate corresponds to imposing boundary conditions on another sublattice. A more detailed picture is the following. In the {\bf K}$^{\prime}$ valley one finds ZEM solutions $\Psi^{\mathbf{K}^{\prime}}_{+} = \left(\matrix{1 \cr 0}\right) \sqrt{\frac{\kappa_y}{\sinh (\kappa_y L)}}\, \exp\left(\kappa_y x\right)$ and $\Psi^{\mathbf{K}^{\prime}}_{-} = \left(\matrix{0 \cr 1}\right) \sqrt{\frac{\kappa_y}{\sinh (\kappa_y L)}}\, \exp\left(-\kappa_y x\right)$, so that in general we have $\Cal{T}\Psi^{\mathbf{K}}_{-} = \Psi^{\mathbf{K}^{\prime}}_{+}$ and $\Cal{T}\Psi^{\mathbf{K}}_{+} = -\Psi^{\mathbf{K}^{\prime}}_{-}$, where $\Psi^{\mathbf{K}}_{\pm}$ are given by Eqs.~(\ref{eq:DiracEuationhbGNRJackiwRebbiExponentialGrowthNorm}) and~(\ref{eq:DiracEuationhbGNRJackiwRebbiExponentialGrowthNormA}). In the hbGNR, the TR partner of ZEM is absent due to the inversion of the sublattice polarization by $\Cal{T}$ which provides another argument for supporting  the chiral character of the ZEM.

The formally defined charge conjugation and TR symmetry operators are defined with respect to the sublattice degree of freedom or pseudo-spin. In order to see how this makes them different with respect to the real spin analogs it is useful to consider symmetry classes as we do further.

At the level of continuum model, the system is described by  fundamental symmetries such that $\Cal{T}^2 = -\mathbf{1}_{2 \times 2}$, $\Xi^2 = \mathbf{1}_{2 \times 2}$, and $\Cal{S} = \mathbf{1}_{2 \times 2}$, where $\mathbf{1}_{2 \times 2}$ stands for a  $2 \times 2$ unit matrix. This means the low-energy Hamiltonian of our one-dimensional system belongs to the DIII Altland-Zirnbauer symmetry class~\cite{Hasan2010}. In contrast, the full Hamiltonian from the tight-binding model belongs to BDI symmetry class, where it is characterized by $\Cal{T}^2 = \mathbf{1}_{2 \times 2}$, $\Xi^2 = \mathbf{1}_{2 \times 2}$, and $\Cal{S} = \mathbf{1}_{2 \times 2}$. This symmetry classification can be verified by taking the graphene Hamiltonian 
$$H(\mathbf{k}) = \left(\matrix{0 & h(\mathbf{k})\cr
h^{\ast}(\mathbf{k}) & 0}\right)$$
and noticing that $\mathrm{Re}[h(\mathbf{k})] = \mathrm{Re}[h(-\mathbf{k})]$ and $\mathrm{Im}[h(\mathbf{k})] = - \mathrm{Im}[h(-\mathbf{k})]$. Then we clearly see that $\Cal{T}H(\mathbf{k})\Cal{T}^{-1} = H(-\mathbf{k})$ for $\Cal{T} = \mathbf{1}_{2\times2} K$, which is the definition of the TR symmetry operator for a spinless particle since $\Cal{T}^2 = \mathbf{1}_{2\times2}$. The definition of the chiral symmetry operator does not change for the full Hamiltonian, $\Cal{S} = \sigma_z$. However, the charge conjugation symmetry operator becomes $\Xi = \Cal{S} \Cal{T} = \sigma_z K$ so that $\Xi^2 = \mathbf{1}_{2\times2}$. Therefore, the BDI class is identified. In the above-mentioned identification of the DIII symmetry class, the TR symmetry operator acts on the sublattice pseudo-spin instead of the real spin of an electron. This important difference between the continuum and tight-binding models must be taken into account in interpretation of the $\Xi$- and $\Cal{T}$-breaking results. For instance, the $\Cal{T}$ acting on the real spin is unbroken for the tight-binding model.

Having now solutions for both {\bf K} and {\bf K}$^{\prime}$ valleys, we notice that if the intervalley and inter-sublattice (i.e. inter-ribbon) mixings were achievable simultaneously, one would be able to construct a Majorana pseudo-fermion (i.e. characterized by a pseudo-spin in lieu of a real spin), since there would then be a way of exploiting the solutions as follows: $\Psi_{\mathrm{M}} = \Psi^{\mathbf{K}}_{+} + \Psi^{\mathbf{K}^{\prime}}_{-} = \Psi^{\mathbf{K}}_{+} + \Xi \Psi^{\mathbf{K}}_{+}  = \left(\matrix{1 \cr 1}\right) \sqrt{\frac{\kappa_y}{\sinh (\kappa_y L)}}\, \exp\left(-\kappa_y x\right)$, where $\left(\matrix{1 \cr 1}\right)$ is an eigenvector of $\Xi \sim \sigma_x$. This relation between the hbGNR ZEM and a Majorana pseudo-fermion is similar to the connection between the ZEM of a 2D Dirac fermion bounded to a real space vortex and a Majorana fermion, cf. with Eq.~(35) in Ref.~\cite{Yanagisawa2020}. The hbGNR ZEM also resembles the edge ZEM hosted in a finite piece of a square lattice with a real space vortex, see Fig.~2e of Ref.~\cite{Chamon2008}. Since similar spatial vortexes have been constructed for an infinite honeycomb lattice~\cite{Hou2007}, the ZEM analogs must also exist in finite structures on the honeycomb lattice. In some sense hbGNR ZEM can be considered as a 1D analog of those modes to be found on the honeycomb lattice.

A brief comment is needed on the exponential growth in  Eq.~(\ref{eq:DiracEuationhbGNRJackiwRebbiExponentialGrowthNorm}) when $\kappa_y \rightarrow +\infty$ and $x>0$. Formally, $\kappa_y$ is not bounded variable in the continuum model. This unbounded $\kappa_y$ is the source of potential misinterpretations. In reality, the range of electron wave vectors is limited by the Brillouin zone. Therefore, the assumption $\kappa_y \rightarrow +\infty$ is not a valid physical case.

{\it To conclude this subsection}, the analytical description based on the Dirac equation has revealed a range of distinctive features making the ZEM of hbGNR very different from all previously reported ZEMs in condensed matter systems and quantum field theory.

\subsection{Topological  properties analysis: non-linear dynamics and index theory perspective\label{app:IndexTheoryOfZEM}}
A general intuition tells us that whenever one encounters an edge localized state, it must be topological and characterized by some non-zero topological invariant. It happens, however, that depending on a chosen topological perspective, the situation may be quite the opposite. By noticing that the Eqs.~(\ref{eq:DiracEquation4GNR1}) and~(\ref{eq:DiracEquation4GNR2}) depend only on one independent variable $x$, we can draw an analogy with typical dynamic equations featuring time as an independent variable. This allows one to make use of the whole apparatus of non-linear dynamics and chaos theory~\cite{BookStrogatz1994} for the analysis of the given equations. In what follows, we analyse both bulk and edge eigenstates of hbGNRs via phase portraits and indexes of their phase space trajectories.

The phase portrait of a dynamic system is the graphical representation of a {\it vector field} in phase space. The phase space is the space featuring the coordinate and momentum as independent variables. Generally speaking, for any second order ordinary differential equation, the phase space coordinates can be defined as the sought function and its first derivative. 

The principle difference between classical dynamic equations and Eqs.~(\ref{eq:DiracEquation4GNR1}) and~(\ref{eq:DiracEquation4GNR2}) is that classical coordinates are always real while wave functions admit complex values. However, the wave functions can always be chosen to be real provided that the magnetic field is absent in the system.
For the equations in question this can be achieved by the following unitary transform of the Hamiltonian: $$U = \left(\matrix{
  1 & 0\cr
  0 & -i
}\right)$$ and $\Cal{H}(\kappa_y) = U H(\kappa_y) U^{\dagger}$, so that the new set of equations for analysis is
\begin{eqnarray}
    \partial_x \phi_B - \kappa_y \phi_B &=& \varepsilon \phi_A \, , \label{eq:RealEquation4GNR1} \\
    - \partial_x \phi_A - \kappa_y \phi_A &=& \varepsilon \phi_B \, . \label{eq:RealEquation4GNR2} 
\end{eqnarray}
It is possible to reduce these two equations to a single second order equation $\partial_{xx} \phi_A + \kappa_y^2 \phi_A = \varepsilon^2 \phi_A$ and, following the standard prescription of the non-linear dynamic theory, to define the phase space with coordinates $(\phi_A,\partial_x \phi_A)$. This form of phase space is convenient for studying stability of fixed points, when the dynamics occurs along a one-dimensional manifold. For our case, it is, however, more natural to define the phase space in terms of coordinates $(\phi_A, \phi_B)$. The latter is also possible since the two spaces are related by the following geometric transform: $$\left(\matrix{\phi_A \cr \phi_B }\right) = \left(\matrix{ 1 & 0 \cr -\kappa_y \varepsilon^{-1} & -\varepsilon^{-1}  }\right) \left(\matrix{\phi_A \cr \partial_x \phi_A }\right)\,. $$
Another important difference between classical dynamic systems and the hbGNR eigenproblem is that the former is the Cauchy problem while the latter is a boundary condition problem, i.e. Dirichlet, Neumann or the mixture of the two, which sometimes referred to as the Robin problem. The solution of the Cauchy problem is uniquely determined by the initial conditions, for example $\phi_A(0) = a$ and $\partial_x \phi_A (0) = b$, where $a$ and $b$ are arbitrary constants. On the other hand, the solution of a boundary condition problem is specialized by the boundary values. For $\phi_A$'s equation presented above, the boundary conditions are $\phi_B(-L/2) = \phi_B(L/2)$, which can be recast as 
\begin{eqnarray}
    -\varepsilon^{-1} \partial_x \phi_A(-L/2) - \kappa_y \varepsilon^{-1} \phi_A(-L/2) &=& 0 \, , \nonumber \\
    -\varepsilon^{-1} \partial_x \phi_A(L/2) - \kappa_y \varepsilon^{-1} \phi_A(L/2) &=& 0 \, . 
    \label{eq:RobinsBCexample}
\end{eqnarray}
Equations~(\ref{eq:RobinsBCexample}) is the Robin boundary condition. The difference between the types of problems, nevertheless, does not prevent us from presenting the solution as a parametric trajectory in the phase space. The final ingredient we need in order to perform an analysis in terms of phase portraits is a {\it vector field}. In the non-linear dynamics, the vector field is introduced by an analogy with the field of velocities in a fluid. In other words, this field is defined by the gradient of independent variables. Thus, if the field of velocities is $\mathbf{v} = \partial_t ( x (t), y(t))$ in the real space, then it is $\mathbf{v}_{A} = \partial_x (\phi_A (x), \partial_x \phi_A(x))$ in the classical phase space with $\phi_A$'s from the above problem. Continuing this analogy, we can easily define the vector field for the modified phase space presented above, namely $\mathbf{v}_{AB} = \partial_x (\phi_A(x), \phi_B(x)) = (-\kappa_y \phi_A - \varepsilon \phi_B, \varepsilon \phi_A + \kappa_y \phi_B)$.

Solving the eigenvalue problem given by Eqs.~(\ref{eq:RealEquation4GNR1}) and~(\ref{eq:RealEquation4GNR2}) in the same way as in~\ref{app:LowEnergyTheoryDEBulkBands}, we obtain the normalized solution
\begin{eqnarray}
    \left(\matrix{
    \phi_A (x) \cr
    \phi_B (x)
    }\right) &=& \frac{1}{\sqrt{L}} \left(\matrix{
    \cos \left[\left(L/2 + x\right) \kappa_x  + \varphi \right] \cr
    \pm \sin \left[ \left( L/2 + x \right)\kappa_x\right]
    }\right) \, ,
\end{eqnarray}
where `$+$' and `$-$' stand for conduction and valence band states, respectively, and where $\kappa_x$, $\varphi$, and $\varepsilon$ are the same as have been found previously. The ZEM solutions are still given by Eq.~(\ref{eq:DiracEuationhbGNRJackiwRebbiExponentialGrowthNorm}), since unitary matrix $U$ does not affect them owing to their sublattice polarization.

Figure~S6 shows the phase portraits of the vector field and the typical phase space trajectories of bulk eigenstates. We can observe in fig.~S6A that the vector field features a {\it fixed point} at the origin and a counterclockwise flow around the origin. The direction of this flow is opposite for the valence (clockwise) and conduction (counter-clockwise) band states, while the shape varies from circular for $\kappa_y = 0$ to elliptic for $\kappa_y \neq 0 $. For conduction bands, the eccentricity of the ellipse increases with increasing absolute values of $\kappa_y$. Furthermore, the major axis is positioned within the 2nd(1st) and 4th(3rd) quadrants of the phase space for $\kappa_y < 0$ ($\kappa_y > 0$). This picture is reversed for the valence band. 

As shown in fig.~S6B, when the parameter $x$ changes from $-L/2$ to $L/2$, the point representing a value of the given bulk eigenstate in the phase space follows the direction of the vector field thereby drawing either circular or elliptic trajectory. The main effect of increasing the ribbon width $L$ is the decreasing of size of the trajectory, i.e. diameter of a circle or major axis of an ellipse. This is followed by some effect on the eccentricity, when $\kappa_y \neq 0$. When $x$ is changing over the full $x$-domain, the angle swept out by the line connecting the origin of the phase space and the point representing the wave function at $x$ is a function of the state number $n$. This angle quantizes as $\pi n$.

The fixed points in the phase portraits of dynamical systems are topological structures. They are characterized by an index that is defined as the vector field winding number for a closed curve $\Omega$ oriented counterclockwise:
\begin{equation}
    I_{\nu} = \frac{1}{2 \pi }\oint_{\Omega} d\nu \, , \label{eq:IndexDefinition}
\end{equation}
where $\nu = \arctan\left(v_y/v_x \right)$. 
Before proceeding, let us note that Eq.~(\ref{eq:IndexDefinition}) is a special case of the residue theorem and Cauchy integral formula in complex analysis. Namely, Eq.~(\ref{eq:IndexDefinition}) can be re-written in other equivalent forms by introducing a complex variable 
$v = v_x + i v_y = v_0 e^{i \nu}$, where $v_0$ and $\nu$ is the module and phase of the variable, respectively. Then $\nu = (1/i) \ln\left(v/v_0\right)$, so that $I_{\nu} = (1 / (2 \pi i)) \oint \partial_{\nu} \ln\left(v\right) \, d\nu = (1/(2 \pi i)) \oint dv/v$, where we have used that $d\nu = (1 / i)  d \ln\left(v/v_0\right) = (1 / i) (v_0/v) \, (dv/v_0) = (1 / i) dv/v = (1 / i) d \ln\left(v\right)$. Defining the winding number via logarithm seems to be more popular throughout modern physics literature, see for instance Eq.~(5) in Ref.~\cite{Kane2005}. However, it obscures interpretation in terms of the familiar complex analysis results. Equation~(\ref{eq:IndexDefinition}) might suffer from the same problem, but at least its interpretation is clear with respect to the vector field. Hereafter, we shall stick to the index definition as the winding number given by the Eq.~(\ref{eq:IndexDefinition}).

The index is a non-zero integer if $\Omega$ contains a fixed point inside. For many types of fixed points including the {\it center} in fig.~S6 $I_{\nu} = +1$. Only for a saddle fixed point $I_{\nu} = -1$. When $\Omega$ encloses no fixed points, $I_{\nu} = 0$. The index is additive. For several fixed points inside the $\Omega$, it is a sum of indexes for these points. Being topological structures, the fixed points cannot arise and disappear on their own. They can be created and annihilated only in pairs conserving the value of the total index, which represents the index conservation law for fixed point collisions.

The index definition given by Eq.~(\ref{eq:IndexDefinition}) can be extended to non-closed curves -- strings, but its quantization in this case is not guaranteed for such strings. Using this extended definition, it is possible to calculate a string index of the bulk eigenstate trajectories $\Omega_n$ as follows:
\begin{equation}
    I_{n} = \frac{1}{2 \pi} \int_{-L/2}^{L/2} \frac{\kappa_x \left[\varepsilon \left(\phi_A^2 + \phi_B^2\right) + 2 \kappa_y \phi_A \phi_B\right]}{2 \left(\varepsilon \phi_A + \kappa_y \phi_B\right)^2}\, dx \, , \label{eq:IndexDiracEquation}
\end{equation}
where the integration path is parameterized by $x$ and is oriented according to this parameterization. Thus, the new string index depends on the orientation of the vector field. If the vector field changes its direction in each point of the phase space to the opposite one, like this happens for conduction and valence band energies, then the string index acquires the opposite sign. The bulk state string index is quantized by $1/2$, i.e. it is half-integer for odd $n$ and integer for even $n$. Then, the string index is positive for the conduction band states and is negative for the valence band states. At this point, we can assume that the edge states of hbGNR ZEM must be characterized by a zero string index, which is indeed the case. Therefore, one can see that bulk states are topological from the index theory point of view.

When both $\varepsilon = 0$ and $\kappa_y = 0$, the vector field is identically vanishing throughout the whole phase space since $\mathbf{v}_{AB} = (-\kappa_y \phi_A - \varepsilon \phi_B, \varepsilon \phi_A + \kappa_y \phi_B) = (0,0)$. This means the vector field is non-defined for the zero-energy state at the Dirac point. The string index, however, can still be evaluated by Eq.~(\ref{eq:IndexDiracEquation}), which vanishing integrand tells us that the index is identically zero. Provided that $\varepsilon = 0$ and $\kappa_y \neq 0$, the vector field does exist but it features now a saddle fixed point at the origin. Hence, we see that the bulk and the edge states reside in the vicinity of the fixed points of different type. In principle, the saddle point could allow us to have a non-zero index for trajectories winding around it. However, we can see that this is not the case and the string index is zero even for $\kappa_y \neq 0$. The ZEM given by  Eq.~(\ref{eq:DiracEuationhbGNRJackiwRebbiExponentialGrowthNorm}) is not affected by the unitary transform $U$. Therefore, as have been mentioned above, it is a valid solution for the present analysis. Since the ZEM is sublattice polarized, i.e. $\phi_B = 0$, the ZEM trajectory in the phase space is a horizontal axis. It happens that this axis is also one of the principle manifolds of the saddle point. Since the vector field direction is constant along principle manifolds in any half-plane of the phase space, the index could be non-zero only if the trajectory were passing through the saddle point. Because the latter is not permissible, we conclude that the ZEM trajectory must be a straight horizontal line in either half-plane of the phase space as shown in  fig.~S7.

{\it To summarize our analysis in this subsection}, we have seen that the hbGNR ZEM is characterized by an index with `trivial' zero value from the index theory perspective. In contrast to this, all bulk states indexes are either integers or half-integers quantized by $1/2$. The vector field fixed point located at the origin of the phase space transforms from a center fixed point for bulk states into a saddle fixed point for ZEM. This transformation breaks the index conservation law for fixed point collisions and it is reminiscent of the degenerate Hopf bifurcation, whereby the fixed point stability can be changed without colliding fixed points.

\subsection{Zero-energy mode in the external electric field\label{app:LowEnergyTheoryDEZEMinEfield}}
Similar to the Jackiw-Rebbi soliton in 2D~\cite{BookShen2012}, the hbGNR ZEM can be used for charge transport. Although the reported ZEM is motionless, i.e. $\varepsilon = 0 \neq \varepsilon(\kappa_y)$ so that its group velocity $v_{\mathrm{gr}} \sim \partial_{\kappa_y}\varepsilon = 0$, it can be made dispersive in the presence of an external electric field. This subsection presents the analytical treatment of the external field effect.

The hbGNR in an external in-plane electric field can be described by the modified Dirac equation supplemented with the boundary conditions that we have revealed considering the hbGNR without the external field. The Dirac equation with the external electric field is
\begin{equation}
    \left( - i \sigma_x \partial_x - \sigma_y \kappa_y - F x\right) \Psi = \varepsilon \Psi \, ,
    \label{eq:DiracEquation4GNRPauliMatricesEfield}
\end{equation}
where $\Psi = \left(\phi_A, \phi_B \right)^\mathrm{T}$ and $F$ is the field strength. By multiplying both side of Eq.~(\ref{eq:DiracEquation4GNRPauliMatricesEfield}) by $i \sigma_x$, we get
\begin{equation}
    \left( \partial_x + \sigma_z \kappa_y - i \sigma_x (F x + \varepsilon)\right) \Psi = 0 \, .
    \label{eq:DiracEquation4GNRPauliMatricesEfield0}
\end{equation}
Hence, it is easy to see that the external field $F$ leads to the term proportional to $\sigma_x$ thereby sublattice mixing is introduced in the system. In terms of the spinor components $\phi_{A,B}$ the equation is
\begin{eqnarray}
    \partial_x \phi_A + \kappa_y \phi_A - i \left(\varepsilon + F x\right) \phi_B &=& 0 \,, 
    \label{eq:DiracEquation4GNRPauliMatricesEfield1}\\
    \partial_x \phi_B - \kappa_y \phi_B - i \left(\varepsilon + F x\right) \phi_A &=& 0 \,.
    \label{eq:DiracEquation4GNRPauliMatricesEfield2}
\end{eqnarray}
When $\varepsilon + F x = 0$ and $\kappa_y \neq 0$, we get back to the flat ``zero"-energy mode residing in the light blue region of fig.~S8A and described by Eq.~(\ref{eq:DiracEquation4GNRPauliMatricesZEM}). The eigenenergy of this state seems to be position dependent. Recall, however, that the ZEM wave function is highly localized at either edge of the ribbon (see main text Fig.~2D). Therefore, the energy of the flat region is $\varepsilon = \pm F L/2$ for zigzag and bearded edge localized wave functions. This describes well the flat flanks of the dispersive in-gap mode in the main text Figs.~2D and~2E.

For further analysis, a new variable $\alpha = \varepsilon + F x$ is introduced and considered for the case $\alpha \neq 0$. From Eq.~(\ref{eq:DiracEquation4GNRPauliMatricesEfield2}), we express $\phi_A$ as $\phi_A = \left(\partial_x \phi_B - \kappa_y \phi_B\right)/\left(i \alpha \right)$ and insert this into Eq.~(\ref{eq:DiracEquation4GNRPauliMatricesEfield1}). We obtain the following equation for the spinor component $\phi_B$:
\begin{equation}
    \partial_{xx} \phi_B - \frac{F}{\alpha} \partial_x \phi_B + \left(\frac{F \kappa_y}{\alpha} - \kappa_y^2 + \alpha^2\right) \phi_B = 0 \, .
    \label{eq:DiracEquation4GNRSingleComponentphiBEfield}
\end{equation}
Then, changing of the independent variable $\xi = \alpha / F = \left(\varepsilon + F x\right) / F$ in Eq.~(\ref{eq:DiracEquation4GNRSingleComponentphiBEfield}) yields
\begin{equation}
    \partial_{\xi \xi} \phi_B - \frac{1}{\xi} \partial_{\xi} \phi_B + \left(\frac{\kappa_y}{\xi} - \kappa_y^2 + F^2 \xi^2\right) \phi_B = 0 \, .
    \label{eq:DiracEquation4GNRSingleComponentphiBEfield1}
\end{equation}
The derived equation is quite involved for analysis, and it does not have a general solution in terms of elementary functions. Therefore, its study will be presented later. 
However, the said equation can be solved exactly at the Dirac point, i.e. within the light red region in  fig.~S8A. Indeed,  we have 
\begin{equation}
    \partial_{\xi \xi} \phi_B - \frac{1}{\xi} \partial_{\xi} \phi_B + F^2 \xi^2 \phi_B = 0 \, ,
    \label{eq:DiracEquation4GNRSingleComponentphiBEfieldinK}
\end{equation}
for the case of $\kappa_y = 0$, with the solution space
\begin{equation}
    \phi_B (\xi) = C_1 \cos \left(\frac{F\xi^2}{2}\right) + C_2 \sin \left(\frac{F\xi^2}{2}\right) \, .
    \label{eq:DiracEquation4GNRinEfieldComponentphiBinKSol}
\end{equation}

Unlike Eq.~(\ref{eq:DiracEquation4GNRZEMscalar}),  Eq.~(\ref{eq:DiracEquation4GNRSingleComponentphiBEfieldinK}) is a second-order differential equation, therefore the general solution given by Eq.~(\ref{eq:DiracEquation4GNRinEfieldComponentphiBinKSol}) carries two arbitrary constants. The two constants can be determined as for the bulk bands. The two boundary condition equalities are $\phi_B(\xi_1) = 0$ and $\phi_B(\xi_2) = 0$, where $\xi_1 = (\varepsilon/F) - (L/2)$ and $\xi_2 = (\varepsilon/F) + (L/2)$. These two boundary conditions lead to a set of simultaneous equations given by
\begin{eqnarray}
     C_1 \cos \left(\frac{F\xi_1^2}{2}\right) + C_2 \sin \left(\frac{F\xi_1^2}{2}\right) &=& 0 \, , \nonumber \\
     C_1 \cos \left(\frac{F\xi_2^2}{2}\right) + C_2 \sin \left(\frac{F\xi_2^2}{2}\right) &=& 0 \, .
     \label{eq:DiracEquation4GNRinEfieldComponentphiBinKSolBC}
\end{eqnarray}
The set of Eqs.~(\ref{eq:DiracEquation4GNRinEfieldComponentphiBinKSolBC}) has non-zero solutions only if the determinant of its matrix is zero. This condition then defines the dispersion equation:
\begin{eqnarray}
      \cos \left(\frac{F\xi_1^2}{2}\right) \sin \left(\frac{F\xi_2^2}{2}\right) - \sin \left(\frac{F\xi_1^2}{2}\right) \cos \left(\frac{F\xi_2^2}{2}\right)  &=& 0 \, , \nonumber \\
     \sin \left[\frac{F}{2}\left(\xi_1^2 - \xi_2^2\right)\right] &=& 0 \,  , \nonumber \\
     \sin\left(\varepsilon L\right) &=& 0 \, ,   \label{eq:DiracEquation4GNRinEfieldComponentphiBinKSolDispersionEquation}
\end{eqnarray}
where we have substituted the definitions of $\xi_{1}$ and $\xi_2$ in the last line. Solving Eq.~(\ref{eq:DiracEquation4GNRinEfieldComponentphiBinKSolDispersionEquation}), a set of field-independent energy states $\varepsilon = \pi n/ L$ is found for $n = 0, \pm 1, \pm 2, \ldots$, which explains the unchanging bulk band gap for the the numerical results in the main text Fig.~2.

When the dispersion equation given by Eq.~(\ref{eq:DiracEquation4GNRinEfieldComponentphiBinKSolDispersionEquation}) is satisfied, the two equations from the set of  Eqs.~(\ref{eq:DiracEquation4GNRinEfieldComponentphiBinKSolBC}) are dependent. Thus, the arbitrary constants $C_1$ and $C_2$ cannot be uniquely defined from this set of equations. The only information we can get is the relation between the two constants. For our purpose either of the two equations can be used. From the second equation, we have $C_1 = - C_2 \tan \left(F \xi_2^2/2\right)$, therefore $\phi_A$ and $\phi_B$ can be expressed in terms of $C_2$'s only. The remaining constant $C_2$ can be fixed by normalizing the whole spinor $\Psi$: $C_2 = \cos\left(F \xi_2^2/2\right)/\sqrt{L}$. Thus, the normalized spinor can be expressed as
\begin{eqnarray}
    \Psi = \left(\matrix{
    \phi_A \cr
    \phi_B
    }\right) &=&  \frac{1}{\sqrt{L}} \left(\matrix{
    i \cos \left[\frac{F}{2} \left(\frac{2 \varepsilon}{F} + \frac{L}{2} + x\right)\left(\frac{L}{2} - x \right)\right]    \cr
    \phantom{i} \sin \left[\frac{F}{2} \left(\frac{2 \varepsilon}{F} + \frac{L}{2} + x\right)\left(\frac{L}{2} - x \right)\right]
    }\right)  \, .
    \label{eq:DiracEquation4GNRinEfieldSpinorPsiInK}
\end{eqnarray}
This spinor $\Psi$ is in agreement with that in Eq.~(\ref{eq:DiracEquation4GNRSpinorPsi}) derived for the case $\varepsilon \neq 0$ without accounting for the external field. By setting $F \rightarrow 0 $ in Eq.~(\ref{eq:DiracEquation4GNRinEfieldSpinorPsiInK}) and making $\kappa_y \rightarrow 0$ in  Eq.~(\ref{eq:DiracEquation4GNRSpinorPsi}), the normalized spinor becomes
\begin{eqnarray*}
    \Psi &=& \frac{1}{\sqrt{L}} \left(\matrix{
    i \cos \left[\frac{\pi n}{2} - \frac{\pi n x}{L} \right]    \cr
    \pm \sin \left[\frac{\pi n}{2} - \frac{\pi n x}{L} \right]
    }\right)
    \qquad \mathrm{from~(\ref{eq:DiracEquation4GNRinEfieldSpinorPsiInK})}\\
    \tilde{\Psi} &=& \frac{1}{\sqrt{L}} \left(\matrix{
    \pm \cos \left[\frac{\pi n}{2} - \frac{\pi n x}{L}  \right]    \cr
    -i \sin \left[\frac{\pi n}{2} - \frac{\pi n x}{L}  \right]
    }\right)
    \qquad \mathrm{from~(\ref{eq:DiracEquation4GNRSpinorPsi})} \, .
\end{eqnarray*}
These two spinors $\Psi$ and $\tilde{\Psi}$ are related by $i$ factor, $\Psi = i \tilde{\Psi}$, for the conduction bands and by $-i $ factor, $\Psi = - i \tilde{\Psi}$, for the valence bands.

For the ZEM with $\varepsilon = 0$, the argument of sine and cosine functions in Eq.~(\ref{eq:DiracEquation4GNRinEfieldSpinorPsiInK}) gains an explicit form, $(F/2) \left(L/2 + x \right) \left(L/2 - x \right)$. Then, it is seen that at the edges of the ribbon $x = \pm L/2$, the argument vanishes and leads to $\phi_B = 0$ as required by the boundary conditions. Concurrently, we have $\phi_A = i/\sqrt{L}$.
The obtained spinor provides a good quantitative description of the ZEM electron density around the Dirac point when the external field is applied. In Fig.~2D of the main text, the electron density on a sublattice that supplies the outer edge atoms to the hbGNR, drops in value at the middle of the ribbon, while the density on another sublattice has a peak value. As presented in  fig.~S8B, a similar feature is found in Eq.~(\ref{eq:DiracEquation4GNRinEfieldSpinorPsiInK}), where $A$ sublattice supplies the edge atoms of the hbGNR. This result is an evidence of $\varepsilon = 0$ as a valid solution for the ZEM energy in an external field. Thus, the ZEM dispersion is an odd function of the electron momentum, since it passes through the coordinate origin $\kappa_y =0$, and it is characterized by opposite signs of energy to the left and right of this origin when $\kappa_y \neq 0$. Hence, the ZEM dispersion in the external field must be linear in the first approximation. It is a unidirectional mode with either right or left moving, i.e. it is chiral mode.

{\it In summary}, we have shown in this subsection that the ZEM transforms to a band once placed in the external field. The band width scales linearly with the field strength. In this band, the ZEM has a linear dispersion around the Dirac point which allows us to use it for charge transport.

\subsection{hbGNR as non-stationary dynamical system: string index in a turbulent vector field\label{app:IndexTheoryOfZEMinEfield}}
When properties of a dynamical system change over a given time interval, the system is referred to as {\it non-stationary} one. From a mathematical point of view, non-stationary systems are described by differential equations with time-dependent coefficients. This results in the vector field being non-constant so that the phase space point representing the state of the system can follow some wiggly phase trajectory somewhat similar to a particle moving in a turbulent fluid flow (see Movie~S1). Hereafter we show for hbGNR in an external in-plane electric field how the eigenproblem secular equation can be constructed and solved and how corresponding eigenstates can be numerically found and presented as phase space trajectories. Finally, we demonstrate the impact of the external electric field on the index description of the hbGNR eigenstates. Other questions of interest such as ZEM transformation and persistent chirality in the external field are also discussed in detail.

As one can see from Eq.~(\ref{eq:DiracEquation4GNRPauliMatricesEfield}), by placing hbGNR in the external in-plane electrostatic field $F$, we introduce a coefficient that is $x$-dependent. The $x$-parameter plays the role of effective time of a dynamical system. Using the same unitary transform as before, we transform the Hamiltonian $H(\kappa_y) = -i \sigma_x \partial_x - \sigma_y \kappa_y - F x$ as $\Cal{H}(\kappa_y) = U H(\kappa_y) U^{\dagger}$ to work on a purely real wave function given by the following set of equations:
\begin{eqnarray}
    \partial_x \phi_B - \kappa_y \phi_B - \alpha(x) \phi_A &=& 0 \, , \label{eq:RealEquation4GNRInEfield1} \\
    - \partial_x \phi_A - \kappa_y \phi_A - \alpha(x) \phi_B &=& 0 \, . \label{eq:RealEquation4GNRinEfield2} 
\end{eqnarray}
where $\alpha(x) = \varepsilon + F x $. Since $\phi_B(x) = -\alpha^{-1}(x) \partial_x \phi_A(x) - \kappa_y \alpha^{-1}(x) \phi_A(x)$, the single working equation for $\phi_A$ is 
\begin{equation}
    \partial_{xx} \phi_A - \frac{F}{\alpha(x)} \partial_x \phi_A - \left(\kappa_y^2 + \frac{F \kappa_y}{\alpha(x)}  - \alpha^2(x) \right) \phi_A = 0 
\end{equation}
and the boundary conditions $\phi_B(-L/2) = \phi_B(L/2) = 0$ takes the form of the Robin boundary conditions:
\begin{equation}
    \left.-\alpha^{-1}(x) \partial_x \phi_A (x) - \kappa_y \alpha^{-1}(x) \phi_A(x)\right|_{x=\pm\frac{L}{2}} =0  \, .
\end{equation}
Thus, the classical phase space can be defined as $(\phi_A, \partial_x \phi_A)$. In contrast, we are going to work as before in the Hilbert {\it psi-space} $(\phi_A,\phi_B)$, which is related to the classical phase space as 
\begin{equation}
\left(\matrix{\phi_A \cr \phi_B }\right) = \left(\matrix{ 1 & 0 \cr -\kappa_y \alpha^{-1} & -\alpha^{-1}  }\right) \left(\matrix{\phi_A \cr \partial_x \phi_A }\right)\, .
\end{equation}
To avoid any confusion with the phase diagrams in our future discussions, we use term psi-space instead of phase space. The vector field in the psi-space is $\mathbf{v}_{AB} = \partial_x (\phi_A,\phi_B) = \left(-\kappa_y \phi_A - \alpha(x) \phi_B, \alpha(x) \phi_A + \kappa_y \phi_B \right)\, .$
It is seen that the field can be treated as a function of two independent parameters $\kappa_y$ and $\alpha$. By investigating the vector field portraits as the function of $\kappa_y$ and $\alpha$ parameters and paying attention to the fixed points characterized by the index, it is possible to obtain an {\it index phase diagram} that characterises possible types of the vector field in the psi-space.

Figure~S9A shows that there are two distinct regions with index $I_{\nu} = \pm 1$ in the index phase diagram. When $\left|\kappa_y\right|>\left|\alpha\right|$, the fixed point at the origin of psi-space is a saddle point, $I_{\nu} = \mathrm{Ind} =-1$ (pink region), otherwise it is a center point (cyan region). By linearizing Eqs.~(\ref{eq:RealEquation4GNRInEfield1}) and~(\ref{eq:RealEquation4GNRInEfield1}), we obtain the Jacobian $\left(\matrix{
    -\kappa_y & -\alpha \cr
    \alpha & \kappa_y
}\right)$ 
with eigenvalues $\lambda_{1,2} = \pm \sqrt{\kappa_y^2-\alpha^2}$. These eigenvalues are always symmetrically positioned with respect to the origin. When both eigenvalues are purely imaginary, the fixed point at the origin is a center. Alternatively, as $\lambda_{1,2}$ become purely real numbers with opposite signs, the fixed point becomes a saddle. Thus, the eigenvalues of the Jacobian move from imaginary to real axis (or vice versa) via coalescence at the origin resulting in a phase transition. This phase diagram is quite general and it also applies to the hbGNR when the electric field is absent, then $\alpha \equiv \varepsilon$. If $\alpha \equiv \alpha(x)$, then the point of psi-space representing the wave function at $x$ moves in a vector field that gradually changes as a function of $x$. For this case, the index phase diagram can be re-plotted in the $(x,\kappa_y)$-coordinates to see how much `time' the system spends in each of the vector field phases, i.e. how large is the fraction of the trajectory affected by each of the vector field phases. The bulk state dynamics can fully reside within the region $I_{\nu}=+1$ (cyan region). On the other hand, as seen from fig.~S9A, the states resulting from a hybridization between bulk states and ZEM due to the effect of the electric field $F \neq 0$, experience switching between the vector field portraits characterized by $I_{\nu}=\pm1$. In the $(x,\kappa_y)$-phase diagram with given $\varepsilon$ and $F$, the eigenstate trajectory can be presented as a horizontal line positioned at a specific $\kappa_y$. As the dynamical parameter $x$ changes from $-L/2$ to $L/2$, the point representing the value of the eigenstate on this line can cross the interface between the two index phases. Thus, the index phase diagram provides a general outlook on the expected dynamics in the psi-space.

Having revealed the general features of the expected dynamics in psi-space, we can proceed to the numerical treatment of the problem. In contrast to plotting trajectories in psi-space for the analytically derived wave function, numerical solution requires elimination of some free-parameters. For example, all the input parameters to start the 4th order Runge-Kutta algorithm are readily available in the Cauchy problem. Dealing, however, with the boundary value eigenspace problem we do not have information about energy $\varepsilon$ that enters the problem. Likewise, we lack the information about first order derivatives or the second component of the wave function at the initial point of the trajectory. Moreover, we cannot get trajectory for the normalized wave function even starting the algorithm as this requires information about the function in the whole domain of the dynamic $x$. All this implies usage of a {\it shooting method}. In the shooting method, we guess the required values of the parameters to start the numerical algorithm and then after finishing evaluation compare the outcome with the boundary condition. 

The eigenproblem has two adjustable parameters, namely $\phi_A(-L/2)$ and $\varepsilon$. By normalizing the trajectory, we eliminate the dependence on the initial $\phi_A$ value. Here, we will show visual interpretation of the normalization procedure in psi-space. Since we deal with real functions, normalization is the integration of $\phi_A^2 + \phi_B^2$. In the psi-space, this integrand can be interpreted as the squared radius vector of the point $(\phi_A,\phi_B)$. Then replacing integration with summation $\sum_{n \rightarrow \infty} \left(\phi_{A,n}^2 + \phi_{B,n}^2\right) dx_n$, where $dx_n = L/n$, one can see that $L \sum_{n \rightarrow \infty} r_n^2/n = L \langle r^2\rangle$ and it is proportional to the averaged square of the trajectory radius vector. The normalization constant requires taking an inverse of a square root of this expression. Hence, up to a scaling by $\sqrt{L}$, the normalization constant is an inverse of a {\it standard deviation} of the trajectory from the origin of the psi-space. For the second adjustable parameter, the proper energy $\varepsilon$ must be found from the condition that the final point of the normalized trajectory satisfies the second boundary condition, i.e. $\phi_B(L/2) =0$. In other words, the $\phi_B$-component of the radius vector must vanish at the final point of the normalized trajectory. This condition represents the {\it secular equation} of the eigenproblem in question. In this way, we obtain a numerical analog of  Eq.~(\ref{eq:DiracEquation4GNRinEfieldComponentphiBinKSolDispersionEquation} that works for all values of $\kappa_y$ and $F$. Solving this secular equation numerically, we find a set of allowed energies for any given $\kappa_y$, $F$ and $L$. Then we can run the differential equation solver with the found $\varepsilon$'s to restore the trajectories representing valid solutions of the eigenproblem. Thus obtained trajectories allow for evaluation of the string index $I_{n}$. In the remainder of this section, we present the results obtained within the formalism described above.

In fig.~S9B the radius vector $\phi_B$-component is plotted as a function of $\varepsilon$. For $\kappa_y=0$, the allowed values of $\varepsilon$ are independent of the field $F$, e.g. compare panels of $F=0.04$ and~$F=0.07$. For $\left|\kappa_y \right| \gg 0$, the ZEM energies asymptotically approach $\varepsilon = \pm FL/2$ depicted by the vertical dashed red lines as seen from panels $\kappa_y=\pm 0.6$. Both these results have been derived analytically in the previous section. For the numerical secular equation, the processes of isolating and finding the root are non-trivial tasks due to two main reasons. Firstly, the well-known techniques of root isolation for polynomials, based for instance on Sturm's theorem or Descartes' rule of signs, are not applicable to the given case. Secondly, the behaviour of the function around the root may be pathological, namely preventing convergence of fast root finding methods due to rapid variations of the function and singular values of its derivatives. To deal with the latter issue, methods based on bisection root finding can be used to guarantee convergences. For the former issue, a valid procedure can be the following. We begin at the analytical $\varepsilon$'s at $\kappa_y = 0$ that are well-separated within the intervals $\left(\frac{\varepsilon_i + \varepsilon_{i-1}}{2}, \frac{\varepsilon_i + \varepsilon_{i+1}}{2}\right) =\left(\frac{\pi i}{L} - \frac{\pi }{2L}, \frac{\pi i}{L} + \frac{\pi}{2L}\right)$ and independent of $F$'s. For small $\kappa_y \neq 0$, the roots should stay within the same intervals as those of $\kappa_y = 0$. Thus, the energies can be easily computed and can be readily used to redefine the root isolating intervals. This iterative root isolation algorithm faces issues only when the roots coalesce. If two roots do coalesce and leave their dedicated intervals, we know that each of them must appear within the neighboring interval and they both must be located close to the point that separates these neighboring intervals. This knowledge allows efficient tracking of the first issue and immediate correction of the error with a suitable routine in the root finding algorithm. In this way, the analytical results facilitate full numerical solution of the eigenproblem in question. 

The band structure of a hbGNR in the in-plane electrostatic field is presented in fig.~S9C together with the selected eigenstate trajectories in the psi-space and the vector field corresponding to the final point of the trajectory, i.e. when $x=L/2$. In the band structure plot, the red points with $\kappa_y = 0$ represent analytical eigenvalues, while the horizontal dashed red lines represent the asymptotic values of the flat flanks of the ZEM. The colored and numbered points represent the selected  eigenstates. For these selected states, the trajectory, vector field and indexes are presented on the right in the panels marked with the same number and color as the selected eigenstates. Moreover, the selected states are also visualized with same color lines tagged with the corresponding number in fig.~S9A. As observed from the eigenstates of No.1 to 3, the bulk states are partially affected by the saddle point $I_{\nu} \equiv \mathrm{Ind}=-1$ due to the external field. The string index of the bulk state trajectory, $I_{n} \equiv \mathrm{Ind}_{\mathrm{tr}}$, is not precisely quantized and it noticeably deviates from the expected value, i.e. $\mathrm{Ind}_{\mathrm{tr}}=-0.9$ instead of $-1$ for eigenstate No.~2. Likewise, the ZEM is influenced by the $\mathrm{Ind}=1$ phase, which leads to $\mathrm{Ind}_{\mathrm{tr}}=-0.2$ instead of the expected $0$ for the eigenstate No.~3. Thus, it becomes clear that the dynamics in two vector field phases yeilds a defect in the string index of the eigenstate trajectory. 

The defect in the index of the eigenstate trajectory mixes the index of trajectories between the bands. This phenomenon is investigated further in fig.~S10, where the band structure is plotted for a range of external fields. In fig.~S10 the ZEM transformation from fully flat to folded cubic (cubic II) dispersion is presented, while the band color scheme is chosen to be aligned with the string index of the eigenstate trajectory $\mathrm{Ind}_{\mathrm{tr}}$. 

Before getting into details of index mixing and other related questions, we have seen in~\ref{app:IndexTheoryOfZEM} that the trajectory string index $\mathrm{Ind}_{\mathrm{tr}}$ was a unique descriptor of hbGNR eigenstates. In the $F=0$ panel of fig.~S10, the eigenstate trajectory string index $\mathrm{Ind}_{\mathrm{tr}}$ is also a good descriptor of hbGNR bands. When $F=0$, the trajectory index becomes a band invariant. Thus, the trajectory string index provides a unique ZEM signature resulting from the saddle fixed point of the vector field. Here, it is worth noting the difference of the given topological description from the winding number defined in the reciprocal space to describe the zero-energy edge states in GNRs~\cite{Ryu2002}. The method presented in Ref.~\cite{Ryu2002} is based on the tight-binding lattice model, which requires choosing a proper gauge related to the unit cell of the ribbon. Specifically, the gauge is set by eliminating the complex phase factor of the intra-cell hopping integral~\cite{Ezawa2014,Grujic2016}. This recipe is straight forward for three main geometries of the edges in honeycomb lattice: zigzag, armchair and bearded. However, for a combination of zigzag and bearded edges in a single ribbon, two different unit cells must be discussed for the same ribbon to reconcile with topological description in Ref.~\cite{Ryu2002}. 
Thus, we need switching from bearded to zigzag ribbon gauge at the Dirac point of the hbGNRs. This impossibility to define a single gauge for the whole Brillouin zone points on some obstruction that could be detected by a topological invariant that is similar to the Chern number (see Sec.~3.6 in Ref.~\cite{BookBernevig2013}) is defined as the winding number of the gauge gluing function. Indeed, if the graphene $2\times2$ Hamiltonian $H(\mathbf{k}) = \left(\matrix{0 & h(\mathbf{k})\cr h^{\ast}(\mathbf{k})&0}\right)$, then the winding number  $\nu = (2 \pi i)^{-1} \oint_{\Cal{L}_{k_x}+\Cal{D}} \partial_{k_x} \log\left[g(k_x,k_y) \right] \, d\,k_{x}$, where $g(k_x,k_y) = h_{1}(k_x,k_y) - h_{2}(k_x,k_y)$, with $h_{1}(k_x,k_y)$ and $h_{2}(k_x,k_y)$ being the off-diagonal term $h(\vec{k})$ taken in the ZGNR and BGNR gauges, respectively. It is straightforward to check that for any $k_y$, which is set as a good quantum number for the GNRs, gauge gluing function loops $\Cal{L}_{k_x}$ plotted in the parametric space $(\mathrm{Re}[g],\mathrm{Im}[g])$ with respect to $k_x$ parameter as $\left.(\mathrm{Re}[g(k_x,k_y)],\mathrm{Im}[g(k_x,k_y)])\right|_{k_y = \mathrm{const}}$ are always passing through the origin of the space. Therefore, the invariant is well-defined if a small regularization parameter $\Cal{D}$ shifing the whole loop so that it encompases the origin is introduced. Simultaneously, we have seen the trajectory string index introduced within the continuum model is homogeneously defined throughout the whole reciprocal space, even though it is an invariant based on the real space, i.e. $x$ coordinate is the integration parameter not $\kappa_{y}$ or $\kappa_{x}$.

The trajectory string index will eventually mix between the energy bands when the external electric field is switched on. Despite this mixing, the trajectory string index still remains a good measure of the band hybridization showing a piecewise composite nature of the chiral gapless mode arising from the flat ZEM. In addition to this, a low absolute value of the trajectory string index remains a good indicator of the state localization, which follows from its comparison with the inverse participation ratio in the main text of the paper. 

Figure~S10 is particularly suitable for discussing chirality of the ZEM and its derivatives. In physics, chirality has several definitions, including:
\begin{enumerate}[label=\roman*.]
    \item as the eigenstate of the chiral symmetry operator $\Cal{S}$;
    \item as the imbalance between the left and right moving modes referred to as {\it chiral charge};
    \item as the asymmetric electron density distribution between the edges of the structure.
\end{enumerate}
The first definition here implies a perfect sublattice (pseudo-spin) polarization. As such, it is fully applicable to the flat ZEM when $F=0$. However, this definition fails for $F \neq 0$ when two sublattices are mixed (see  fig.~S8B). On the other hand, the second definition can be readily applied to $F \neq 0$ with careful interpretation.

The imbalance between the left and right moving modes implies the breaking of the TR symmetry $\Cal{T}$. The continuum model redefines the {\bf K} point to the TR invariant $\Gamma$ point, i.e. $\kappa_y=0$. We consider the band structure to correspond a spinless particle so that $\Cal{T}=K$, where $K$ is the complex conjugation. Then, it must be $\varepsilon(\kappa_y) = \varepsilon(-\kappa_y)$ provided that $\Cal{T}\Psi(\kappa_y) = \Psi(-\kappa_y)$. From the  panels $F\neq0$ in fig.~S10, the ZEM derivatives lack TR partners {\it locally} around $\kappa_y = 0$ due to the inequality $\varepsilon(\kappa_y) \neq \varepsilon(-\kappa_y)$. In other words, the right mover does not have the left moving partner given by linearly dispersed in-gap mode in the panel $F=0.07$ of fig.~S10. This imbalance in turn implies the presence of a net right moving current in the system even without an applied electric field along the ribbon axis, which breaks the current conservation law. This situation is similar to an Adler-Bell-Jackiw axial current anomaly in the quantum field theory~\cite{Adler1969,Bell1969}, which can also arise in condensed matter systems~\cite{Nielsen1983}. In a modern language this phenomenon is usually referred to as a chiral anomaly because the difference between left and right movers is usually referred to as {\it chiral charge}. For instance, the chiral anomaly takes place in Weyl semimetals with the zero-energy Landau level, i.e. in the presence of a magnetic field that breaks the TR symmetry (cf. with Fig.~2b in Ref.~\cite{Yan2017}). However, {\it no magnetic field} is involved in producing the chiral anomaly in the given case. Therefore, a {\it global} TR symmetry is preserved, which is similar to the valley Chern insulators considered in the supplementary text~\ref{app:BulkBoundaryCorrespondenceCommVsIncomm}. 

For a high value of the field $F$, the mode develops a wiggly cubic-like structure labeled as cubic~II in fig.~S10. Here, we make a few brief remarks to avoid confusion in counting the left and right moving modes. We refer the term mode to a band that hosts one or several zero-energy states. Thus, panel $F = 0.21$ of fig.~S10 still shows a single mode that provides several transport channels, namely one left transmission and two right transmission channels. The change of the number of transmission channels is a consequential advance in the system's transport properties. Hence, we name this phonomenon as {\it bifurcation} by analogy with the dramatic changes in the dynamics of the non-linear systems. Creating new channels exhibits a characteristic similar to the fixed point bifurcations. The channels are created and destructed in pairs, i.e. left and right. This analogy will be made more formal in the next section, for now it is important that the difference between left and right transmition channels is a conserved quantity equal to the chiral charge defined within the given valley. In the continuum model, the mode is defined within each valley separately, while it can be a single entity possessing valley transmission channels in the tight-binding lattice model.

The left and right movers cannot be determined for the flat ZEM since the mode is stationary and its group velocity is zero. Formally, the symmetry for energy levels is in place, since $\varepsilon(\kappa_y) = \varepsilon(-\kappa_y)$. Hence, the local TR symmetry seems to be preserved. However, the symmetry for the wave function is broken: $\Cal{T} \Psi(\kappa_y) = \Psi(\kappa_y) \neq \Psi(-\kappa_y)$. The expected relations are violated due to the wave function localization at zigzag and bearded edges for $\kappa_y > 0$ and $\kappa_y <0$, respectively.  
This naturally guides us to the 3rd definition of the mode chirality that readily applies to both flat ZEM and its derivatives.

Around the zero energy the ZEM derivatives (linear, cubic~I and cubic~II) are characterized by trajectory string indexes between $-0.5$ and $0.5$, which corresponds to localized states as shown in fig.~S10. The electron density distribution of these localized states is also asymmetric with respect to the mid-line of a hbGNR when $\kappa_y \neq 0$. Thus, all the modes, i.e. flat, linear, cubic~I and folded cubic~II, are chiral modes.

The numerical results obtained from the continuum $\mathbf{k}\cdot\mathbf{p}$-model, are all in perfect agreement with the full tight-binding model. The ZEM transformation can be tracked from fully flat ZEM to cubic-like, specifically the cubic~I and~II types of dispersion. Also, the {\it local breaking} of the TR symmetry does not contradicts to globally preserved $\Cal{T}$. Indeed, the right moving current is compensated by the left moving current in the second valley. Thus, our numerical results agree with the full tight-binding model when both {\bf K} and~{\bf K}$^{\prime}$ valleys are accounted for in our system. 

{\it In summary}, the eigenvalue problem of hbGNR in the external in-plane field can be mapped to an effective non-stationary dynamical system and treated numerically as turbulent dynamics in an abstract Hilbert psi-space. In addition to the perfect agreement of numerical results with the tight-binding model ZEM transformation, applying an external field leads to the trajectory string index mixing between the bands. Despite this sting index mixing, the low absolute value of the index remains a good measure of the edge localization and chirality of the state. Finally, the string index also clearly reveals the composite nature of the chiral gapless mode arising from the flat ZEM.

\subsection{Analytical treatment of the hbGNR in an external electric field\label{app:LowEnergyTheoryDEZEMinEfieldII}}
In this section, we present an analytical solution for the equation describing hbGNR subject to an external in-plane electric field. In what follows, we show that solution is possible in terms of special functions defined via an infinite power series within the Frobenius method. The found solutions together with the boundary conditions allows one to construct dispersion equation that is to be solved numerically.

We start from Eq.~(\ref{eq:DiracEquation4GNRSingleComponentphiBEfield1}), which is
\begin{equation}
    \partial_{\xi \xi} \phi_B - \frac{1}{\xi} \partial_{\xi} \phi_B + \left(\frac{\kappa_y}{\xi} - \kappa_y^2 + F^2 \xi^2\right) \phi_B = 0 \, 
    \label{eq:DiracEquation4GNRSingleComponentphiBEfield2}
\end{equation}
and the boundary conditions $\phi_B(\xi_1) = 0$ and $\phi_B(\xi_2) = 0$, where $\xi_1 = (\varepsilon/F) - (L/2)$ and $\xi_2 = (\varepsilon/F) + (L/2)$. Similar to the Bessel's equation or a confluent hypergeometric equation, this equation features two singular points $\xi=0$ and $\xi=\infty$. The fist singular point is \emph{regular}, while the second one is \emph{irregular}. Thus, the Frobenius method can be attempted to find the solution of Eq.~(\ref{eq:DiracEquation4GNRSingleComponentphiBEfield2}) around the regular singular point $\xi=0$.  
Within the Frobenius method, we are looking for the solution of a differential equation in terms of infinite power series expansion $S=\sum_{n=0}^{\infty} a_n \xi^n$ and try to find a recursive relation between the series coefficients $a_n$ such that the differential equation is satisfied. This method applied straight forwardly to the above equation fails due to the large variety of terms leading to power series with largely shifted degrees of $\xi$'s. In order to reduce all of these power series to the same form, one has to isolate many terms and in the final series one has to set to zero the first $4$ coefficients which, in turn, breakdowns the recursive relation between $a_n$'s.

In order to fix the above issue, we can try to reduce the variety of the terms. A simple analysis shows that term $\left(\partial_{\xi} \phi_B/\xi\right)$ is not an issue as it matches well with $\partial_{\xi \xi} \phi_B$. Thus, we need to eliminate some terms in $\left(\left(\kappa_y/\xi\right) - \kappa_y^2 + F^2 \xi^2\right) \phi_B$. For example, we can get rid of the $\kappa_y^2$- and $\kappa_y/\xi$-terms by making the following substitution of the function: $\phi_B (\xi) = u(\xi) \exp\left(\kappa_y \xi\right)$. From physical intuition point of view, this substitution can be also considered as the limiting solution of the equation for the zero external field $F$ found previously in supplementary text~\ref{app:LowEnergyTheoryDEZEM} or similarly the large $\kappa_y$ limit for $F\neq 0$. The substitution yields
\begin{equation}
    \partial_{\xi \xi} u + \left(2 \kappa_y - \frac{1}{\xi} \right) \partial_{\xi} u + F^2 \xi^2 u = 0 \, .
    \label{eq:DiracEquation4GNRSingleComponentphiBEfield2uSub}
\end{equation}
Equation~(\ref{eq:DiracEquation4GNRSingleComponentphiBEfield2uSub}) has now only one additional term instead of two. The new term has appeared in front of the derivative which is also important because it means it has shifted closer to the second derivative power expansion series. Despite seeming simplicity and the reduced diversity of terms, the obtained equation is still not good enough because of the term $F^2 \xi^2 u$. The latter gives rise to a power series that is too distant from the other ones due to $\xi^2$ factor. To eliminate this issue, we employ another substitution of the function: $u(\xi) = g(\xi) \exp\left(-i F \xi^2 / 2 \right)$, where one can immediately recognize the solution found for the case $\kappa_y = 0$ in supplementary text~\ref{app:LowEnergyTheoryDEZEMinEfield}, cf. with Eq.~(\ref{eq:DiracEquation4GNRinEfieldComponentphiBinKSol}). The resulting equation is
\begin{equation}
    \partial_{\xi \xi} g + \left(2 i F \xi + 2 \kappa_y - \frac{1}{\xi} \right) \partial_{\xi} g + 2 i \kappa_y F \xi  g = 0 \, .
    \label{eq:DiracEquation4GNRSingleComponentphiBEfield2ugSub}
\end{equation}
Although the obtained Eq.~(\ref{eq:DiracEquation4GNRSingleComponentphiBEfield2ugSub}) look is more formidable, this first impression is delusive. In fact, it is only now when the series expansion will work. By setting $g(\xi) \equiv S = \sum_{n=0}^{\infty} a_n \xi^n$, we get
\begin{equation}
    \sum_{n=0}^{\infty} a_n n (n-2) \xi^{n-2} + \sum_{n=0}^{\infty} 2 i F a_n n \xi^{n} + \sum_{n=0}^{\infty} 2 \kappa_y a_n n \xi^{n-1} - \sum_{n=0}^{\infty} a_n n \xi^{n-2} + \sum_{n=0}^{\infty} 2 i \kappa_y F a_n \xi^{n+1} = 0 \, .
    \label{eq:DiracEquation4GNRSingleComponentphiBEfield2ugSubPE}
\end{equation}
Next we change the dummy summation variable $n$ in each of the power series so that $\xi$-exponents turn into new summation variables $k$'s:
\begin{equation}
    \sum_{k=-2}^{\infty} a_{k+2} k (k+2) \xi^{k} + \sum_{k=-1}^{\infty} 2 \kappa_y a_{k+1} (k+1) \xi^{k} + \sum_{k=0}^{\infty} 2 i F a_k k \xi^{k} + \sum_{k=1}^{\infty} 2 i \kappa_y F a_{k-1} \xi^{k} = 0 \, ,
    \label{eq:DiracEquation4GNRSingleComponentphiBEfield2ugSubPE1}
\end{equation}
where we have merged power series with $\xi^{n-2}$ before the change of the summation variable. Now the only problem is that all the given power series have different summation starting points, e.g. $k=-2$, $k=-1$ etc. Note, however, that each of these power series contains terms equal to zero. Those few terms that are not zero can be collected in front. Thus, all summations can be started from $k=1$:
\begin{equation}
    \left(2 \kappa_y -\frac{1}{\xi}\right) a_1 + \sum_{k=1}^{\infty} \left[a_{k+2} k (k+2) + 2 \kappa_y a_{k+1} (k+1) + 2 i F a_k k + 2 i \kappa_y F a_{k-1} \right] \xi^{k} = 0 \, ,
    \label{eq:DiracEquation4GNRSingleComponentphiBEfield2ugSubPE2}
\end{equation}
Now it is seen that the equation is satisfied if $a_1 \equiv 0$ and 
the square braket $\left[ \ldots \right]$ is equal to zero too. The later condition implies the following recursive relation between the coeffients $a_k$:
\begin{equation}
    a_{k+2} = -\frac{ 2 \kappa_y a_{k+1} (k+1) + 2 i F a_k k + 2 i \kappa_y F a_{k-1}}{k (k+2)} \, \label{eq:DiracEquation4GNRSingleComponentphiBEfieldRR}
\end{equation}
or equivalently by going back to the $n$ summation variable and introducing a set of parameters $\mu = 2 \kappa_y $, $\nu = 2 i F$ and $\lambda = 2 i \kappa_y F$:
\begin{equation}
    a_{n} = - \mu \frac{n-1}{n (n-2)} a_{n-1} - \frac{\nu}{n} a_{n-2} - \frac{\lambda}{n (n-2)} a_{n-3}\, ,  \label{eq:DiracEquation4GNRSingleComponentphiBEfieldRR1}
\end{equation}
where $n=3, 4, \ldots$. Here we imply $F \neq 0$, i.e. $\nu \neq 0$ and $\lambda \neq 0$, because otherwise by definition $\xi = \infty$ and the Frobenius method may be not valid around this irregular singular point. By careful inspection of this recurrence relation we notice that we can restore the full series once we know first three coefficients: $a_0$, $a_1$ and $a_2$. It is already known that $a_1 \equiv 0$. Thus, we are left with two unknown coefficients $a_0$ and $a_2$. Note, however, that the initial differential equation is of the second order, which means its general solution is defined up to two arbitrary constants. These arbitrary constants will be later fixed by the boundary conditions. For now, by collecting separately the power series terms with $a_0$ and $a_2$, we can split the full series into two special functions $R(\xi,\mu,\nu,\lambda)$ and $V(\xi,\mu,\nu,\lambda)$. Each of these functions is presented by a power series, where coefficients are defined by the recursive relation in Eq.~(\ref{eq:DiracEquation4GNRSingleComponentphiBEfieldRR1}) but with two different initial conditions. Namely, for $R$-function we set $a_0=1$ and $a_2=0$ while for $V$-function we set $a_0=0$ and $a_2 = 1$. Then the general solution of Eq.~(\ref{eq:DiracEquation4GNRSingleComponentphiBEfield2ugSub}) can be given in the form
\begin{equation}
    g(\xi) = C_1 R(\xi,\mu,\nu,\lambda) + C_2 V(\xi,\mu,\nu,\lambda) \, ,
    \label{eq:DiracEquation4GNRSingleComponentphiBEfieldGenSol}
\end{equation}
where $C_{1,2}$ are the arbitrary constants to be found from the boundary conditions; they are basically our free parameters $a_0$ and $a_2$.

Using the newly introduced special functions $R$ and $V$, we have
\begin{equation}
    \phi_B(\xi) = \exp\left(\kappa_y \xi \right) \exp\left(\frac{i F \xi^2}{2} \right) \left[C_1 R(\xi,\mu,\nu,\lambda) + C_2 V(\xi,\mu,\nu,\lambda)\right] \, .
    \label{eq:DiracEquation4GNRSingleComponentphiBEfieldGenSol1}
\end{equation}
for the general solution of Eq.~(\ref{eq:DiracEquation4GNRSingleComponentphiBEfield2}). Now the boundary condition requires, $\phi_B(\xi_2) = \phi_B(\xi_1) = 0 $, so that
\begin{eqnarray}
     0 &=& \exp\left(\kappa_y \xi_2 \right) \exp\left(\frac{i F \xi_2^2}{2} \right) \left[C_1 R(\xi_2,\mu,\nu,\lambda) + C_2 V(\xi_2,\mu,\nu,\lambda)\right] \, , \\
    0 &=& \exp\left(\kappa_y \xi_1 \right) \exp\left(\frac{i F \xi_1^2}{2} \right) \left[C_1 R(\xi_1,\mu,\nu,\lambda) + C_2 V(\xi_1,\mu,\nu,\lambda)\right]\, ,
    \label{eq:DiracEquation4GNRSingleComponentphiBEfieldDispEq}
\end{eqnarray}
where $\xi_{1,2} = (\varepsilon/F) \pm (L/2)$ with `$+$' being taken for $1$. Since exponents cannot lead to zero, they can be safely cancelled for each of the boundary equations. It is, however, worth noting here that the complex exponent is better be kept in the dispersion equation as it nicely works as \textit{regularizing factor} ensuring that the zeros of the dispersion equation are given only by the real part of that equation. Thus, we have
\begin{eqnarray}
     0 &=&  \exp\left(\frac{i F \xi_2^2}{2} \right) \left[C_1 R(\xi_2,\mu,\nu,\lambda) + C_2 V(\xi_2,\mu,\nu,\lambda)\right] \, , \\
    0 &=& \exp\left(\frac{i F \xi_1^2}{2} \right) \left[C_1 R(\xi_1,\mu,\nu,\lambda) + C_2 V(\xi_1,\mu,\nu,\lambda)\right] \, .    \label{eq:DiracEquation4GNRSingleComponentphiBEfieldDispEq1}
\end{eqnarray} 

The homogeneous system of linear equations with respect to $C_{1,2}$ has non-trivial solutions only if the determinant of its matrix is zero, which is the sought dispersion equation:
\begin{equation}
    \mathrm{Disp} = \exp\left(\frac{i F (\xi_2^2+\xi_1^2)}{2} \right) \left[ R(\xi_2,\mu,\nu,\lambda) V(\xi_1,\mu,\nu,\lambda) - V(\xi_2,\mu,\nu,\lambda) R(\xi_1,\mu,\nu,\lambda)\right] = 0 \, .
\label{eq:DiracEquation4GNRSingleComponentphiBEfieldDispEq2}
\end{equation}
For given $F$ and $L$, Eq.~(\ref{eq:DiracEquation4GNRSingleComponentphiBEfieldDispEq2}) fixes a relation between $\varepsilon$ and $\kappa_y$.

In fig.~S11, we present two plots of the real part of a regularized $\mathrm{Disp}$ function defined by Eq.~(\ref{eq:DiracEquation4GNRSingleComponentphiBEfieldDispEq2}). The plots are overlaid by purely numerical (blue points) and exact analytical (red points) solutions. The blue points are obtained from the dispersion equation from supplementary text~\ref{app:IndexTheoryOfZEMinEfield}, while the red ones are eigenenergies from supplementary text~\ref{app:LowEnergyTheoryDEZEMinEfield}. It is seen that there is a good agreement between the derived dispersion function and the two other approaches to the problem.

The obtained analytically dispersion equation allows for some basic analysis. Namely, by expanding the special $R$- and $V$-functions into series with respect to $\xi$ up to a few terms, say up to $\xi^6$, and substituting those expression into the dispersion function without exponent, i.e. taking only square brakets in Eq.~(\ref{eq:DiracEquation4GNRSingleComponentphiBEfieldDispEq2}), one can see that the even order terms with respect to $\varepsilon$ (or $\kappa_y$) turn into zeros once $\kappa_y \rightarrow 0$~(or $\varepsilon \rightarrow 0$). Since dispersion equation is given by dispersion function equated to zero, this means that the spectrum of hbGNR in the external in-plane electric field contains a persistent $\varepsilon(\kappa_y=0)=0$ solution. This further supports the results obtained within the tight-binding lattice model in supplementary text~\ref{app:TBtheoryofhbGNRZEMwithEfield}. Moreover, around $\varepsilon(\kappa_y=0)=0$ point the dispersion can have only $\kappa_y$-terms of odd powers which is in agreement with the numerical results within both continuum and the lattice tight-binding models showing linear to cubic-like dispersion relation transformation that shall be latter also justified by the fitting the dispersion with a cubic functional form in supplementary text~\ref{app:FittingZEMwithCubicFunctional}.

In summary, the eigenproblem of hbGNR subject to an external in-plane electric field can be solved analytically in terms of new $R$- and $V$- special functions defined as power series with coefficients obtained recursively via Eq.~(\ref{eq:DiracEquation4GNRSingleComponentphiBEfieldRR1}).

\subsection{Supercritical pitchfork bifurcation \label{app:SupercriticalPitchforkBifurcation}}
We have already discussed the ZEM transport properties by tackling the difference between various types of energy dispersion in fig.~S10. In particular, we have called transition from cubic~I to cubic~II dispersion a {\it bifurcation of transmission channels}. For those familiar with non-linear dynamics, this terminology may be confusing and even not properly justified, since no precise relation was given to the fixed points. Indeed, important question about bifurcation may arise such as why bifurcation takes place only for ZEM and not for the bulk bands, where the effective $x$-dynamics is also affected by the saddle point of the vector field. Another important question to consider is how the bifurcation can be distinguished from a simple increase of transmission. To avoid inconsistencies and possible confusion, we present here a precise definition of what we mean by the ZEM bifurcation.

In the non-linear dynamics, a {\it pitchfork bifurcation} is a trippling of the fixed point, which is common in systems with some symmetry. The total index of all the points involved is preserved prior and after the bifurcation. Thus, the stable node fixed point can give rise to one saddle fixed point and two stable nodes. 
Alternatively, a stable node can merge with two saddles leaving only one saddle. The first case is referred to as supercritical, while the second is subcritical. The name pitchfork arises from the look of the bifurcation diagram. 

Let us consider the supercritical pitchfork bifurcation given by the equation
\begin{equation}
    \partial_t x = r x - x^3\, 
    \label{eq:SupercriticalPitchfork}
\end{equation}
where $x$ and $r$ are the coordinate and bifurcation parameter, respectively. The fixed points are $x^{\ast}=0$ and $x^{\ast} = \pm \sqrt{r}$, i.e. $\partial_t x = 0$. When $r<0$, only one fixed point exist, since unlike in quantum mechanics imaginary values are not permissible, while $r>0$ gives rise to the three fixed points. By analysing the sign of the vector field $v = \partial_t x$ on the one-dimensional $x$ manifold, the fixed point at the origin is a stable node when $r<0$. However, the same point becomes unstable once $r>0$, while two stable nodes appear symmetrically on both sides: $x^{\ast} = \pm \sqrt{r}$. Plotting the positions of the fixed points as functions of $r$, we obtain a {\it bifurcation diagram} that looks like a pitchfork, so is the name of the bifurcation. Figure~S12A shows the pitchfork bifurcation diagram for the non-linear system given by Eq.~(\ref{eq:SupercriticalPitchfork}). Since the style of the curves tracking stable and unstable fixed points is chosen differently, the diagram contains all essential information about the given bifurcation.

The behaviour of the zero-energy states  of hbGNR ZEM can be analogously described by the pitchfork bifurcation diagram. Namely, we consider zero-energy states as fixed points and replace their stability type by the group velocity sign corresponding to the left and right transmission. By inspection, the transition form linear to cubic~II dispersion corresponds to a pitchfork bifurcation. However, the analogy is quite deeper than just behavioral pattern. For instance, we can identify the {\it critical slowing down} taking place in ZEM, which is evocative of the standard pitchfork bifurcation at $r=0$. The merging of the zero-energy states corresponding to the left and right transmissions proceeds via cubic~I dispersion, that is with $\sim k^3$ behaviour. The group velocity for this cubic dispersion vanishes at $k=0$, which then blocks the electronic transport in the energy range around $\varepsilon = 0$ resulting in a notch in the transmission spectrum. In fig.~S12B, we plot the positions $\kappa_y^{\ast}$ of the zero-energy states in the reciprocal space as functions of the applied external field $F$. Here, we use the sign of the group velocity $v_{\mathrm{gr}}$ as our color scheme. The plot shows that below a critical value of the field, $F \approx 0.165$, the ZEM features only one zero-energy state. The sole zero-energy state is located at $\kappa_y^{\ast} = 0$ and propagates to the right, i.e. only one right transmission channel. Passing through the critical value, the zero-energy state changes the direction of transmission from the right to the left. Simultaneously, two side channels transmitting to the right appear symmetrically around the $\kappa_y^{\ast} = 0$. Hence, this bifurcation diagram has a clear pitchfork structure. This allows us to conclude that the bifurcation of the ZEM is a supercritical pitchfork bifurcation.

Finally, the energy $\varepsilon$ in our system is allowed to deviate from zero when an electrostatic doping is introduced via a back gate, which then results in an imperfection parameter. Thus, the pitchfork bifurcation becomes an {\it imperfect bifurcation}. Similar to the standard imperfect bifurcation of non-linear systems~\cite{BookStrogatz1994}, we can also define {\it stability diagram} in $(\varepsilon,F)$ parameter space, including a codimension-2 bifurcation in the cusp point or {\it cusp catastrophe} surface in $(F,\varepsilon,\kappa_y)$ space. 

\section{Fitting the ZEM dispersion with cubic functional\label{app:FittingZEMwithCubicFunctional}}
In this section, we verify that ZEM in hbGNRs can be transformed from a linear to cubic dispersion.
Consider for each non-zero value of the electrostatic field used in Fig.~3C the following functional form: $\alpha \left(k - \delta \right)^3 + \beta \left(k - \delta\right) + \gamma$, where $\alpha$, $\beta$, $\gamma$ and $\delta$ are the fitting parameters. The fitting curves with this functional are presented in fig.~S13A. A cubic polynomial is observed to be sufficient functional form for fitting all ZEM curves in the given range of the fields. The dependence of each best fit parameter as a function of the applied field is then investigated to reveal a character of the cubic polynomial. Such dependences are depicted for each best fit parameter in fig.~S13B. The inflection point of the cubic dispersion is pinned to the Dirac point as shown in fig.~S13B by negligible and overlapping horizontal and vertical shifts $\delta$ and $\gamma$, respectively.
The $\alpha$ and $\beta$ coefficients gain opposite signs in the two intervals of the field, which signify the possibility of the pitchfork bifurcation onset. Indeed, both intervals correspond to a cubic polynomial maintaining two extrema. Only one of the intervals, however, is associated with a cubic dispersion of a true pitchfork bifurcation. The light gray area in  fig.~S13B denotes the interval of pseudo-cubic dispersion that does not contain a bifurcation because this area contain a linear dispersion that smoothly transforms to the flat bands at the flanks as seen from Fig.~2D. On the other hand, the light blue area provides a cubic dispersion exhibiting the pitchfork bifurcation. The non-shaded central area in fig.~S13B presents an interval, where $\alpha$ and $\beta$ have the same signs. In this area, the ZEM continuously transforms from the linear to the cubic dispersion without the extrema. Our data substantiate the initial hypothesis that the hbGNR ZEM in external electric field attains a cubic functional form.

\section{Phenomenological theory of the cubic dispersion\label{app:PhenomenologicalTheory}}
In this section, we investigate the theories that could describe the folded cubic dispersion $\omega = a k^3 + c k$, where $c < 0$ and $a > 0$. We start from the wave equation analysis and its generalizations to the dispersive and dumping solutions. Then the same generalization are considered with respect to the Schr{\"o}dinger equation.

\subsection{Wave equation\label{app:LowEnergyTheoryWaveEquation}}
Consider the wave equation in 1D:
\begin{equation}
    \frac{\partial^2 u}{\partial t^2} = c^2\frac{\partial^2 u}{\partial x^2}
    \label{eq:1DWaveEquation}
\end{equation}
where $c$ is the phase velocity. Formally, the solution of this equation is obtained by making the replacement of the variables $\xi = x \pm c t$, that turns the above equality into an identity:
$\frac{\partial^2 u}{\partial \xi^2} \equiv \frac{\partial^2 u}{\partial \xi^2}$.
Thus, the general (D'alambert's) solution of this equation is described in terms of two perturbations travelling in opposite directions (right and left):
\begin{equation}
    u = f(x-ct) + g(x+ct)
\end{equation}
with $f$ and $g$ being arbitrary functions that can be found from the boundary and initial conditions in any particular case of a given problem. Each of the unidirectionally propogating solutions $f$ and $g$ can be obtain from a separate equation by decomposing Eq.~(\ref{eq:1DWaveEquation}) as follows:
\begin{equation}
    \frac{\partial^2 u}{\partial t^2} - c^2\frac{\partial^2 u}{\partial x^2} = \left( \frac{\partial u}{\partial t} + c \frac{\partial u}{\partial x} \right) \left( \frac{\partial u}{\partial t} - c \frac{\partial u}{\partial x} \right) = 0 \, ,
\end{equation}
whence
\begin{eqnarray}
    \frac{\partial u}{\partial t} + c \frac{\partial u}{\partial x} &=& 0 \,, \label{eq:1DUnidirectionalWaveF}\\
    \frac{\partial u}{\partial t} - c \frac{\partial u}{\partial x} &=& 0\, . \label{eq:1DUnidirectionalWaveG}
\end{eqnarray}

Let us focus on the unidirectional Eq.~(\ref{eq:1DUnidirectionalWaveF}). Further insight about this equation can be obtained by making assumptions about the form of the function $f$. We set $f = \exp\left(i k \xi \right) = \exp\left[i \left(k x - k c t\right) \right]$, so that the dispersion relation of a plane wave $\omega = k c$ can be easily identified. By setting $u = f = \exp\left[i \left(k x - \omega t\right) \right]$ and inserting this $u$ into Eq.~(\ref{eq:1DUnidirectionalWaveF}), we arrive at
\begin{eqnarray}
    \frac{\partial u}{\partial t} + c \frac{\partial u}{\partial x} &=& 0 \, \nonumber \\
    (-i \omega) u + c (i k)u &=& 0\, , \nonumber \\
    \omega - c k &=& 0 \, , \nonumber \\
    \omega &=& c k \, .
\end{eqnarray}
The unidirectional wave equation can be generalized to a higher order partial derivative with respect to $x$. By adding a term proportional to the second order derivative and setting $u$ as above, we get
\begin{eqnarray}
    \frac{\partial u}{\partial t} + c \frac{\partial u}{\partial x} - b \frac{\partial^2 u}{\partial x^2} &=& 0 \, \nonumber \\
    (-i \omega) u + c (i k)u - b (i k)^2 u &=& 0\, , \nonumber \\
    \omega - c k + i b k^2 &=& 0 \, , \nonumber \\
    \omega &=& c k - i b k^2 \, .
\end{eqnarray}
Now, substituting the found $\omega$ into initial plane wave yields a dumping wave:
\begin{equation}
    u = \exp\left[i \left(k x - \omega t\right) \right] = \exp\left\{i \left[k x - \left(c k - i b k^2\right) t\right] \right\} = \exp\left[i \left(k x - c k t\right) \right] \exp\left(- b k^2 t \right) \, .
\end{equation}
From this consideration, even order derivatives introduce only time decays into the unidirectional wave equation.

The next generalization is to consider the odd order derivatives. We begin from the lowest possible 3rd order (1st order is already considered above). Repeating the same manipulations we obtain:
\begin{eqnarray}
    \frac{\partial u}{\partial t} + c \frac{\partial u}{\partial x} - a \frac{\partial^3 u}{\partial x^3} &=& 0 \, \nonumber \\
    (-i \omega) u + c (i k)u - a (i k)^3 u &=& 0\, , \nonumber \\
    \omega - c k + i^2 a k^3 &=& 0 \, , \nonumber \\
    \omega &=& c k + a k^3 \, .
    \label{eq:CubicDispersionFromWaveEq}
\end{eqnarray}
In contrast to the previous case, this additional term $ak^3$ in the dispersion relation does not cause the decay of the plane wave solution wave-form. At the same time, Eq.~(\ref{eq:CubicDispersionFromWaveEq}) is qualitatively different form the case of linear dispersion of the plane waves. Namely, the phase and group velocities of the newly obtained waves are $k$-dependent:
\begin{eqnarray}
    v_{\mathrm{ph}} &=& \frac{\omega}{k} = c + a k^2 \, \\
    v_{\mathrm{gr}} &=& \frac{d\,\omega}{d\,k} = c + 3 a k^2\, .
\end{eqnarray}
Thus, all terms containing the odd order derivatives introduce a dispersion of the plane waves.

The discussion above of the unidirectional wave equation emphasizes that adding terms with any even order derivative with respect to the spatial coordinate introduces a time-dependent dumping of the wave, while adding terms with any odd order derivatives introduces dispersion of the wave. In other words, it is impossible to have waves with dispersion $\omega (k)$ that is quadratic with respect to the wave number $k$. The cubic dispersive equation considered above can be seen as the Korteweg–De Vries equation:
\begin{equation}
    \frac{\partial u}{\partial t}  + \frac{\partial^3 u}{\partial x^3} + \left(1 - u\right)\, \frac{\partial u}{\partial x} = 0
\end{equation}
with omitted non-linear term $\sim u_x\,u$. An alternative view point is that the cubic dispersive equation is a generalization of the unidirectional massless Klein-Gordon equation. Note that a massive Klein-Gordon equation reads
\begin{equation}
    \frac{1}{c^2}\,\frac{\partial^2 u}{\partial t^2} - \frac{\partial^2 u}{\partial x^2} + \frac{m^2 c^2}{\hbar^2} u = 0 \, .
    \label{eq:1DKleinGordonEquation}
\end{equation}
It describes spin-$0$ particles such as a pion or a Higgs boson.

\subsection{Schr{\"o}dinger equation\label{app:LowEnergyTheorySchrodingerEquation}}
For completeness, we present a proof that the cubic dispersion cannot be obtained from the Schr{\"o}dinger equation. The 1D Schr{\"o}dinger equation for a free spinless particle reads
\begin{equation}
    i \frac{\partial u}{\partial t} = - \alpha \,  \frac{\partial^2\, u}{\partial\, x^2} \,
    \label{eq:1DSchrodingerEquation}
\end{equation}
with $\alpha = \frac{\hbar^2}{2 m}$. Unlike  Eq.~(\ref{eq:1DWaveEquation}),  Eq.~(\ref{eq:1DSchrodingerEquation}) contains a first order derivative in time, therefore it corresponds to Eqs.~(\ref{eq:1DUnidirectionalWaveF}) and~(\ref{eq:1DUnidirectionalWaveG}). Substituting the plane wave $u = \exp\left[i \left(k x - \omega t\right) \right]$ into  Eq.~(\ref{eq:1DSchrodingerEquation}), we obtain the following: 
\begin{eqnarray}
    i \frac{\partial u}{\partial t} + \alpha\,  \frac{\partial^2\, u}{\partial\, x^2} &=& 0 \, \nonumber \\
    i (-i \omega) u + \alpha (i k)^2 u &=& 0\, , \nonumber \\
    \omega - \alpha k^2 &=& 0 \, , \nonumber \\
    \omega &=& \alpha k^2 \, .
\end{eqnarray}
Thus, the dispersion of the wave solution from the Schr{\"o}dinger equation can be obtained by introducing terms with even order derivatives with respect to the spatial coordinate, i.e. $\frac{\partial^{(2n)}\, u}{\partial\, x^{(2n)}}$.

Generalizing Eq.~(\ref{eq:1DSchrodingerEquation}) to the case containing odd order derivatives $\frac{\partial^{(2n-1)}\, u}{\partial\, x^{(2n-1)}}$ yields
\begin{eqnarray}
    i \frac{\partial u}{\partial t} + \alpha\,  \frac{\partial^2\, u}{\partial\, x^2} - \beta\,  \frac{\partial^3\, u}{\partial\, x^3} &=& 0 \, \nonumber \\
    i (- i \omega) u + \alpha (i k)^2 u - \beta (i k)^3 u &=& 0\, , \nonumber \\
    \omega - \alpha k^2 + i \beta k^3 &=& 0 \, \nonumber \\ \omega &=& \alpha k^2 - i \beta k^3 \, .
\end{eqnarray}
This dispersion relation introduces an exponential decay into the wave function solution:
\begin{equation}
    u = \exp\left[i \left(k x - \omega t\right) \right] = \exp\left\{i \left[k x - \left(\alpha k^2 - i \beta k^3\right) t\right] \right\} = \exp\left[i \left(k x - \alpha k^2 t\right) \right] \exp\left(- \beta k^3 t \right) \, .
\end{equation}
Thus, the Schr{\"o}dinger equation terms containing derivatives of $2n-1$ order contribute into the time-dependent decay of the wave function, whereas the terms with even order derivatives contribute into the dispersion leading to the phase and group velocity dependence on $k$.
In other words, a free non-relativistic spinless particle cannot have dispersion proportional to an odd power of $k$. The free particle Schr{\"o}dinger equation is widely used in an effective mass low-energy description of semiconductor carriers near the band edge. However, the Schr{\"o}dinger equation is not suitable to describe the cubic dispersion of ZEM in hbGNRs.

\section{Dissipationless transport on a square lattice\label{app:SquareLattice}}
To show peculiarity of the reported hbGNR case, we compare our results with an iconic example of the square lattice in this section. We shall see that some features of hbGNR ZEM can be indeed reproduced on a square lattice structures. However, the full range of properties attainable on a honeycomb lattice is inaccessible on a square lattice.
Consider two cases of a square lattice ribbon: (i) a regular ribbon with an odd number of atoms in the unit cell, and (ii) a similar ribbon but with one edge atom being detached from its neighbors at the ribbon edge, i.e. with one bearded edge. These two cases will also be considered with and without an applied in-plane electric field across the ribbon width. 

A regular ribbon band structure is a series of cosine-like bands with energy separations controlled by an external field. By specializing to the unit cell with just three atoms in it, the Hamiltonian of the system becomes
\begin{equation}
    H = \left(\matrix{
        2 t \cos(ka) - F & t & 0 \cr
        t & 2 t \cos(ka) & t \cr
        0 & t & 2 t \cos(ka) + F
    }\right) \, ,
    \label{eq:SLRibbonFull}
\end{equation}
where $t$, $F$ and $a$ are the hopping integral, the strength of the external field and the lattice constant, respectively. The corresponding energy bands are $E_0 =  2 t \cos(ka)$ and $E_{\pm} = 2 t \cos(ka) \pm \sqrt{2 t^2+F^2}$. Let us first analyse the case of $F = 0$. Assuming $t<0$, for $E_0$-band there is a left-mover at $k=-\pi/(2a)$ and a right-mover at $k=\pi/(2a)$. Then, if the disorder potential is smooth on the scale of $a$, no backscattering and unit transmission occur for this mode. This unit transmission, however, coexists with zero-energy transmission channels of $E_{-}$ at $k_{-}=\pm \pi/(4a)$ and $E_{+}$ at $k_{+} = \pm 3 \pi/(4a)$. These channels have a group velocity that is lower than that of $E_0$-band channel by a factor of $\sqrt{2}$, which affects coherence of the charge transport in this system. As the number of atoms in the unit cell increases, the number of channels increases too, thereby no coherent transport is possible due to the dispersion of the wave packet in space over time. In order to achieve a coherent single channel transmission, we can apply the in-plane electric field $F\neq 0$. The main effect of the applied electric field is to increase energy separations between the bands without changing their shapes. Therefore, the gap between $E_{\pm}$-bands can be opened, so that only one left and one right transmission channels exist at the intrinsic Fermi level, i.e. at $E=0$. These results qualitatively remain the same as the width of the ribbon increases. In comparison to hbGNR ZEM, the $E=0$ band transmission channel has no tunability. Specifically, the group velocity cannot be controlled by the external field. In addition, large values of the field are needed to open the single channel, and these values cannot be reduced by choosing the larger width of the ribbon. All this makes the given system impractical for dissipationless transport devices.

Next let us consider the previous ribbon with a slight modification. To obtain a tunable flat band, we disconnect all edge atoms from each other at one of the edges of the ribbon to obtain a bearded edge. The Hamiltonian for such a system is
\begin{equation}
    H = \left(\matrix{
        2 t \cos(ka) - F & t & 0 \cr
        t & 2 t \cos(ka) & t \cr
        0 & t &  F
    }\right) \, .
    \label{eq:SLRibbonDis}
\end{equation}
Despite minute difference between Eqs.~(\ref{eq:SLRibbonFull}) and~(\ref{eq:SLRibbonDis}), the analytical eigenvalues of the latter are cumbersome and not useful for our analysis. Therefore, we investigate the numerical solutions for different fields and ribbon widths in fig.~S14. The partially flat band, indeed, appears in the energy band spectrum of the ribbon. However, this flat band is not a ZEM. Depending on the choice of the sign of the hopping integral, the partially flat band is either the lowest valence or the highest conduction band. As one applies the in-plane electric field the flatness of the band increases, while it moves away from $E=0$. The behaviour of the other cosine-like bands is very similar to the previous case with that exception that the slope of the dispersion of the central $E_0$-band can now be varied for a very limited range. The wave function of the flat band, i.e. at high field $F$, is sharply localized on the bearded edge of the ribbon. When $F\gg 0$, the wave functions of $E=0$ states of the $E_0$-band are totally concentrated at the ribbon center. For $F=0$, the wave functions of $E=0$ states are distributed in the alternating manner -- $(1,0,1,0,...,0,1)$, where $1$'s and $0$'s are lattice site amplitudes of the non-normalized wave function over the whole ribbon width. This shows that the square lattice analog is incapable of bulk transport unlike hbGNR ZEM in similar conditions.

Thus, while the dissipaitonless transport can be realized in principle on square lattice structures, it is impractical due to the large field required to open a coherent single transmission channel. A limited control over dispersion in square lattice structures makes them less interesting from the fundamental point of view. Overall, the full range of the reported critical phenomena such as ZEM bifurcation for hbGNRs cannot be achieved on a square lattice.

\section{Mean free path estimate at a supercritical disorder\label{app:TheEffectOfDisorder}}
Since the dissipationless ballistic transport has been proposed for several carbon-based structures, especially those featuring honeycomb lattice, it is a relevant question to what extend the transport is dissipationless and protected in various schemes (including our proposal) and how they compare with each other.

While all the proposals promise a perfect transmission under specific conditions, neither scheme guarantees a full protection from the backscattering upon increasing temperature via inelastic scattering on the phonons. The level and parameters of static charged impurities disorder, such as the density, strength and range of scatterers, also plays crucial role. Upon reaching a critical level, the disorder shall re-introduce backscattering into the system making the \emph{mean free path} a finite quantity. The behaviour of each system at this critical conditions may be different. A fair comparison implies analysing the transport through the same supercritical disordered potential within the same size scattering region. Taking this into account the two-dimensional graphene may be excluded from this comparison. Tailoring of graphene into a 
nanodevice inevitably transforms a macroscopic honeycomb sheet into one of the above-mentioned nanostructures.

By using a short-range scatterers of moderate strength and density described in methods, we have calculated transmissions for dopped aSWCNT(11,11),  ZGNR(10), our hbGNR(11) and cumulenic carbyne. Here the tube is chosen so that it decomposes into the given ribbons as discussed in supplementary text~S1. All systems are modeled within the single-orbital, nearest neighbor tight-binding model. They were placed on top of the disordered potential produced by scatterers randomly distributed within the $xOy$-plane (see methods). Owing to its cylindrical geometry, the sole tube used in this comparison is placed above $xOy$-plane so that it only touches the plane, other structures are placed within $xOy$-plane and, in principle, shall be more exposed to the disordered potential. Cumulenic carbyne is a chain of equally spaced sites separated by $a_0 = 1.42$~\AA.

The calculated transmissions for an increasing length of the scattering region and the corresponding estimated mean free paths are presented in table~\ref{tab:TransmissionVsScatteringRegionLength4CarbonStructures}. As one can see, the transmission decays as a function of the scattering region length differently for each of the structures in question. Namely, the transmission decay is the smallest and the estimated mean free path is the largest for the aSWCNT(11,11) which may well be a result of its peculiar positioning above the disordered potential due to the tube cylindrical geometry as mentioned above. A larger decay and smaller mean free path is observed for cumulene that provides the largest $k$-space separation of the left and right transmission channels, i.e. ZES divide the first Brillouin zone into equal parts. The thinnest possible geometry among the 1D conductors may also play the role in the exceptional resilience of cumulene to the short-range disorder ensuring it 2nd place among the compared structures. The hbGNR(21) subjected to the in-plane electric field to make its ZEM dispersive takes the 3rd place in this ranking by exhibiting even lower transmission and mean free path for the same conditions. Finally, the electrostatically dopped ZGNR(20) demonstrates the strongest decay of the transmission and the smallest estimated mean free path.

Despite the above mentioned differences, all considered conductors demonstrate comparable performance (accounting their peculiar nature) in the supercritical conditions of the short-range disorder. Their advantages or disadvantages are strongly related to the specific context of their intended usage. It must be noted that cylindrical shape of the tube that is partially responsible for its resilience to the scattering on impurities within the substate, poses a significant challenge for their technological processing and device integration. Similarly, the hypothetical cumulene is prone to the Peierls instability which still poses a serious challenge for chemistry and material science in synthesis of this fascinating structure. Finally, zigzag and half-bearded ribbons offer quite different ranges and regimes of tunability. The latter is the key point for hbGNRs. The broad variety of regimes originating from fully flat to linear and exotic cubic dispersions featuring critical slowing down makes them distinct among other carbon-based ballistic conductors.

\clearpage
{\bf Fig.~S1}
\begin{figure}[hbt!]
    \includegraphics[width=\textwidth]{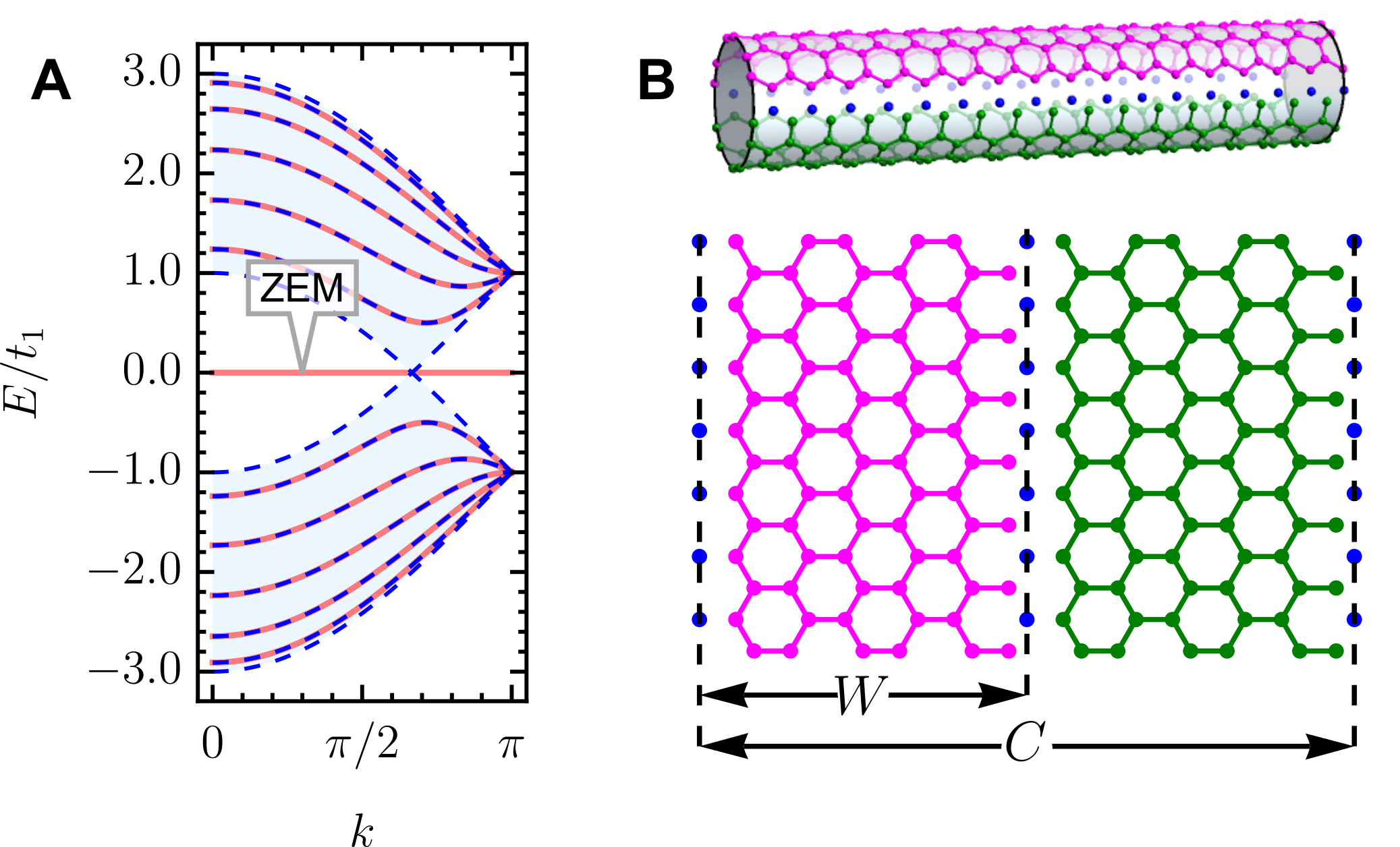}
    \caption{{\bf Flat zero-energy mode.} ({\bf A}) hbGNR $N_r = 11$ (solid, light red) versus aSWCNT$(6,6)$ $N_t=24$ (dashed, blue). The light blue shades the area of a bulk graphene band structure. ({\bf B}) $N_t = 2 N_r + 2$ relation between number of carbon atoms in the unit cell of the hbGNR and aSWCNT (see~\ref{app:TBtheoryofhbGNRZEM} for analytical considerations). $W$ and $C$ are the width and the circumference of the ribbon and tube, respectively.}
    \label{fig:EnergyBandsvsTube}
\end{figure}

\clearpage
{\bf Fig.~S2}
\begin{figure}[hbt!]
    \includegraphics[width=\textwidth]{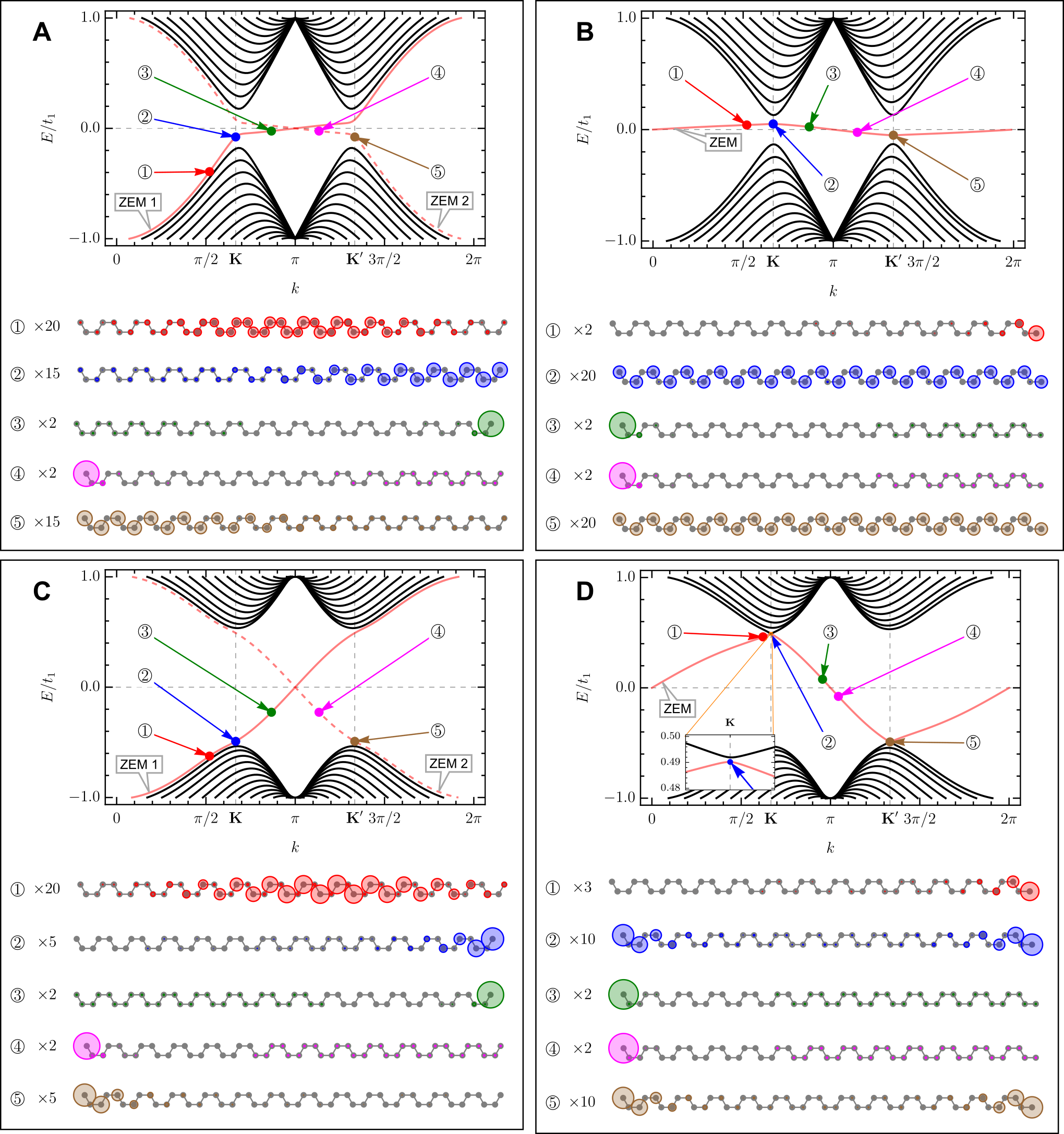}
    \caption{{\bf The $\mathbf{Z}$ topological edge states in commensurate and incommensurate structures.} ({\bf A}), ({\bf C}) Energy bands and electron densities of ZGNR $N = 52$ in Haldane model: $t_2 = 0.01 t_1$ and~$0.1 t_1$, respectively, $\phi = \pi/2$. The staggered sublattice potential is zero. ({\bf B}), ({\bf D}) Same as (A,C) but for hbGNR $N_r = 51$.}
    \label{fig:Haldane_model}
\end{figure}

\clearpage
{\bf Fig.~S3}
\begin{figure}[hbt!]
    \includegraphics[width=\textwidth]{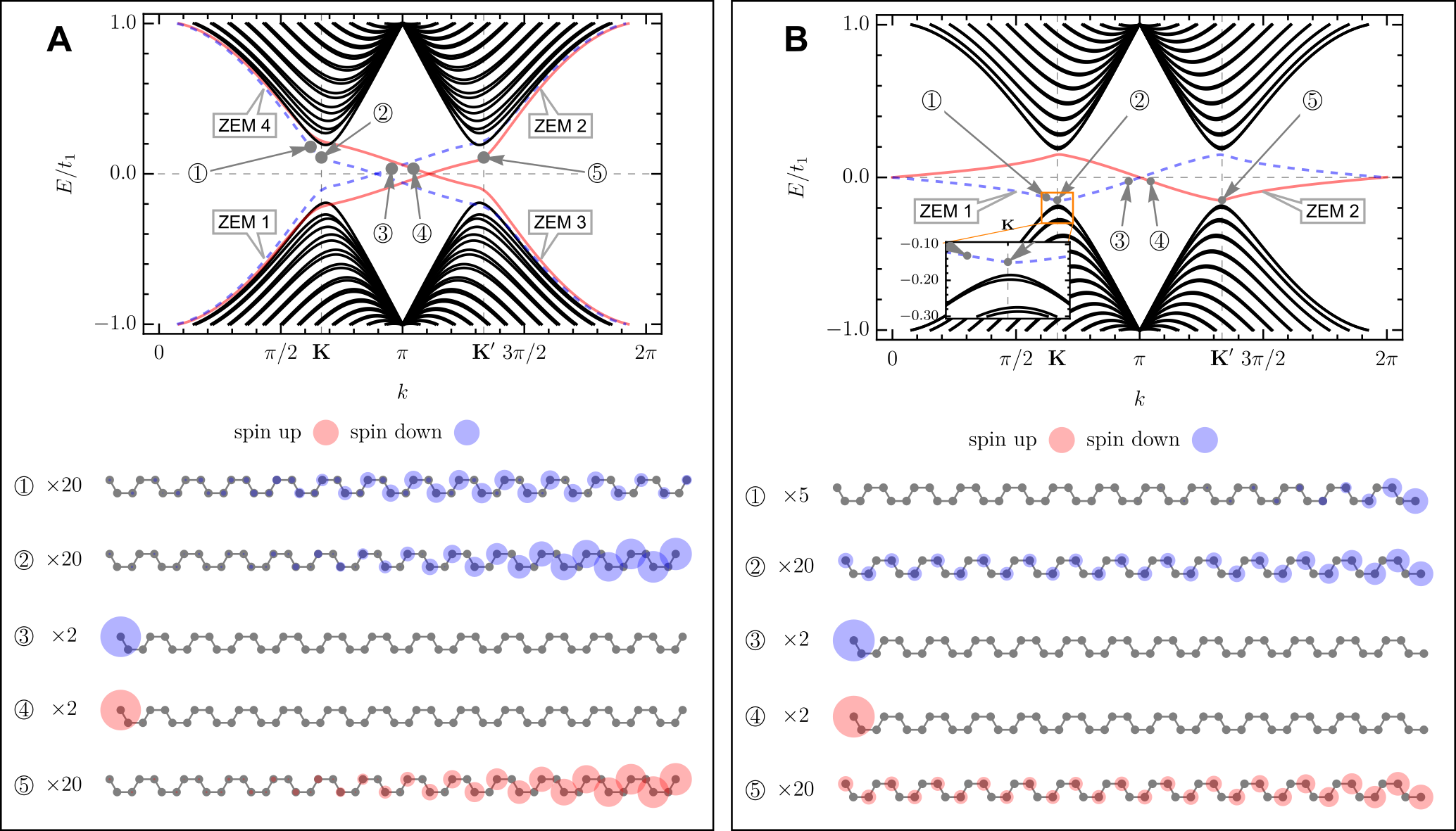}
    \caption{{\bf The $\mathbf{Z}_2$ topological edge states in commensurate and incommensurate ribbons.} ({\bf A}) Energy bands and electron spin polarization densities of ZGNR $N = 52$ in Kane-Mele model: $t_2 = 0.03 t_1$, Rashba spin-orbit coupling is $t_R = 0$, while staggered sublattice potential is $\Delta = 2 t_2$. ({\bf B}) Same as (A) but for hbGNR $N_r = 51$ and $\Delta = 0$.}
    \label{fig:KaneMele_model}
\end{figure}

\clearpage
{\bf Fig.~S4}
\begin{figure}[hbt!]
    \includegraphics[width=\textwidth]{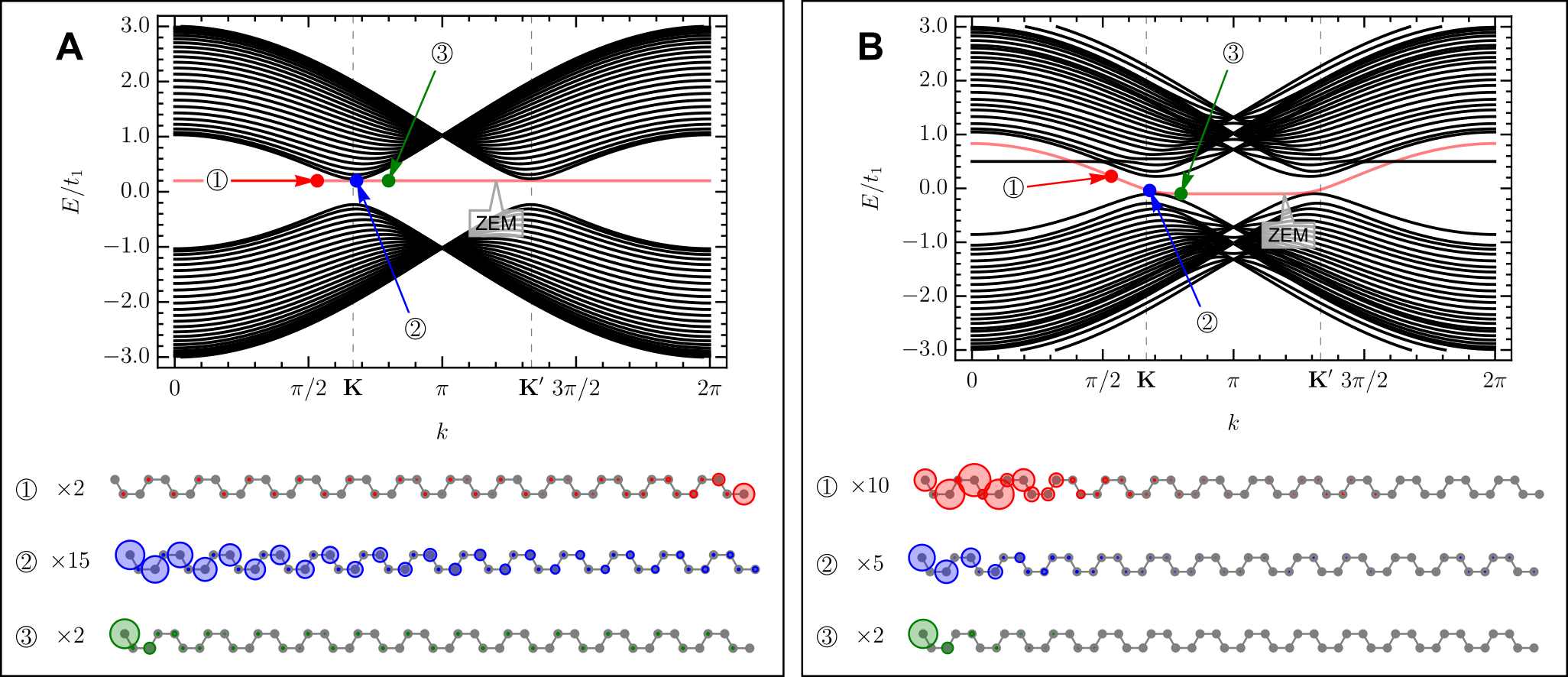}
    \caption{{\bf Topological kink states in incommensurate ribbons.} ({\bf A}) Energy bands and electron densities of hbGNR $N = 51$ with staggered potential being switched on: $\Delta = 0.2 t_1$. ({\bf B}) Same as (A) but for Liu's model of the QVH effect, see Ref.~\cite{Liu2021a}: $\Delta = 0.2 t_1$, $U_l = - U$, $U_r = U$, where $\Delta$ is the amplitude of the on-site staggered potential, $U = 0.3 t_1$ is the side back gate potential, $U_{l,(r)}$ are the left (right) side back gate potentials that is applied to the $12$ atoms on each side of the ribbon. The circles representing electron densities are scaled by a factor shown beside each picture. The ZEM is tracked by sorting the numerical eigenvalue according to the maximum of overlap for the eigenvectors at the neighboring $k$-points.}
    \label{fig:quantumvalleyHallLiu_model}
\end{figure}

\clearpage
{\bf Fig.~S5}
\begin{figure}[hbt!]
    \includegraphics[width=\textwidth]{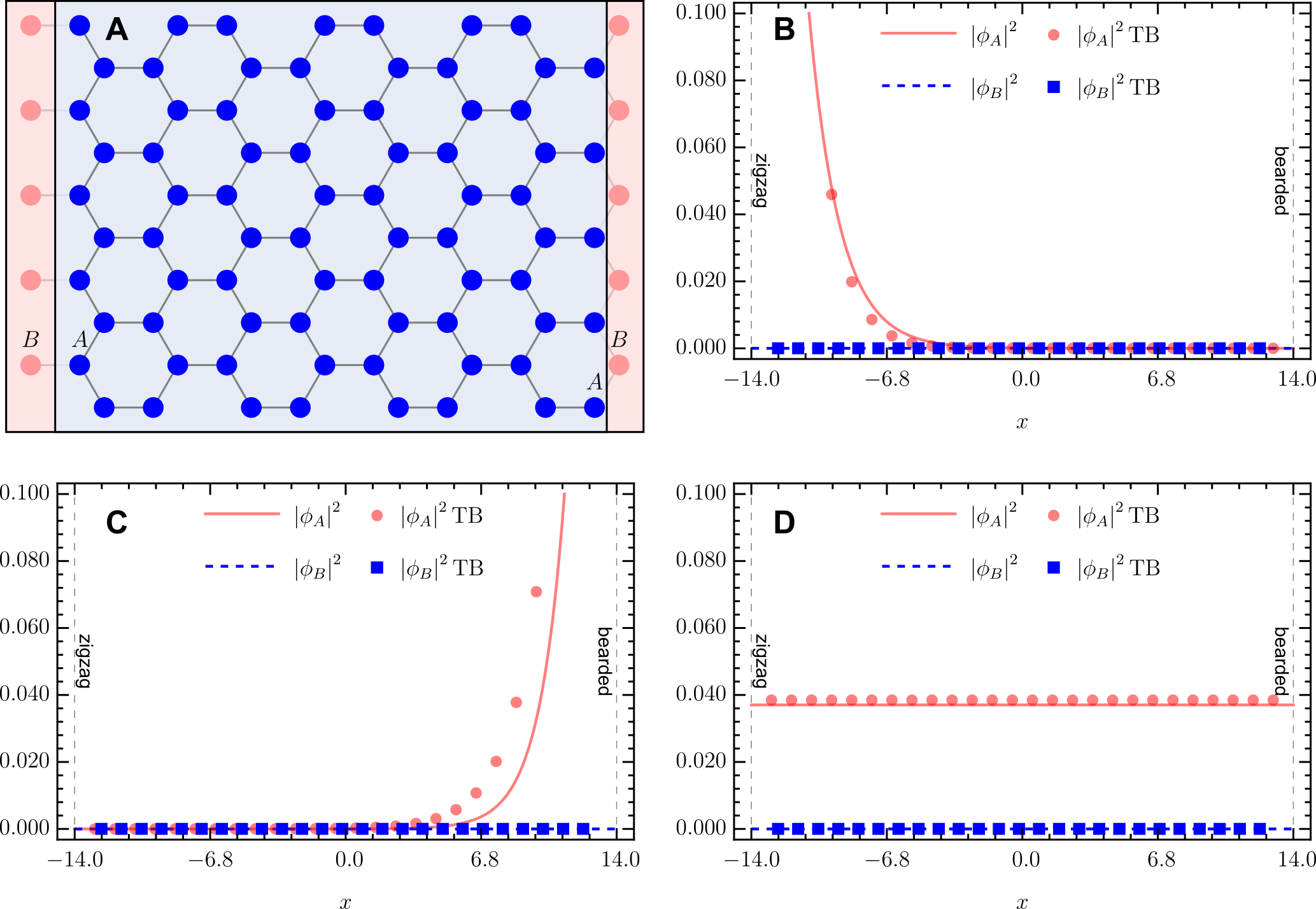}
    \caption{{\bf Zero-energy mode in the continuum model.} ({\bf A}) Schematics of the auxiliary lattice sites, where the electron wave function vanishes. The hbGNR atoms are depicted and highlighted in blue and light blue, respectively, while additional sites are shown and highlited in red and light red.  ({\bf B}) The electron density of the zero-energy mode in the continuum (lines) and tight-binding (markers) models for the point $\kappa_y > 0$, i.e. circled~3 in the main text Fig.~2A. hbGNR is positioned between $-L/2$ and $L/2$, where $L = (N_r + 1)/2 + 1$, with $N_r = 51$. The results for $A$ and $B$ sublattices are shown in red (solid line and points) and blue (dashed line and squares), respectively. ({\bf C}) Same as (B), but for the point $\kappa_y < 0$, i.e. point circled~1 in Fig.~2A. ({\bf D}) Same as (B) but for $\kappa_y = 0$ that is point circled~2 in Fig.~2A.}
    \label{fig:ContinuumModelJackiwRebbiSolution}
\end{figure}

\clearpage
{\bf Fig.~S6}
\begin{figure}[hbt!]
    \includegraphics[width=\textwidth]{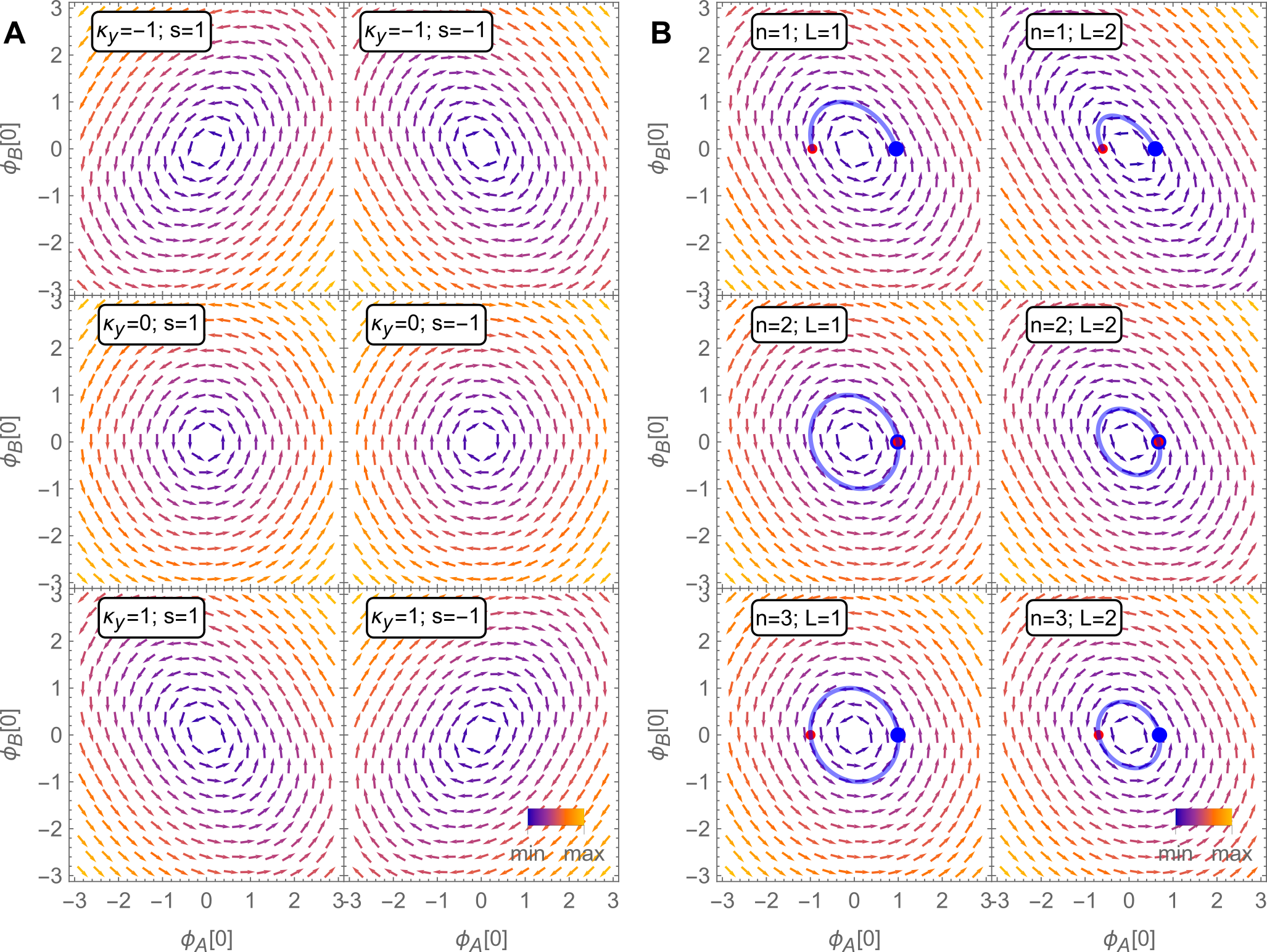}
    \caption{{\bf The phase portraits of a hbGNR in the continuum model.} ({\bf A}) The vector field for conduction, $s=1$, and valence, $s=-1$, bands and several representative values of $\kappa_y$; for all panels $n=1$ and $L=1$. ({\bf B}) The bulk eigenstate trajectories (thick blue) in the phase space for increasing $n$'s and $L$'s; for all panels $s=1$ and $\kappa_y = 1$. The large blue and medium red points are the starting and the finishing points of the trajectory, respectively.}    \label{fig:ContinuumModelBulkStatesPhasePortraitsAndTrajectories}
\end{figure}

\clearpage
{\bf Fig.~S7}
\begin{figure}[hbt!]
    \includegraphics[width=\textwidth]{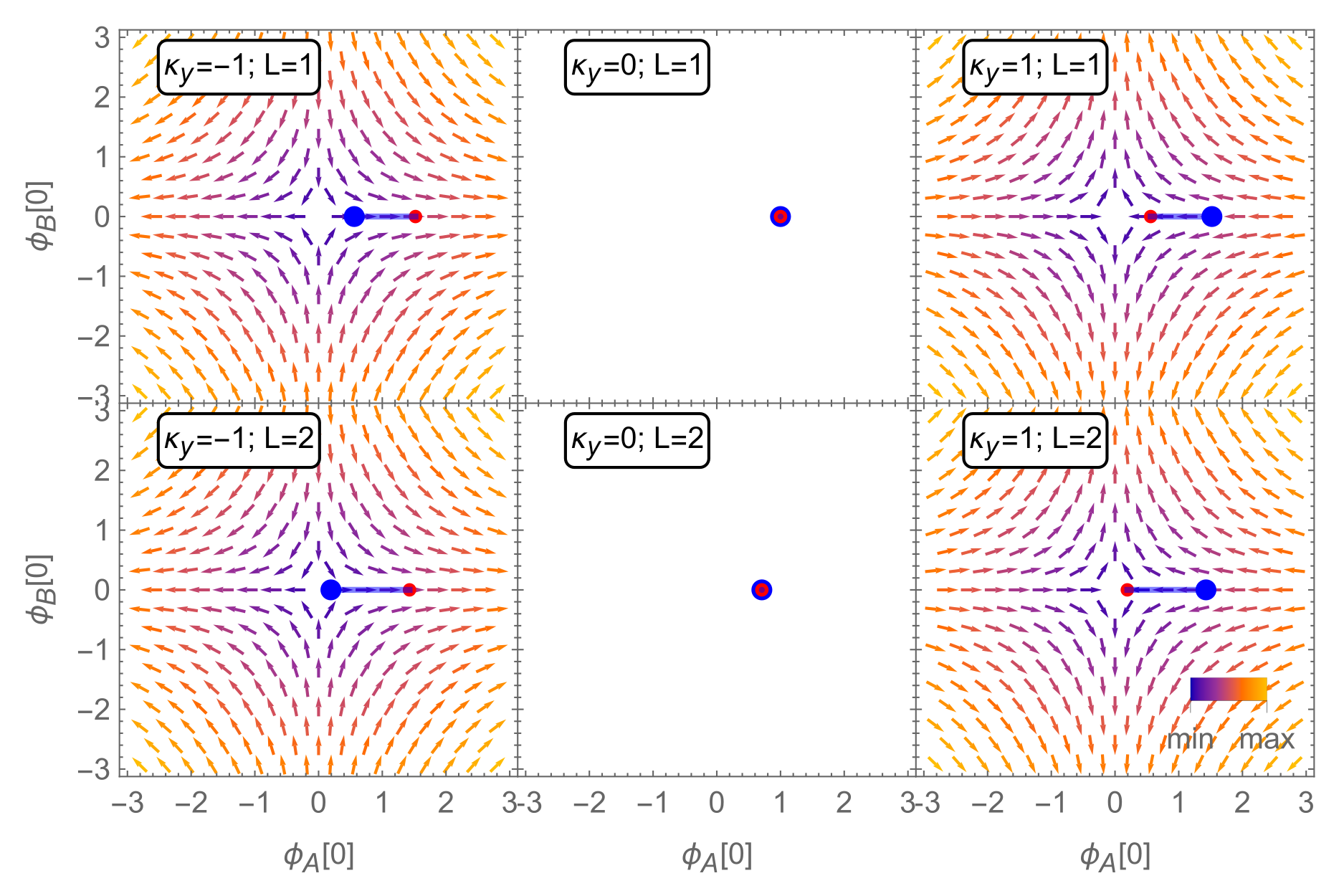}
    \caption{{\bf The phase portraits and ZEM trajectories of a hbGNR.} The trajectories are shown for main representative values of $\kappa_y$'s and $L$'s. The large blue and medium red points are the starting and the finishing points of the trajectory, respectively.}
    \label{fig:ContinuumModelZEMPhasePortraitsAndTrajectories}
\end{figure}

\clearpage
{\bf Fig.~S8}

\begin{figure}[hbt!]
    \includegraphics[width=\textwidth]{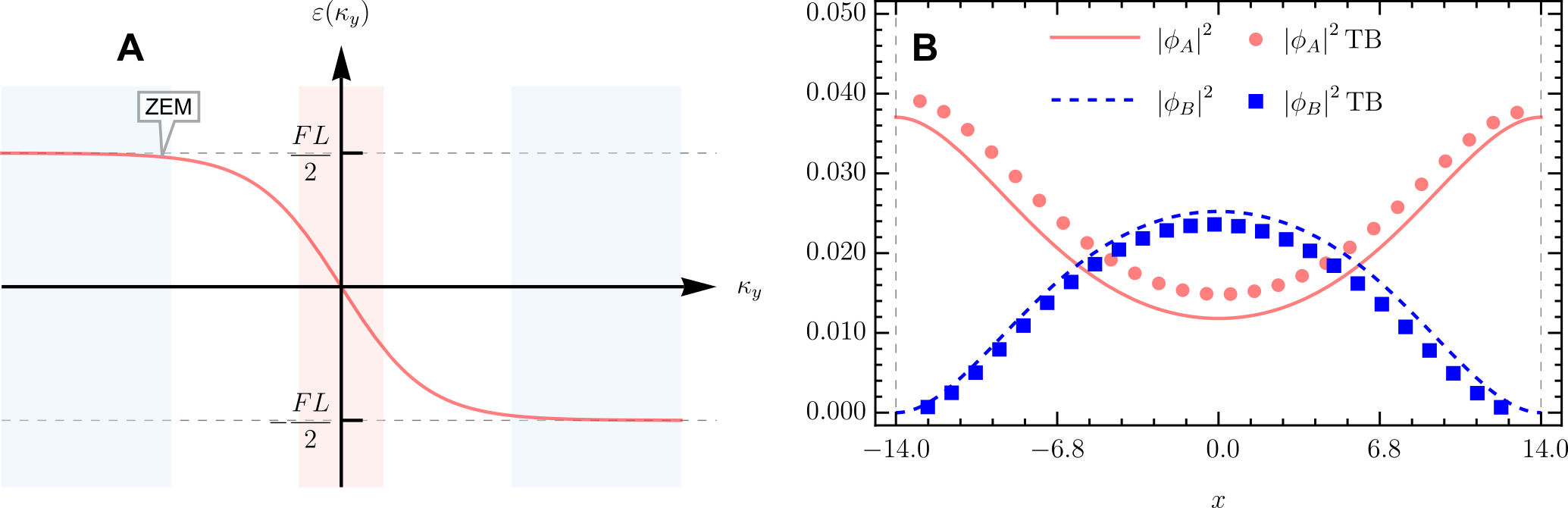}
    \caption{{\bf The in-gap dispersive mode in the continuum model} ({\bf A}) Schematics of the $k$-space regions studied for ZEM in the in-plane electric field: flat flanks (light blue), where $\kappa_y \neq 0$ and $\varepsilon = \pm F L/2$ and a zero-energy region (light red), where $\kappa_y = 0$ and $\varepsilon = 0$. ({\bf  B}) Electron densities for the zero-energy region (light red  $k$-region in (A)) for hbGNR positioned between $-L/2$ and $L/2$, where $L = \left[(N_r + 1)/2\right] +1$, with $N_r = 51$, and for the field $F = 0.005 \sqrt{3} a /2$ with $a = 2. 46$~\AA, applied to the in-plane ribbon. The lines and points show the continuum and tight-binding (TB) model results, respectively. The results for $A$ and $B$ sublattices are shown in red (solid line and points) and blue (dashed line and squares), respectively.}
    \label{fig:ContinuumModel}
\end{figure}

\clearpage
{\bf Fig.~S9}

\begin{figure}[hbt!]
    \includegraphics[width=\textwidth]{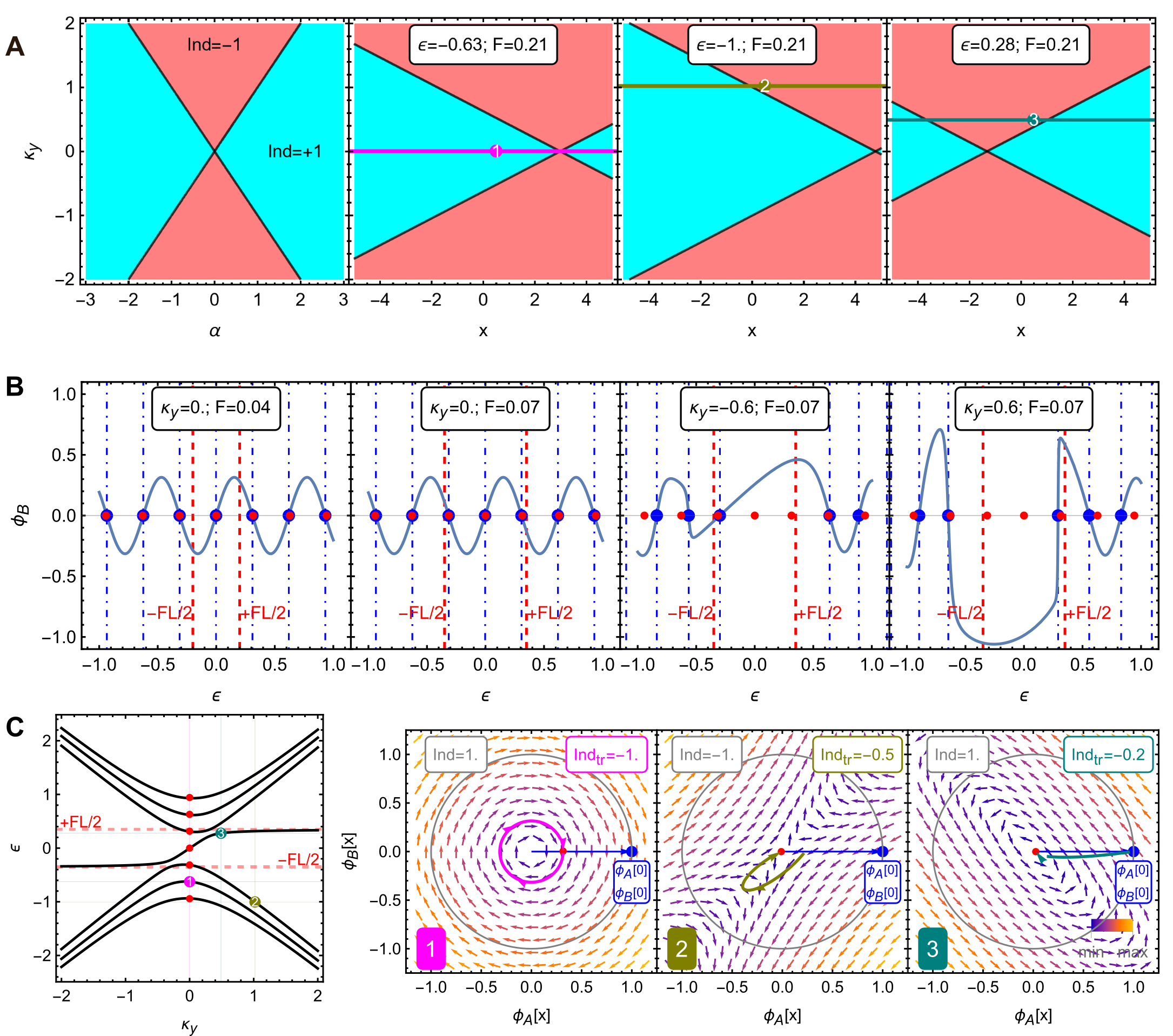}
    \caption{{\bf The hbGNR in the in-plane field as a dynamical system.} ({\bf A}) The index phase diagrams. ({\bf B}) The radius vector $\phi_B$-component plot showing graphical roots (large blue points), analytical roots at $\kappa_y=0$ (medium red points), flat flanks of ZEM (dashed red lines), and the iterative algorithm roots (dot dashed blue lines). ({\bf C}) The band structure and the eigenstate normalized trajectories: initial (large blue) and final (medium red) point of the non-normalized and normalized trajectories, respectively. The red features of the left side band structure are the same as in (B), $F=0.07$. For all panels $L=10$.}
    \label{fig:ContinuumModelNumericalSolution}
\end{figure}

\clearpage
{\bf Fig.~S10}
\begin{figure}[hbt!]
    \includegraphics[width=\textwidth]{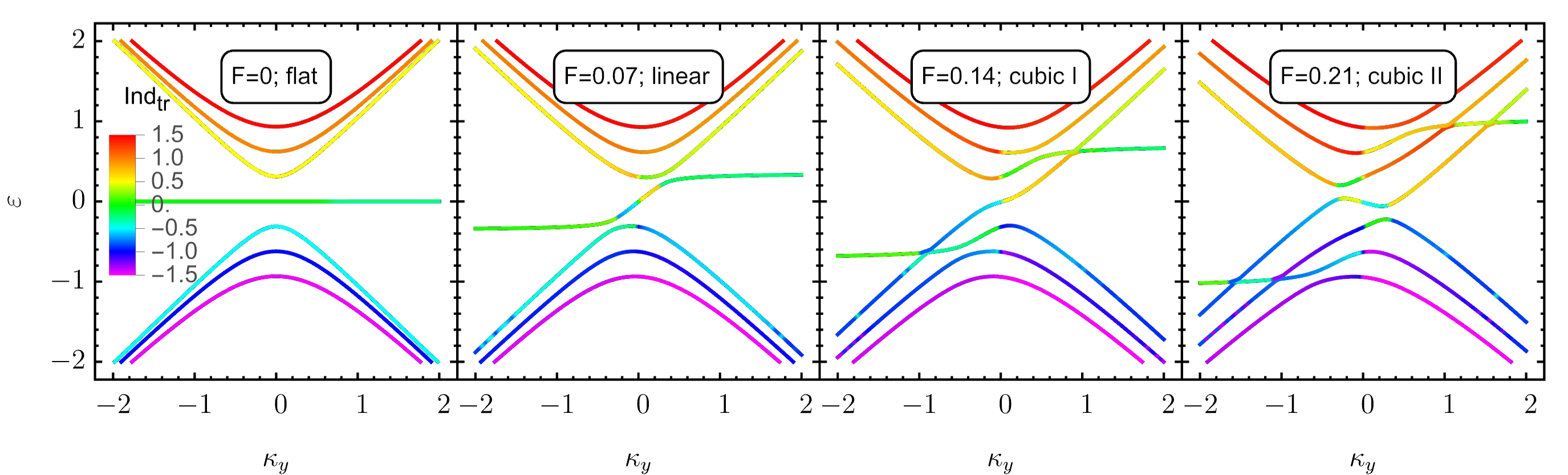}
    \caption{{\bf From flat band to cubic dispersion.} The hbGNR $L=10$ bands with the eigenstate trajectory index $\mathrm{Ind}_{\mathrm{tr}}$ as color scheme for several representative values of the in-plane electrostatic field.}
    \label{fig:ContinuumModelIndexAnalysis}
\end{figure}

\clearpage
{\bf Fig.~S11}

\begin{figure}[hbt!]
    \includegraphics[width=\textwidth]{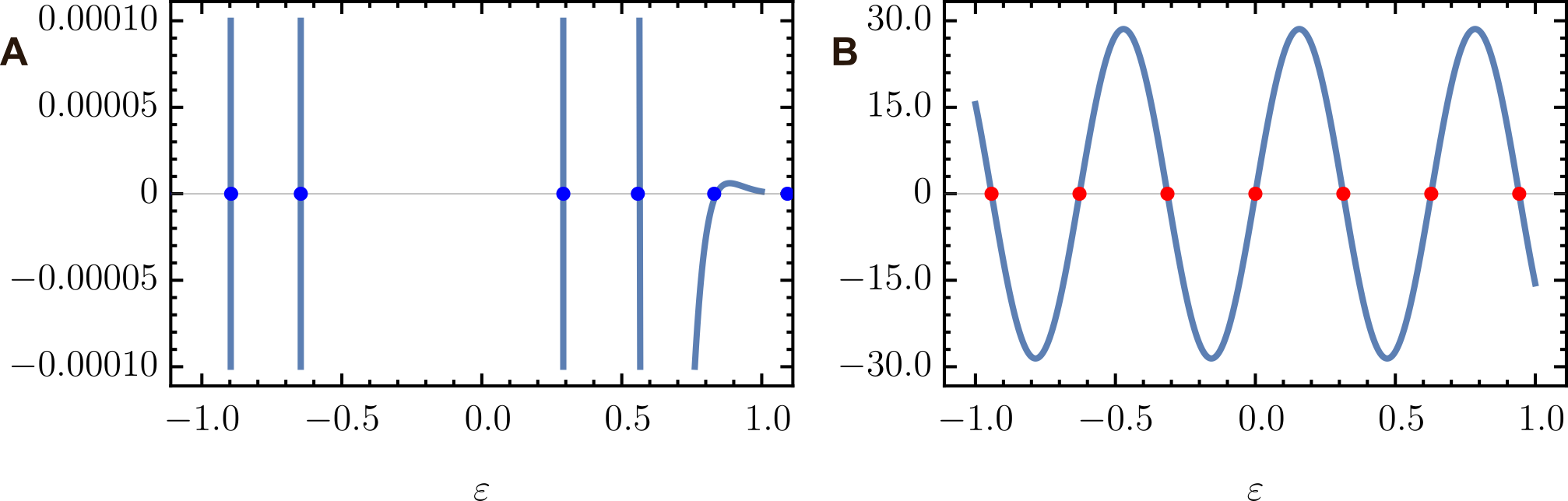}
    \caption{{\bf The hbGNR dispersion function zeros.} ({\bf A}) The real part of the regularized dispersion function given by Eq.~(\ref{eq:DiracEquation4GNRSingleComponentphiBEfieldDispEq2}) for $F=0.07$ and $\kappa_y=0.6$. The blue dots are numerical solutions of the numerical realization of the dispersion equation in supplementary text~\ref{app:IndexTheoryOfZEMinEfield}. ({\bf B}) Same as (A) but for $F=0.07$ and $\kappa_y=0$. The red dots are analytical solutions $\varepsilon = \pi n / L$ from supplementary text~\ref{app:LowEnergyTheoryDEZEMinEfield}. For both panels $L=10$. }
    \label{fig:DispersionFunctionFromExactSolution}
\end{figure}

\clearpage
{\bf Fig.~S12}

\begin{figure}[hbt!]
    \includegraphics[width=\textwidth]{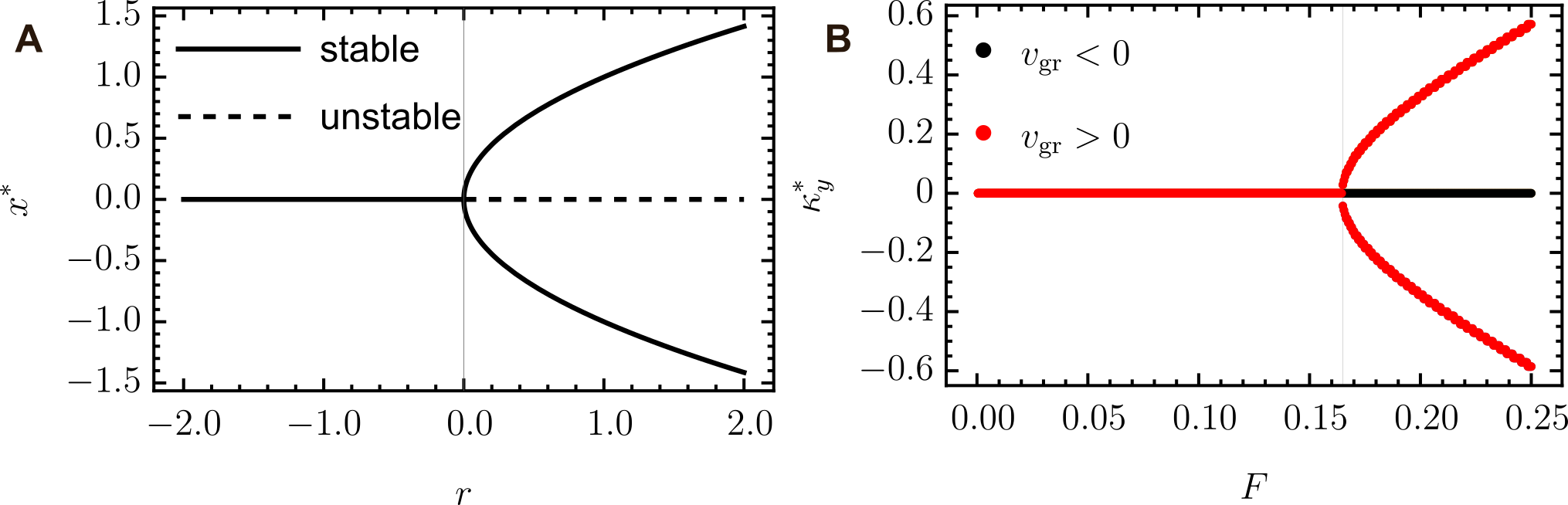}
    \caption{{\bf The bifurcation diagrams.} ({\bf A}) The standard pitchfork bifurcation diagram of a non-linear system described by Eq.~(\ref{eq:SupercriticalPitchfork}). ({\bf B}) The pitchfork bifurcation diagram of the ZEM for hbGNR with $L=10$. Gray vertical line marks the critical value of the bifurcation parameter in both cases.}
    \label{fig:PitchforkBifurcationDiagrams}
\end{figure}

\clearpage
{\bf Fig.~S13}

\begin{figure}[hbt!]
    \includegraphics[width=0.86\textwidth]{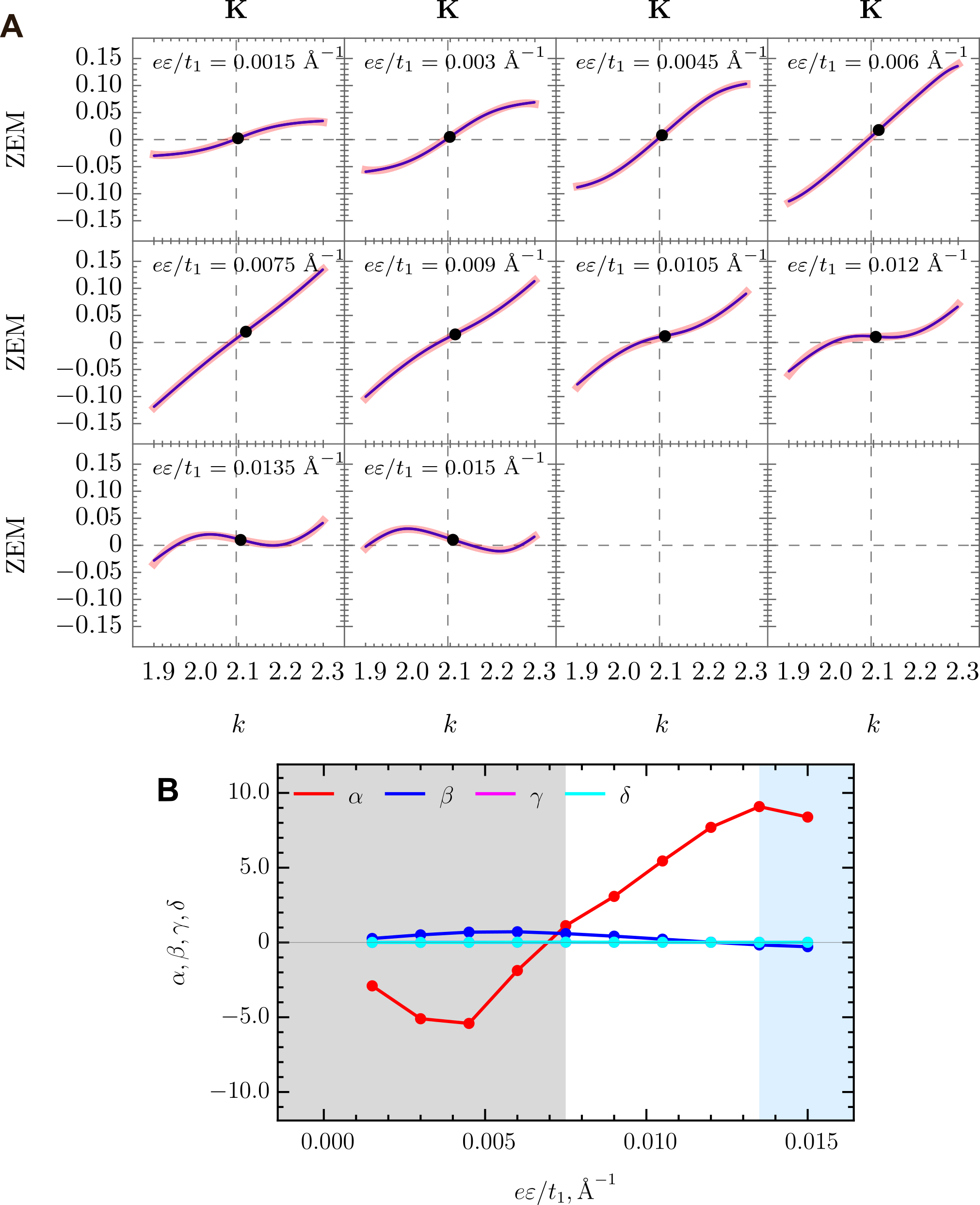}
     \caption{{\bf Fitting the ZEM in the in-plane electric field.} ({\bf A}) ZEM best fit with cubic funtional (light red) compared with the dense numerical data (blue). The large black points with coordinates $\left(\delta, \gamma\right)$ mark the centers of the cubic dispersions. ({\bf B}) Fitting parameters $\alpha$, $\beta$, $\gamma$ and $\delta$ as functions of the in-plane electric field. The lines joining the data points are only used as guide.}
    \label{fig:ZEMCubicFitting}
\end{figure}

\clearpage
{\bf Fig.~S14}

\begin{figure}[hbt!]
    \includegraphics[width=\textwidth]{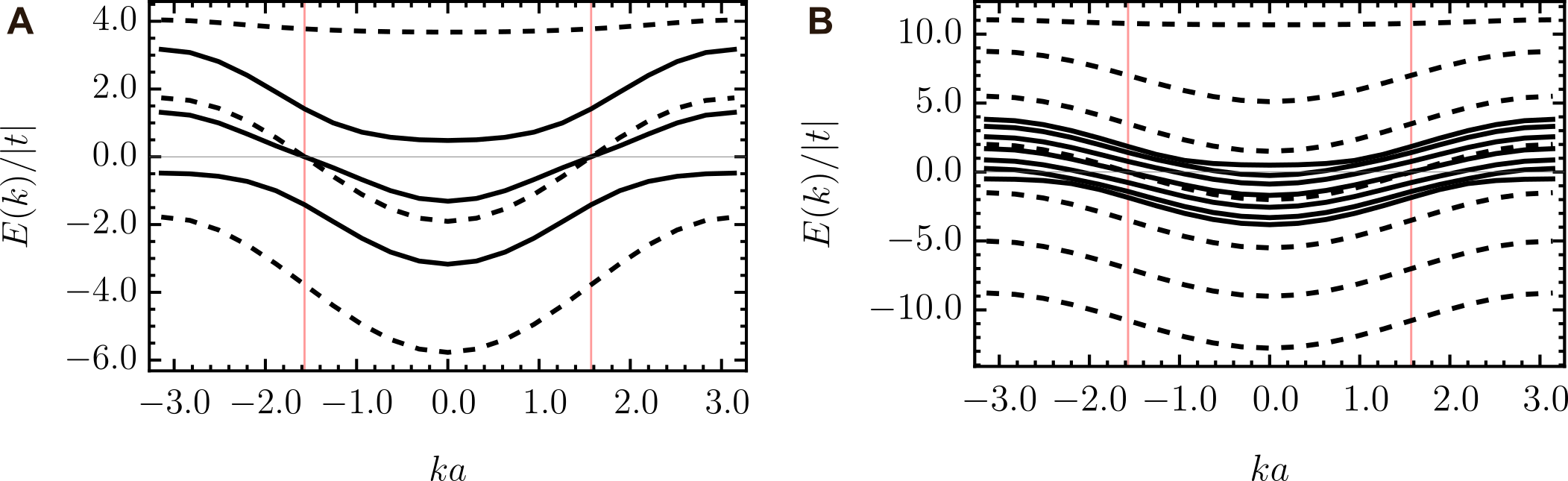}
    \caption{{\bf The energy bands of a square lattice ribbon with one bearded edge.} ({\bf A}) The narrow ribbon with 3~atoms in the unit cell for $F/\left|t\right| = 0$ (black solid) and $F/\left|t\right| = 3.5$ (black dashed). ({\bf B}) The wide ribbon with 7~atoms in the unit cell for $F/\left|t\right| = 0$ (black solid) and $F/\left|t\right| = 3.5$ (black dashed). Here, $t = -1$~eV and $a = 1$~\AA; vertical red lines, $k=\pm \pi/(2a)$, mark the position of $E=0$ left and right transmission channels in the $k$-space.}
\label{fig:SquareLatticeRibbonWithEdgeDefect}
\end{figure}

\clearpage
{\bf Table~S1}

\begin{table}[hbt!]
\centering
\begin{tabular}{|c | c c c c|} 
 \hline
 Length, $a$ &  ZGNR(20) & aSWCNT(11,11) & hbGNR(21) & cumulene \\ [0.5ex] 
 \hline
 1	&  0.972946471	&  1.997831506	&  0.974825803 &	  0.987584710 \\
11	&  0.725904803	&  1.967682615	&  0.758665051 &	  0.877350108 \\
21	&  0.557443835	&  1.937707105	&  0.606546981	&  0.770512279 \\
31	& 0.435032798	&  1.910696267	&  0.489729602	&  0.680317416 \\
41	&  0.332512924	&  1.880055611	&  0.386471792	&  0.614276322 \\
51	&  0.274852415	&  1.854659417	&  0.321465857	&  0.540750536 \\ 
\hline
MFP estimate, nm & 9.3 & 164.7& 10.8 & 20.4\\[0.5ex]
\hline
{\bf Ranking} & {\bf 4th} & {\bf 1st} & {\bf 3rd} & {\bf 2nd} \\[1ex] 
 \hline
\end{tabular}
\caption{{\bf Transmission as a function of length for aSWCNT, ZGNR, hbGNR and cumulenic carbyne.} The length of the scattering region is presented in terms of graphene lattice constant $a=2.46$~\AA. Each transmission value is averaged over $5000$ random impurity configurations. The simulation parameters are impurity strength $u_0 = 1$~\AA$^2$, impurity density $n_{\mathrm{im}}=0.1$~\AA$^2$, impurity scattering range $d=0.12$~\AA~(see methods). The transmission energy is taken to be $0.1t_1$ for aSWCNT(11,11) and ZGNR(20), while for hbGNR(21) and cumulenic carbyne it is set to $0.0t_1$, where $t_1$ is the nearest neighbor hopping integral. Note: hbGNR is subjected to an external in-plane electric field such that ZEM is converted into the dispersive in-gap mode shown in the main text Fig.~2D,E. MFP stands for the mean free path estimated by fitting the transmission data with $A \exp\left(-x/l\right)$ functional, where $A\approx2$ for tube and $1$ otherwise and $l$ is the MFP value.}
\label{tab:TransmissionVsScatteringRegionLength4CarbonStructures}
\end{table}

\clearpage
\setcounter{figure}{0}
\makeatletter
\renewcommand{\fnum@figure}{{\bf Movie \thefigure}}
\makeatother
{\bf Movie~S1}

\begin{figure}[hbt!]
    \caption{{\bf The dynamics of the Dirac equation, i.e. Eqs.~(\ref{eq:RealEquation4GNRInEfield1}) and~(\ref{eq:RealEquation4GNRinEfield2}), solutions in the turbulent vector field.} ({\bf A}) The Hilbert space vector field of a hbGNR in the in-plane external electric field and the Dirac equation solution trajectories. `Ind' is the index of the fixed point located at the origin. The large red dot: the initial condition $(1,1) = (\phi_A(0), \phi_B(0)])$. ({\bf B}) The dynamics of point solutions; Magenta solution is the one starting form large red point in (A). $80$ randomly chosen initial conditions in the interval $(-2,2)$ for both $\phi_A$ and $\phi_B$ are presented by black points. In both (A) and~(B) $\epsilon = 0$, $F=0.2$, $\kappa_y = 0.1$. $x$-coordinate across the ribbon width represents an effective time evolution parameter changing in the interval $(-5,5)$.}
    \label{mov:DynamicsOfDiracEquationSolutionsInVectorField}
\end{figure}

\end{document}